\documentclass[11pt]{article}

\usepackage[margin=1in]{geometry}
\usepackage{amsmath,amssymb,amsthm,mathtools}
\usepackage{booktabs,multirow,array}
\usepackage{tabularx}
\usepackage{enumitem}
\usepackage{comment}
\usepackage{microtype}
\usepackage{xcolor}
\usepackage[numbers,sort&compress]{natbib}
\usepackage[colorlinks=true,allcolors=blue]{hyperref}
\usepackage[nameinlink,noabbrev,capitalize]{cleveref}

\RemoveFromHook{label}[firstaid/cleveref]
\AddToHook{cmd/appendix/after}{\crefalias{section}{appendix}}

\newcommand{\C}{\mathbb C}

\newcommand{\cH}{\mathcal{H}}
\newcommand{\cP}{\mathcal{P}}
\newcommand{\R}{\mathbb R}
\newcommand{\E}{\mathbb E}
\newcommand{\1}{\mathbf 1}
\newcommand{\Gr}{\operatorname{Gr}}
\newcommand{\Tr}{\operatorname{Tr}}
\newcommand{\rank}{\operatorname{rank}}
\newcommand{\Aff}{\operatorname{Aff}}
\newcommand{\TV}{\operatorname{TV}}
\newcommand{\Bias}{\operatorname{Bias}}
\newcommand{\KL}{D_{\mathrm{KL}}}
\newcommand{\Ren}{D_{1/2}}

\newcommand{\Harm}{\mathcal H}
\newcommand{\Obs}{\mathsf{Obs}}

\newcommand{\eps}{\varepsilon}
\newcommand{\op}{\mathrm{op}}
\DeclarePairedDelimiter{\norm}{\lVert}{\rVert}
\newcommand{\dd}{\,\mathrm d}

\newcommand{\Herm}{\mathrm{Herm}}
\newcommand{\Ad}{\mathrm{Ad}}

\DeclareMathOperator{\diag}{diag}
\DeclareMathOperator{\Cov}{Cov}
\DeclareMathOperator*{\Var}{\mathbf{Var}}
\DeclareMathOperator{\Hess}{Hess}
\DeclareMathOperator{\Ric}{Ric}

\theoremstyle{plain}
\newtheorem{theorem}{Theorem}[section]
\newtheorem{lemma}[theorem]{Lemma}
\newtheorem{proposition}[theorem]{Proposition}
\newtheorem{corollary}[theorem]{Corollary}
\theoremstyle{definition}
\newtheorem{definition}[theorem]{Definition}

\theoremstyle{plain}
\newtheorem{remark}[theorem]{Remark}

\title{Optimal Lower Bound on the Sample Complexity of Coherent Classical Shadows
for Mixed States}
\author{Pavel Svetlichnyy}
\date{}

\begin{document}
\maketitle
{\renewcommand{\thefootnote}{}\footnotetext[0]{\texttt{psvetlichny7@gmail.com}}}

\begin{abstract}
We study coherent classical shadows protocols for an unknown
\(d\)-dimensional quantum state of rank at most \(r\), allowing an arbitrary
joint measurement on all available copies. For observables with operator norm
at most one and centered squared Frobenius norm at most \(F\),
we study the minumum number of copies \(T\) needed for the protocol
to successed with propbability \(1 - \delta\).
We prove a lower bound on \(T\), that depends both on \(\delta\) and 
the number \(M\) of post-measurement observables. Denoting
\(\ell=\max\{\log(2M),\log(1/\delta)\}\), we show that
\[
T=\Omega\!\left(\min\left\{\frac{\sqrt{rF\ell}}{\eps},\frac{(F\ell)^{2/3}}{\eps^{4/3}}\right\}+\frac{\ell}{\eps^2}\right)
\]
under the mild assumption \(\log(2M)\lesssim d\min\{r,F\}\).
The lower bound nearly matches the known upper bound and matches the
conjectural strengthening of the known upper bound. 

The proof is inspired by the framework of the proof of the lower bound 
for Boolean Hidden Matching.
We develop a Gaussian-lift noise kernel that transfers forward and 
reverse Gaussian hypercontractivity to the Grassmannian, allowing 
us to bound the distinguishing advantage in an auxiliary game with 
a revealed query observable.
We use Schur--Weyl structure and posterior moment bounds 
to control arbitrary coherent measurements on the source side.
\end{abstract}

\begingroup
\renewcommand{\contentsname}{\hfill Contents\hfill\mbox{}}
\tableofcontents
\endgroup
\clearpage

\section{Introduction}

A fundamental task in quantum learning is to predict properties of an unknown quantum
state or process. For mixed quantum states, the most straightforward and complete approach
is full state tomography, which produces a full classical description of the state from
which arbitrary properties can subsequently be inferred. However, full state tomography
incurs high sample complexity---learning a general \(d\)-dimensional mixed state to
trace-norm error \(\eps\) requires \(\Theta(d^2/\eps^2)\) samples even
with coherent measurements~\citep{10.1145/2897518.2897544,scharnhorst2025optimal}, while 
potentially extracting substantially more information than is required by a particular task.

Classical shadows offer a different paradigm: the state is measured to produce a classical
sketch, which can later be used to estimate expectation values of observables that were not
specified at the time of measurement~\citep{huang2020predicting,Grier_2024}. In this work 
we consider classical shadows in the fully coherent setting, where all available copies of 
the unknown state may be measured jointly by an arbitrary POVM. Although such collective 
measurements can be experimentally demanding, fully coherent protocols provide the strongest 
possible measurement model and therefore form a natural benchmark for the sample 
complexity of more restricted protocols.

Recent works have substantially improved coherent classical shadows protocols for pure and
low-rank mixed states. Let
\[
    \ell
    :=
    \max\left\{
        \log(2M),
        \log\frac1\delta
    \right\},
\]
where \(M\) is the number of observables and \(\delta\) is the allowed failure
probability. For pure states, the observable-concentration results of
\citet{pelecanos2026keylwerner}, combined with the corresponding shadow estimator, give
the upper bound
\[
    T
    =
    O\!\left(
        \frac{\sqrt{F\ell}}{\eps}
        +
        \frac{\ell}{\eps^2}
    \right),
\]
improving the previous result of \cite{Grier_2024}.
For pure states and a single observable at constant confidence,
\citet[Theorem~16]{Grier_2024} established the lower bound
\[
T=\Omega\!\left(
\frac{\sqrt F}{\eps\log(1+F)}
+\frac{1}{\eps^2}
\right)
\]
for \(F\leq\eps d\), under the stronger, uncentered Frobenius-norm constraint.

For states of rank at most \(r\), the estimators developed in
\citep{pelecanos2025mixedstatetomographyreduces,pelecanos2025debiased,
pelecanos2026keylwerner}
provide the upper bound
\[
    T
    =
    O\!\left(
        \min\left\{
            \frac{\sqrt{rF\ell}}{\eps},
            \frac{F^{2/3}\ell}{\eps^{4/3}}
        \right\}
        +
        \frac{\ell}{\eps^2}
    \right).
\]
Motivated by the open concentration extension discussed in
\citep[Footnote~6]{pelecanos2026keylwerner}, we conjecture
that the term \(\frac{F^{2/3}\ell}{\eps^{4/3}}\) may
be improved to \(\frac{(F\ell)^{2/3}}{\eps^{4/3}}\).
Prior to the present work, a matching rank-dependent lower bound 
for mixed states was not known.

In this work we resolve this question.
For states of rank at most
\(r\), and observables \(O\) satisfying
\[
    \|O\|_{\op}\leq 1,
    \qquad
    \|O_0\|_F^2\leq F,
    \qquad
    O_0:=O-\frac{\Tr O}{d}I_d,
\]
we prove that any coherent classical shadows protocol requires
\[
    T
    =
    \Omega\!\left(
        \min\left\{
            \frac{\sqrt{rF\ell}}{\eps},
            \frac{(F\ell)^{2/3}}{\eps^{4/3}}
        \right\}
        +
        \frac{\ell}{\eps^2}
    \right)
\]
copies, provided
\[
    \log(2M)
    \lesssim
    d\min\{r,F\}.
\]
For larger lists, the same argument gives the corresponding bound with the
\(\log M\) dependence capped at this scale (see \cref{app:large-M}).

Our lower bound matches the conjectural strengthening of the upper bound
throughout the parameter regime considered here.
For a single observable at constant confidence, it matches the best
known upper bound up to universal constants. For pure states, our result 
removes the factor \(\log(1+F)\) from the previous single-observable lower bound of
\citet{Grier_2024}, thereby matching the corresponding upper bound.

At a high level, the architecture of our proof resembles that of the one-way communication
lower bound for the Boolean Hidden Matching (BHM) problem%
~\citep{kerenidis2006onewaycommunicationcomplexityboolean,GavinskyKKRW2007}.
We reduce coherent classical shadows estimation to a distinguishing game in which the
measurement is performed before a correlated observable is revealed. The resulting bias is
controlled by combining two complementary ingredients: a source-side bound on how much
information an arbitrary coherent measurement can extract from the planted state family,
and a query-side contraction describing how strongly that information can correlate with
the subsequently revealed observable. The latter is implemented through a Gaussian-lift 
noise operator that we develop, which acts diagonally on harmonic components with 
rapidly decaying multipliers. The source side is controlled via
Schur--Weyl structure and uniform posterior moment bounds.
High-confidence lower bounds are obtained through affinity amplification. The
multiple-observable extension is formulated as a hidden-location problem and analyzed
using information-theoretic arguments.

The rest of the paper is organized as follows. In \cref{sec:notation} we introduce notation
and conventions. In \cref{sec:main-results} we formally state the problem and our main
theorem, and compare our lower bound with the known and conjectured upper bounds.
In \cref{sec:related-works} we briefly review related work. In \cref{sec:overview} we give
a technical overview of the proof and isolate its main ingredients. In
\cref{sec:discussion} we discuss the remaining gap and directions for further work.
The subsequent sections contain the proofs, with auxiliary arguments deferred to the
appendices.

\section{Notation}\label{sec:notation}

For integers \(0\leq k\leq n\), let
\[
    \Gr(k,n)
    :=
    \{V\subseteq\C^n : V \text{ is a linear subspace with } \dim V=k\}
\]
denote the complex Grassmannian of \(k\)-dimensional subspaces of \(\C^n\).
A point \(V\in\Gr(k,n)\) specifies only the subspace, with no choice or
ordering of basis vectors.  Equivalently, fixing
\[
    V_0=\operatorname{span}\{e_1,\ldots,e_k\},
\]
the transitive action of \(U(n)\) gives the identification
\[
    \Gr(k,n)
    \cong
    \frac{U(n)}{U(k)\times U(n-k)},
    \qquad
    U\longmapsto UV_0.
\]
Here \(U(k)\times U(n-k)\) is the stabilizer of \(V_0\); in particular, the
\(U(k)\) factor quotients out the choice of an orthonormal basis within the
subspace. This is the standard compact symmetric-space realization
\citep{HelgasonGGA}.

Let \((X,\mu)\) be a measure space and let \(h:X\to\C\) be measurable.  For
\(1\leq p<\infty\), we write
\[
    \norm{h}_{L^p(\mu)}
    :=
    \left(\int_X |h(x)|^p\dd\mu(x)\right)^{1/p},
    \qquad
    \norm{h}_{L^\infty(\mu)}
    :=
    \operatorname*{ess\,sup}_{x\in X}|h(x)|.
\]
For \(f,g\in L^2(\mu)\), we use the inner product
\[
    \langle f,g\rangle_{L^2(\mu)}
    :=
    \int_X \overline{f(x)}\,g(x)\dd\mu(x).
\]
We keep the underlying measure explicit in the subscript whenever such norms
or inner products are used, with the sole shorthand
\[
    \norm{h}_2:=\norm{h}_{L^2(\nu)},
    \qquad
    \langle f,g\rangle_\nu
    :=
    \int \overline{f}\,g\dd\nu.
\]
Unless stated otherwise, \(\nu\) denotes the normalized invariant probability
measure on the Grassmannian under consideration.

When a formula contains \(L^2\)-norms with different underlying measures,
we write every such norm with its explicit \(L^2\) measure subscript.

Throughout the paper, the logarithms are natural.
\section{Setup and main results}\label{sec:main-results}

For a Hermitian matrix \(O\in\C^{d\times d}\), define its traceless part
\[
    O_0:=O-\frac{\Tr O}{d}I_d
\]
and the centered Frobenius class
\[
    \Obs_d(F)
    :=
    \left\{
        O=O^\dagger:
        \|O\|_{\op}\leq 1,\ 
        \|O_0\|_F^2\leq F
    \right\}.
\]

\begin{definition}[Coherent classical shadows]\label{def:protocol}
Fix integers \(d,r,M,T\geq 1\), a Frobenius budget \(F>0\), an accuracy
parameter \(\eps>0\), and a failure probability \(0<\delta<1\).
A \(T\)-copy coherent classical shadows protocol receives
\(\rho^{\otimes T}\), where \(\rho\) is a density operator on \(\C^d\)
with \(\rank(\rho)\leq r\).  The 
protocol performs an arbitrary POVM \(\{M_z\}_z\) on \(\rho^{\otimes T}\) and records 
the classical outcome \(z\). After obtaining \(z\), the protocol receives a list of 
observables \(O_1,\ldots,O_M\in\Obs_d(F)\).
The protocol outputs estimates \(\widehat o_1,\ldots,\widehat o_M\).
It succeeds if, for every state \(\rho\) with \(\rank(\rho)\leq r\) and every list 
\(O_1,\ldots,O_M\in\Obs_d(F)\),
\[
 \Pr_\rho\!\left[
   \max_{1\leq i\leq M}|\widehat o_i-\Tr(O_i\rho)|\leq\eps
 \right]\geq 1-\delta.
\]
The POVM \(\{M_z\}_z\) may depend on the parameters \(d,r,F,M,\eps,\delta\), 
but not on the future list of observables.
\end{definition}

Denote
\begin{equation}\label{eq:def-ell}
    \ell
    :=
    \max\left\{
        \log(2M),
        \log\frac{1}{\delta}
    \right\}.
\end{equation}
In particular, \(\ell=\Theta(1)\) for a single observable at constant confidence.

For a sufficiently small universal constant \(c_M>0\), we call the
list size \emph{moderate} when
\begin{equation}\label{eq:moderate-M}
    \log(2M)
    \leq
    c_M d\min\{r,F\}.
\end{equation}
This parameter range already allows exponentially many future observables.  
The restriction can be removed at the price of capping the logarithmic
dependence (see \cref{app:large-M}).

The main result of this paper is the following theorem.
\begin{theorem}[Main lower bound]\label{thm:master}
There are universal constants \(c,c_\eps,c_M>0\) such that the following
holds.  Let \(d\ge2\), \(1\leq r\leq d\), \(1\leq F\leq d\), \(M\ge1\),
\(0<\delta\le1/3\), and \(0<\eps\leq c_\eps\).  If
\eqref{eq:moderate-M} holds, then any 
coherent classical shadows protocol that succeeds for 
all \(d\)-dimensional states of rank at most \(r\)
with accuracy \(\eps\) and failure probability \(\delta\) 
requires
\begin{equation}\label{eq:master-lb}
    T
    \geq
    c\left[
        \min\left\{
            \frac{\sqrt{rF\ell}}{\eps},
            \frac{(F\ell)^{2/3}}{\eps^{4/3}}
        \right\}
        +
        \frac{\ell}{\eps^2}
    \right].
\end{equation}
\end{theorem}
Note that the hard instances used in the proof may 
have rank strictly smaller than \(r\).

The lower bound has two conceptually different contributions.
The term \(\ell/\eps^2\) is the precision term. It follows from
standard binary and multiple-hypothesis constructions of the same
type used in \citep{Grier_2024}. Our main contribution is the geometric term
\[
    \min\left\{
        \frac{\sqrt{rF\ell}}{\eps},
        \frac{(F\ell)^{2/3}}{\eps^{4/3}}
    \right\}.
\] 
It is rank-sensitive until
\[
    r
    \asymp
    \left(\frac{F\ell}{\eps^2}\right)^{1/3},
\]
after which it saturates at \((F\ell)^{2/3}/\eps^{4/3}\).

For a single observable and constant confidence, our result simplifies to:
\begin{corollary}[Single observable, constant confidence]\label{cor:single}
For \(M=1\) and \(\delta=1/3\), every protocol in
\cref{def:protocol} requires
\begin{equation}\label{eq:single-lb}
    T
    =
    \Omega\!\left(
        \min\left\{
            \frac{\sqrt{rF}}{\eps},
            \frac{F^{2/3}}{\eps^{4/3}}
        \right\}
        +
        \frac{1}{\eps^2}
    \right).
\end{equation}
In particular, for pure states,
\begin{equation}\label{eq:single-pure-lb}
    T
    =
    \Omega\!\left(
        \frac{\sqrt F}{\eps}
        +
        \frac{1}{\eps^2}
    \right).
\end{equation}
\end{corollary}

\paragraph{Comparison with upper bounds.}
Combining the classical shadows application of the debiased Keyl
estimator~\citep[Theorem~1.9 and Lemma~7.3]{pelecanos2025debiased}
with the observable-concentration bound for the random-purification
estimator~\citep[Proposition~4.7]{pelecanos2026keylwerner} gives
\footnote{%
The rank-sensitive term follows from
\citep[Proposition~4.7]{pelecanos2026keylwerner}.
The saturated term follows by applying median-based confidence
amplification to
\citep[Theorem~1.9 and Lemma~7.3]{pelecanos2025debiased}.
Both results extend to our centered Frobenius constraint by
applying them to \(O_0/2\) and adding back \(\Tr O/d\).
}
\begin{equation}\label{eq:combined-shadow-ub}
    T
    =
    O\!\left(
        \min\left\{
            \frac{\sqrt{rF\ell}}{\eps},
            \frac{F^{2/3}\ell}{\eps^{4/3}}
        \right\}
        +
        \frac{\ell}{\eps^2}
    \right).
\end{equation}

Hence our lower bound is optimal up to universal
constants for constant confidence and a single observable.  For general
\(\ell\), it is also optimal whenever the rank-sensitive geometric branch
is active in the lower bound, namely
\(r\lesssim(F\ell/\eps^2)^{1/3}\). In this regime the upper bound has
the same rank-sensitive rate, up to universal constants.

The only remaining discrepancy is a factor
\(\ell^{1/3}\) in the saturated geometric branch.
We conjecture, motivated by the concentration extension left open in
\citep[Footnote~6]{pelecanos2026keylwerner}, that the
saturated upper-bound term can be improved from
\[
    \frac{F^{2/3}\ell}{\eps^{4/3}}
    \quad\text{to}\quad
    \frac{(F\ell)^{2/3}}{\eps^{4/3}},
\]
which would match our lower bound throughout the moderate-\(M\)
regime.

\begin{table}[tp]
\centering
\footnotesize
\setlength{\tabcolsep}{4pt}
\renewcommand{\arraystretch}{1.4}
\renewcommand{\tabularxcolumn}[1]{m{#1}}
\begin{tabularx}{\textwidth}{
    @{}
    >{\centering\arraybackslash}m{2.0cm}
    >{\centering\arraybackslash}m{2.55cm}
    >{\centering\arraybackslash}X
    >{\centering\arraybackslash}X
    @{}
}
\toprule
States & Regime & Upper bound & Lower bound \\
\midrule

Pure & \multicolumn{3}{c@{}}{%
\begin{tabularx}{\dimexpr\textwidth-2.0cm-2\tabcolsep\relax}{
    @{}>{\centering\arraybackslash}m{2.55cm}
    >{\centering\arraybackslash}X
    >{\centering\arraybackslash}X@{}
}
Moderate \(M\)\textsuperscript{\(\ddagger\)}, \(0<\delta\le1/3\)
&
\(\displaystyle
O\!\left(
    \frac{\sqrt{F\ell}}{\eps}
    +\frac{\ell}{\eps^2}
\right)
\)
\par\smallskip
\citep{pelecanos2026keylwerner}
&
\(\displaystyle
\Omega\!\left(
    \frac{\sqrt F}{\eps\log(1+F)}
    +\frac{\log(1/\delta)}{\eps^2}
\right)
\)
\par\smallskip
\citep{Grier_2024}\textsuperscript{\(\dagger\)}
\par\medskip
\(\displaystyle
\Omega\!\left(
    \frac{\sqrt{F\ell}}{\eps}
    +\frac{\ell}{\eps^2}
\right)
\)
\par\smallskip
\cref{thm:master}
\\

\cmidrule{1-3}
\(M=1,\ \delta=1/3\)
&
\(\displaystyle
O\!\left(
    \frac{\sqrt F}{\eps}
    +\frac1{\eps^2}
\right)
\)
\par\smallskip
\citep{Grier_2024}
&
\(\displaystyle
\Omega\!\left(
    \frac{\sqrt F}{\eps\log(1+F)}
    +\frac1{\eps^2}
\right)
\)
\par\smallskip
\citep{Grier_2024}\textsuperscript{\(\dagger\)}
\par\medskip
\(\displaystyle
\Omega\!\left(
    \frac{\sqrt F}{\eps}
    +\frac1{\eps^2}
\right)
\)
\par\smallskip
\cref{cor:single}
\\
\end{tabularx}}
\\

\midrule

Mixed & \multicolumn{3}{c@{}}{%
\begin{tabularx}{\dimexpr\textwidth-2.0cm-2\tabcolsep\relax}{
    @{}>{\centering\arraybackslash}m{2.55cm}
    >{\centering\arraybackslash}X
    >{\centering\arraybackslash}X@{}
}
Moderate \(M\)\textsuperscript{\(\ddagger\)}, \(0<\delta\le1/3\)
&
\emph{Known:}
\[
O\!\left(
    \min\left\{
        \frac{\sqrt{rF\ell}}{\eps},
        \frac{F^{2/3}\ell}{\eps^{4/3}}
    \right\}
    +\frac{\ell}{\eps^2}
\right)
\]
\citep{pelecanos2025debiased,pelecanos2026keylwerner}
\par\smallskip
\emph{Conjectured:}
\[
O\!\left(
    \min\left\{
        \frac{\sqrt{rF\ell}}{\eps},
        \frac{(F\ell)^{2/3}}{\eps^{4/3}}
    \right\}
    +\frac{\ell}{\eps^2}
\right)
\]
&
\(\displaystyle
\Omega\!\left(
    \min\left\{
        \frac{\sqrt{rF\ell}}{\eps},
        \frac{(F\ell)^{2/3}}{\eps^{4/3}}
    \right\}
    +\frac{\ell}{\eps^2}
\right)
\)
\par\smallskip
\cref{thm:master}
\\

\cmidrule{1-3}
\(M=1,\ \delta=1/3\)
&
\(\displaystyle
O\!\left(
    \min\left\{
        \frac{\sqrt{rF}}{\eps},
        \frac{F^{2/3}}{\eps^{4/3}}
    \right\}
    +\frac1{\eps^2}
\right)
\)
\par\smallskip
\citep{pelecanos2025debiased}
&
\(\displaystyle
\Omega\!\left(
    \min\left\{
        \frac{\sqrt{rF}}{\eps},
        \frac{F^{2/3}}{\eps^{4/3}}
    \right\}
    +\frac1{\eps^2}
\right)
\)
\par\smallskip
\cref{cor:single}
\\
\end{tabularx}}
\\

\bottomrule
\end{tabularx}

\par\smallskip
\begin{minipage}{\textwidth}
\footnotesize
\raggedright
\textsuperscript{\(\dagger\)}
\citet{Grier_2024}
assumes \(B\leq\eps d\), where \(B\) bounds \(\Tr(O^2)\): \(\Tr(O^2) \le B\). 
\par
\textsuperscript{\(\ddagger\)}
The general-\(M\) rows assume the moderate-list condition
\eqref{eq:moderate-M}. See \cref{app:large-M} for arbitrary \(M\).
\end{minipage}
\caption{
Sample complexity of coherent classical shadows for states of rank at
most \(r\). The conjectured entry is the improved saturated mixed-state
upper bound conjectured here.
}
\label{tab:comparison}
\end{table}

For pure states, the lower bound \eqref{eq:single-pure-lb} 
removes the factor \(\log(1+F)\) in the previous single-observable lower bound of
\citet{Grier_2024} and matches the corresponding upper bound.
\section{Related work}\label{sec:related-works}

\paragraph{Classical shadows.}
Classical shadows were introduced by Huang, Kueng, and Preskill~\citep{huang2020predicting} 
for predicting properties of an unknown quantum state as an alternative 
to full state tomography. Subsequent work improved measurement
efficiency by using locally biased and derandomized 
schemes~\citep{hadfield2022locallybiased,huang2021derandomization}, 
locally scrambled dynamics, and shallow 
circuits~\citep{hu2023locallyscrambled,bertoni2024shallow,ippoliti2023operator}. 
Classical shadows were extended to fermionic and continuous-variable 
systems~\citep{zhao2021fermionic,wan2023matchgate,west2026particle,becker2024continuous}, 
gauge symmetries and more general group 
actions~\citep{bringewatt2025gauge,west2026representations,chang2026symmetric}, 
measurement noise~\citep{koh2022noise,chen2021robust,brieger2025stability,farias2025robust}, 
and non-i.i.d. sources~\citep{zambrano2026noniid}. Other variants of the protocol 
estimate nonlinear properties~\citep{vermersch2024common,liu2026replica} or properties of 
the principal eigenstate~\citep{grier2024principaleigenstateclassicalshadows}. 
Classical shadows of quantum channels address predicting the results
of later queries to them~\citep{kunjummen2023process,levy2024process}, 
with nearly query-optimal protocols for unitary channels~\citep{li2025unitaryshadows}
and optimal Heisenberg scaling~\citep{he2026optimal}.
A review of randomized measurement methods may be found in~\citep{elben2023toolbox}.

The closest results relevant to the setting of this paper concern collective 
measurements. Grier, Pashayan, and Schaeffer~\citep{Grier_2024} 
established nearly matching upper and lower bounds for pure-state shadows. 
Grier, Liu, and Mahajan~\citep{grier2024localsymmetries} obtained rank-dependent 
improvements for mixed states, followed by sharper guarantees from the debiased 
Keyl estimator~\citep{pelecanos2025debiased} and the reduction to pure-state 
tomography through random purification~\citep{pelecanos2025mixedstatetomographyreduces}. 
Related advances established optimal high-precision estimation~\citep{chen2024highprecision} 
and stronger observable concentration~\citep{pelecanos2026keylwerner}. 
Our work complements these algorithms with mixed-state lower bounds accounting for rank, 
observable size, and simultaneous estimation at high confidence.

The importance of measurement resources was also demonstrated by exponential 
separations between learning with and without quantum 
memory~\citep{huang2021information,chen2022exponential}, 
lower bounds for restricted families of adaptive 
single-copy measurements~\citep{lowe2022lower},
and resource tradeoffs and efficient two-copy protocols for 
Pauli estimation~\citep{chen2024paulitradeoffs,king2025triply}. 
A complementary line of work studies classical compression for future expectation-value
queries~\citep{gosset2019compressed,sankar2025relative}.

\paragraph{Shadow tomography.}
Shadow tomography is the original problem that was 
introduced by Aaronson in~\citep{aaronson2018shadow}, and subsequently 
received an improved analysis via gentle measurement and quantum 
data analysis in~\citep{aaronson2019gentle,badescu2021improved}.
The problem considers a related setting in which the measurements are allowed 
to depend on the target observables. 
Recent algorithms removed dimension dependence~\citep{sinha2024dimension,jeronimo2026dimensionfree} 
and achieved classical rates for online queries~\citep{chen2026online}. 
Other works studied the advantages of adaptivity~\citep{chen2024adaptivity}, 
efficient few-copy protocols~\citep{king2025triply}, and instance-dependent 
high-precision estimation~\citep{chen2026instanceoptimal}. 

\paragraph{Quantum state tomography.}
Full state reconstruction has been studied through compressed 
sensing~\citep{gross2010compressed,flammia2012compressed}, projected 
least squares~\citep{GKKT20}, and collective 
measurements~\citep{Haah_2017,10.1145/2897518.2897544,o2017efficient}. 
Recent works sharpen lower bounds in trace distance and 
fidelity~\citep{scharnhorst2025optimal,yuen2023improved} 
and obtain optimal estimation guarantees through random 
purification~\citep{pelecanos2025mixedstatetomographyreduces}. 
Restrictions on joint measurements lead to additional costs, 
quantified by single-copy lower bounds~\citep{lowe2022lower} and 
increasingly sharp entanglement--sample-complexity 
tradeoffs~\citep{chen2024entanglementtradeoff,pelecanos2025debiased,keskin2026tight}. 
These results provide natural comparison benchmarks for 
the classical shadows protocols.

\paragraph{Coupling-based noise operators and hypercontractivity.}
Our Gaussian-lift kernel is obtained using a general coupling-based approach 
for constructing hypercontractive operators on non-product spaces. 
Given a coupling to a simpler probability space carrying a suitable 
noise operator, one transfers the operator back to the non-product 
space by conditional expectation. This framework is developed
for the symmetric group by \citet{filmus2024hypercontractivity}. 
Related coupling arguments have been used to transfer 
hypercontractivity from product spaces to non-product domains. 
In particular, \citet{keevash2023sharp} couple the symmetric 
group \(S_n\) to the product space \([n]^n\) to obtain a sharp 
hypercontractive inequality on \(S_n\). Even closer to our 
construction, \citet{ellis2024productmixing} couple Haar 
orthogonal matrices to Gaussian matrices via the Gram--Schmidt
map and transfer Ornstein--Uhlenbeck noise from Gaussian space 
to the group. This Gaussian-to-Haar coupling is the classical 
QR construction~\citep{mezzadri2007generate}.

Our construction applies the hypercontractivity transfer 
principle to the Grassmannian through the spectral map 
sending a Hermitian matrix to its top-half eigenspace. This 
places the resulting equivariant Markov kernel alongside intrinsic 
Grassmannian operators such as Brownian/heat semigroups~\citep{baudoin2021windings} 
and Berezin transforms on compact Hermitian symmetric spaces~\citep{zhang1998berezin}.

\section{Proof Strategy and Main Reductions}\label{sec:overview}

We first consider the version of the protocol \cref{def:protocol}
for a single observable. Hence we may assume \(\ell=\log(1/\delta)\) 
throughout this part of the overview. 

It is convenient to formulate the lower bound as a game between a dealer and
a player. Any coherent classical shadows protocol induces a strategy for the
player. We show that no such strategy can win with the required advantage
unless the player is given sufficiently many copies of the planted mixed state.

\subsection{Parameter range reduction}
\label{subsec:moderate-error-overview}
Before proceeding further and formulating the game, 
we first do some technical work and reduce 
the range of parameters to the regime that actually requires non-trivial 
treatment.

Notice that the lower bound \eqref{eq:master-lb} consists of two terms:
the precision term \(\Omega(\ell/\eps^2)\) and the geometric term 
\(\min\left\{
     \frac{\sqrt{rF\ell}}{\eps},
     \frac{(F\ell)^{2/3}}{\eps^{4/3}}
 \right\}\).
As in \cite{Grier_2024}, a two-state testing argument, using the
Holevo--Helstrom theorem and the fidelity bound
\citep[Theorems~3.4 and 3.33]{watrous2018theory}, proves the lower bound
\(T = \Omega(\ell/\eps^2)\).
Hence the precision term dominates the geometric contribution outside 
the \emph{moderate-error regime}
\begin{equation}\label{eq:overview-window}
 \ell\leq c_*\eps^2F\min\{r,F\},
\end{equation}
where \(c_*>0\) is a sufficiently small universal constant.
Indeed, the ratio of the geometric target to the precision term is
\[
\frac{\min\{\sqrt{rF\ell}/\eps,(F\ell)^{2/3}/\eps^{4/3}\}}
     {\ell/\eps^2}
=
\min\left\{
    \eps\sqrt{\frac{rF}{\ell}},
    \left(\frac{\eps^2F^2}{\ell}\right)^{1/3}
\right\}.
\]
Outside \eqref{eq:overview-window}, the ratio is at most
\(\max\{c_*^{-1/2},c_*^{-1/3}\}\).
Thus it suffices to prove the geometric lower bound only in the 
regime \eqref{eq:overview-window}. 
For constant confidence, the same condition reduces to 
\(\eps^2 F\min\{r,F\} \gtrsim 1\).

Notice also that since \(\ell\geq\log 3\), the regime reduction
\eqref{eq:overview-window} forces \(F\) to be sufficiently large. We may
therefore set
\[
    f=\lfloor F/2\rfloor=\Theta(F)
\]
and consider the subspace
\[
    \cH=\C^{2f}\subseteq\C^d.
\]
We call \(\cH\) the active space, since all the objects in the hard
instance construction will be supported on it.

For the fixed-rank analysis below, we let
\[
    1\leq R\leq \min\{r,f/2\}
\]
be arbitrary. We choose the effective rank \(R\) only after deriving the
fixed-rank game bound.

\subsection{Single-observable game}

Before defining the game we need to introduce some terminology.

Let \(\nu\) denote the Haar probability measure on \(\Gr(f,2f)\),
the space of \(f\)-dimensional complex subspaces of \(\cH\).
To define the game we first need to define  
a \emph{planting kernel}
\[
 K_\alpha:\Gr(f,2f) \longrightarrow \cP(\Gr(f,2f)),
 \qquad V\longmapsto K_\alpha(V,\cdot).
\]
Here \(\cP\) denotes Borel probability measures and \(0 \leq \alpha < 1\)
controls the strength of planting. For a subspace \(V \subset \cH\), 
which may be considered to be a point in \(\Gr(f,2f)\), the 
kernel specifies a distribution of another point (subspace) \(W \in \Gr(f,2f)\). 

For the moment this is all that is needed from \(K_\alpha\) to define the game. 
We explain the choice of kernel and the motivation behind it 
in \cref{subsec:planting-kernel}.

\begin{definition}[The planted-sign game]\label{def:sign-game}
Fix \(f,R,T\) and \(\alpha\). The dealer and the player perform the following experiment.
\begin{enumerate}[label=(\roman*)]
\item The dealer chooses Haar-random \(V\in\Gr(f,2f)\),
and then chooses an \(R\)-dimensional subspace \(P \subseteq V\) 
Haar-randomly on \(\Gr(R,V)\).
\item The dealer provides the player with \(T\) copies of \(\rho_P=\Pi_P/R\). 
The player performs an arbitrary joint measurement and records the classical outcome \(Z\). 
No quantum system is retained.
\item The dealer uniformly samples a sign \(b\in\{+1,-1\}\), 
independently of everything sampled so far. 
Denote \(V_+ = V\) and \(V_- = V^\perp\).
\item The dealer samples \(W \sim K_\alpha(V_b,\cdot)\) and reveals the observable
\begin{equation}\label{eq:query-observable-overview}
 A_W=2\Pi_W-I_{\cH} \in \Obs_d(F),
\end{equation}
extended by zero on \(\cH^\perp\). 
\item Using \(Z\) and \(A_W\), the player outputs a sign 
\(\widehat b\). The player wins if \(\widehat b=b\).
\end{enumerate}
\end{definition}

Notice that indeed \(A_W \in \Obs_d(F)\): \(\Tr A_W = 0\), 
\(\norm{A_W}_{\op} = 1\), and \(\norm{A_W}^2_F = \Tr I_{\cH} = 2f \leq F\). 
The state \(\rho_P\) has rank \(R \leq r\). 
We will use the same letter for a subspace and 
its projector when the meaning is unambiguous.

The state \(\rho_P\) and the observable \(A_W\) separately reveal nothing about the sign \(b\).
\begin{proposition}[Haar marginals]\label{prop:Haar-marginals}
In \cref{def:sign-game}, for each value of \(b\), the marginal distribution 
of \(P\) is Haar measure on \(\Gr(R,2f)\) and the marginal distribution of \(W\) 
is \(\nu\). In particular, each marginal is independent of \(b\).
\end{proposition}
The proposition follows from the properties of the planting kernel presented in \cref{subsec:planting-kernel}. 
We defer the proof to \cref{sec:gaussian}.
Thus the player must possess a classical description \(Z\) that is good enough
to correctly guess the sign \(b\) from the correlation between the query \(A_W\) and the state \(\rho_P\).

\subsection{Posterior distributions and bias}\label{sec:posterior-bias}

Fix the player's measurement. For an outcome \(z\) of positive probability,
let \(g_z\) be the posterior density of the latent subspace \(V \sim \nu\) relative to 
Haar measure \(\nu\) on \(\Gr(f, 2f)\). Explicitly, for
\[
 q_z=\int \E_{P\mid V}\Tr(M_z\rho_P^{\otimes T})\dd\nu(V),
\]
by Bayes' rule,
\begin{equation}\label{eq:posterior-overview}
 g_z(V)=\frac{\E_{P\mid V}\Tr(M_z\rho_P^{\otimes T})}{q_z},
 \qquad \int g_z\dd\nu=1.
\end{equation}
Since the input of the planting kernel is \(V\), we choose to condition on it,
rather than conditioning directly on \(P\).

The kernel \(K_\alpha\), which we construct in the next subsection, is
self-adjoint on \(L^2(\nu)\). This allows us to express the conditional
distribution of the query directly in terms of the posterior density \(g_z\).

For \(b=+1\), the kernel starts from \(V\), and the conditional density of \(W\)
given \(Z=z\), relative to \(\nu\), is
\[
    p_+(W\mid z)=(K_\alpha g_z)(W).
\]
For \(b=-1\), the kernel instead starts from \(V^\perp\). Since the complementation 
\((Ch)(V)=h(V^\perp)\) preserves \(\nu\), the density of the starting subspace is \(Cg_z\). 
Hence
\[
    p_-(W\mid z)=(K_\alpha Cg_z)(W).
\]

After observing \((z,W)\), distinguishing the two signs is therefore a binary
hypothesis-testing problem between the densities \(K_\alpha g_z\) and
\(K_\alpha Cg_z\). The optimal decision rule is then the likelihood-ratio test.

For any decision rule, define its bias by
\begin{equation}\label{eq:bias-def}
 \Bias(\widehat b)=\E[b\widehat b]
 =2\Pr[\widehat b=b]-1.
\end{equation}

Let \(\mathsf P_+\) and \(\mathsf P_-\) denote the joint laws of \((Z,W)\) under
\(b=+1\) and \(b=-1\), respectively.
For the optimal decision rule,
\begin{align}
 \sup_{\widehat b}\Bias(\widehat b)
 &=\TV(\mathsf P_+,\mathsf P_-)
 \label{eq:bias-TV}\\
 &=\frac12\sum_z q_z
   \int
   \left|
       (K_\alpha g_z)(W)-(K_\alpha Cg_z)(W)
   \right|\dd\nu(W).
 \label{eq:posterior-bias-overview}
\end{align}
We present the derivation in \cref{app:bias}.

\subsection{Planting kernel}\label{subsec:planting-kernel}

The planting kernel \(K_\alpha\) has two roles in the construction.

The first is to produce a hard instance of the game at the target accuracy
scale \(\eps\). If the dealer simply revealed \(V_b\), then the player could estimate
\[
    \Tr(A_{V_b}\rho_P)=b
\]
to constant accuracy and recover the sign. Indeed, in this case the source is
entirely contained in either the positive or the negative eigenspace of the
query. We therefore use a noisy planting kernel to weaken the correlation
between the planted subspace \(V_b\) and the revealed query, while preserving
the Haar marginal of the query.

The second role of the planting kernel is to provide spectral control over the
conditional query distributions that arise from the posterior. In particular,
we want the noise to preserve a tunable degree-one component, which carries
the signal, while suppressing the higher harmonic components that appear in
the proof.

A useful way to think about the planting kernel is as a noisy Markov transition
on the Grassmannian. Starting from a subspace \(V\), the transition produces a
random subspace whose distribution remains correlated with \(V\). The amount of
noise controls the strength of this correlation: with little noise the output
remains strongly correlated with the starting subspace, while increasing the
noise gradually erases this dependence. At the fully noisy endpoint the output
is independent of \(V\) and distributed according to Haar measure. We choose an 
intermediate noise level for which the remaining
correlation carries a signal of the desired strength.

The construction of our planting kernel follows the coupling-based approach to 
transfer of hypercontractive noise from a simpler probability space to a non-product domain
\citep{filmus2024hypercontractivity,ellis2024productmixing,keevash2023sharp}.
We discuss this general mechanism in \cref{sec:related-works}, and develop the
specific spectral and representation-theoretic properties needed for the lower
bound in \cref{sec:gaussian}.

We realize such a transition by lifting the Grassmannian to a Gaussian space,
applying Gaussian noise there, and projecting back. The ambient Gaussian space
is the real Euclidean space \(\Herm(2f)\) of Hermitian \(2f\times 2f\)
matrices. Let
\[
    \Herm_f^\circ(2f)
    :=
    \left\{
        H\in\Herm(2f):
        x_f(H)>x_{f+1}(H)
    \right\},
\]
the set of Hermitian matrices with a spectral gap across the middle of the
spectrum. On this set the top-half spectral subspace is uniquely defined.

We define
\[
    \pi:
    \Herm_f^\circ(2f)
    \longrightarrow
    \Gr(f,2f),
    \qquad
    \pi(H)
    :=
    \operatorname{span}\{u_1(H),\ldots,u_f(H)\},
\]
where
\[
    x_1(H)\geq\cdots\geq x_{2f}(H)
\]
are the eigenvalues of \(H\), with corresponding orthonormal eigenvectors
\(u_1(H),\ldots,u_{2f}(H)\). Thus \(\pi(H)\) is the top-half spectral
subspace of \(H\).

When convenient, we identify a Grassmannian point \(V\in\Gr(f,2f)\) with its
orthogonal projector \(\Pi_V\). In particular,
\[
    \Pi_{\pi(H)}
    =
    \sum_{i=1}^{f}
    |u_i(H)\rangle\langle u_i(H)|.
\]
Since a Hermitian Gaussian matrix has simple spectrum almost surely,
\(\Herm_f^\circ(2f)\) has full Gaussian measure, so \(\pi(H)\) is almost surely
well-defined under the Gaussian measure used below.

We therefore define the transition from the planted subspace
\(V_b\) to the query subspace \(W\) by the chain of transformations
\begin{equation}\label{eq:gaussian-lift-schematic}
    V_b
    \ \xrightarrow{\ \substack{\text{Gaussian}\\ \text{lift}}\ }\
    H\sim
    \gamma_{2f}\bigl(\,\cdot\,\bigm|\pi(H)=V_b\bigr)
    \ \xrightarrow{\ \substack{\text{Gaussian}\\ \text{noise}}\ }\
    H_\alpha
    :=
    \alpha H+\sqrt{1-\alpha^2}\,Z
    \ \xrightarrow{\ \pi\ }\
    W:=\pi(H_\alpha),
\end{equation}
where \(Z\) is an independent Hermitian Gaussian matrix. We define
\(K_\alpha(V_b,\cdot)\) to be the distribution of the final subspace
\(W\).

There is also a useful operator description of the same construction. If
\(\nu\) denotes Haar measure on the Grassmannian and \(\gamma_{2f}\) denotes the
Hermitian Gaussian measure, define
\[
    (Jh)(H):=h(\pi(H)).
\]
Then
\begin{equation}\label{eq:gaussian-lift-operator-schematic}
    L^2(\nu)
    \xrightarrow{\ J\ }
    L^2(\gamma_{2f})
    \xrightarrow{\ T_\alpha\ }
    L^2(\gamma_{2f})
    \xrightarrow{\ J^*\ }
    L^2(\nu),
    \qquad
    K_\alpha=J^*T_\alpha J,
\end{equation}
where \(T_\alpha\) is the Gaussian Ornstein--Uhlenbeck noise operator
\citep[Sec.~2.7]{bakry2014analysis}.

This factorization is the reason for using the Gaussian lift. The
Ornstein--Uhlenbeck operator has explicit decay on Gaussian chaos, and the
lift transfers this decay to the Cartan harmonic decomposition on the
Grassmannian. At the same time, the degree-one multiplier remains continuously
adjustable and determines the mean planted signal. We use only the resulting
one-step reversible Markov kernel \(K_\alpha\) on the Grassmannian; the family
\(\{K_\alpha\}\) does not need to form a semigroup.

We summarize the properties of the planting kernel needed in the proof in the 
following proposition:
\begin{proposition}[Properties of the planting]
\label{prop:planting-overview}

There exists a universal constant \(\eta_0>0\) such that for every
\(0<\eta\leq\eta_0\) we can choose \(\alpha\) for which the following properties
hold.

\begin{enumerate}[label=(\roman*),ref=(\roman*)]

\item\label{item:planting-geometry}
\emph{Invariance and reversibility.}
The kernel is unitary-equivariant and reversible with respect to Haar measure \(\nu\):
\[
    \nu(\mathrm{d}V)K_\alpha(V,\mathrm{d}W)
    =
    \nu(\mathrm{d}W)K_\alpha(W,\mathrm{d}V).
\]
Its operator on functions,
\[
    (K_\alpha h)(V)
    =
    \int h(W)K_\alpha(V,\mathrm{d}W),
\]
is self-adjoint on \(L^2(\nu)\) and commutes with complementation
\[
    (Ch)(V)=h(V^\perp).
\]

\item\label{item:planting-signal}
\emph{The mean answer encodes the sign with strength \(\eta\).}
If \(W\sim K_\alpha(V_b,\cdot)\), then
\begin{equation}\label{eq:soft-signal-overview}
    X_P(W)
    :=
    \Tr(A_W\rho_P)
    =
    \frac{2}{R}\Tr(\Pi_P\Pi_W)-1,
    \qquad
    \E[X_P(W)\mid V,P,b]
    =
    b\eta.
\end{equation}
Equivalently, \(K_\alpha\) has multiplier \(\eta\) on the degree-one harmonics:
\begin{equation}\label{eq:deg-one-mult-overview}
    \tau_{(1)}(\alpha)=\eta.
\end{equation}

\item\label{item:planting-harmonics}
\emph{Higher harmonics are suppressed.}
If \(\Harm_\mu\) is a Cartan harmonic component of degree \(|\mu|\), then
\begin{equation}\label{eq:query-multipliers-overview}
    0\leq \tau_\mu(\alpha)\leq \alpha^{|\mu|},
    \qquad
    \beta_*\alpha\leq\tau_{(1)}(\alpha)\leq\alpha,
    \qquad
    \beta_*=\frac{9}{128}.
\end{equation}
The Cartan harmonic decomposition and the multipliers \(\tau_\mu(\alpha)\) are
introduced below.

\item\label{item:planting-margin}
\emph{The planted answer has the correct sign with a margin with high
probability.}
There exist universal constants \(c,C>0\) such that
\begin{equation}\label{eq:PT-overview}
    \Pr\!\left[
        bX_P(W)<\frac{\eta}{4}
        \,\middle|\,
        V,P,b
    \right]
    \leq
    C\exp(-cRf\eta^2).
\end{equation}

\end{enumerate}
\end{proposition}

We set
\[
    \eta=a\eps
\]
with a sufficiently large universal constant \(a>4\). Whenever
\(bX_P(W)\geq\eta/4\), an \(\eps\)-accurate estimate of \(X_P(W)\) necessarily has
sign \(b\). Thus, outputting the sign of the shadow estimate gives a strategy for
the planted-sign game whose failure probability satisfies
\begin{equation}\label{eq:shadow-to-game}
    p_e
    \leq
    \delta+C\exp(-cRf\eta^2).
\end{equation}

The proof of \cref{prop:planting-overview} is the most technically complicated 
part of the paper. Once the properties of the planting kernel are established, 
the remaining lower-bound argument proceeds through more standard testing 
and information-theoretic reductions.

\subsection{The lower bound for the game}

We first show that the classical shadows protocol induces a strategy 
for the planted-sign game. Set
\[
    \eta=a\eps
\]
for a sufficiently large universal constant \(a>4\). By
\cref{item:planting-margin}, on the event
\(\left\{bX_P(W)\geq \frac{\eta}{4}\right\}\),
any \(\eps\)-accurate estimate of \(X_P(W)\) has sign \(b\). Hence a shadow
protocol with failure probability \(\delta\) induces a strategy for the
planted-sign game with
\[
    p_e
    \leq
    \delta+C\exp(-cRf\eta^2).
\]
For the effective rank in \eqref{eq:effective-rank-overview}, the
moderate-error condition \eqref{eq:overview-window} gives
\[
    \log(1/\delta)\lesssim Rf\eta^2.
\]
After adjusting the universal constants, the second term is therefore at most
\(\delta/10\), hence \(p_e\leq 11\delta/10\leq11/30<2/5\),
within the range of \cref{thm:sign-overview}.
We may therefore apply \cref{thm:sign-overview} with the error
parameter \(\delta':=11\delta/10<2/5\), for which
\[
    \log\frac1{\delta'}
    =
    \Theta\!\left(\log\frac1\delta\right).
\]
It therefore suffices to lower-bound the sample complexity of the
planted-sign game.
Hence, we need to bound how well a \(T\)-copy measurement can distinguish the two
distributions \(\mathsf P_+\) and \(\mathsf P_-\) induced by the hidden sign.

\begin{theorem}\label{thm:sign-overview}

Set \(f\geq 2\) and \(1\leq R\leq f/2\). Let \(0<\eta\leq\eta_0\), and choose
\(\alpha\) according to \cref{prop:planting-overview}.
There exists a universal constant \(C>0\) such that every \(T\)-copy player
strategy in \cref{def:sign-game} has error probability at least
\begin{equation}\label{eq:game-error-overview}
 p_e
 \geq
 \frac{1}{4}
 \exp\left[
    -C\frac{\eta^2}{f}
    \left(
        RT+\frac{T(T-1)}{R}
    \right)
 \right].
\end{equation}

Consequently, for \(0<\delta\leq 2/5\), achieving error probability at most
\(\delta\) requires
\begin{equation}\label{eq:game-general-overview}
 T
 =
 \Omega\left(
    \min\left\{
        \frac{f\log(1/\delta)}{\eta^2R},
        \frac{\sqrt{Rf\log(1/\delta)}}{\eta}
    \right\}
 \right).
\end{equation}

In particular, whenever
\[
    \eta^2R^3\lesssim f\log(1/\delta),
\]
the second term is the smaller one, and
\begin{equation}\label{eq:fixedR-overview}
 T
 =
 \Omega\left(
    \frac{\sqrt{Rf\log(1/\delta)}}{\eta}
 \right).
\end{equation}

\end{theorem}

The proof of \cref{thm:sign-overview} follows the general template of the
lower bound for Boolean Hidden Matching (BHM)
\citep{kerenidis2006onewaycommunicationcomplexityboolean,GavinskyKKRW2007}.
In BHM, a message is produced before the recipient's query is known.
Conditioning on the message changes the original distribution into a
posterior distribution, and a Fourier expansion is then used to control how
strongly that posterior can correlate with the subsequently revealed query.

The same structure appears here. The measurement outcome \(Z\) plays the role
of the message: conditioning on \(Z=z\) produces the posterior density \(g_z\) of
the latent subspace \(V\). The later query probes this posterior through the
planting kernel \(K_\alpha\). Grassmannian harmonics take the place of the
Fourier characters of the Boolean cube. The proof then separates into two
estimates. On the source side, the Schur--Weyl argument bounds how concentrated
the posterior produced by the measurement can be. On the query side, Gaussian
hypercontractivity shows that the planting kernel strongly contracts this
posterior information before it reaches the revealed query.

\subsection{Separating the query from the source}

Fix a transcript \(Z=z\), and let \(g_z\) denote the posterior density of the
latent subspace \(V\). Under the two possible signs, the conditional query
densities are
\[
    p_+(\,\cdot\mid z)=K_\alpha g_z,
    \qquad
    p_-(\,\cdot\mid z)=K_\alpha Cg_z,
\]
where \((Ch)(V)=h(V^\perp)\). Thus the difference between the two query laws is
\[
    p_+(\,\cdot\mid z)-p_-(\,\cdot\mid z)
    =
    K_\alpha(g_z-Cg_z).
\]
To separate the effect of the planting kernel from the information contained
in the posterior \(g_z\), we decompose this expression into Grassmannian
harmonics.

The unitary group \(U(2f)\) acts transitively on \(\Gr(f,2f)\) by
\[
    V\longmapsto UV,
    \qquad U\in U(2f).
\]
Since the Haar measure \(\nu\) is invariant under this action, it induces a
unitary representation on \(L^2(\nu)\),
\[
    h(V)\longmapsto h(U^{-1}V).
\]
By the Cartan--Helgason theory for compact symmetric spaces
\citep{HelgasonGGA,goodman2009symmetry}, this representation is
multiplicity-free and decomposes orthogonally as
\[
    L^2(\nu)
    =
    \C\1
    \oplus
    \widehat{\bigoplus}_{\mu\ne0}\Harm_\mu.
\]
The irreducible components are indexed by partitions
\[
    \mu=(\mu_1\geq\cdots\geq\mu_f\ge0),
\]
and we call
\[
    |\mu|:=\sum_i\mu_i
\]
their Cartan degree. We denote the orthogonal projection onto
\(\Harm_\mu\) by \(\Pi_\mu\).

The degree-one component consists of the linear functions
\[
    V\longmapsto \Tr(A\Pi_V),
    \qquad \Tr A=0,
\]
while the higher-degree components describe higher-order dependence on
the projector \(\Pi_V\).

By property~\ref{item:planting-geometry} of \cref{prop:planting-overview}, \(K_\alpha\) is
\(U(2f)\)-equivariant. The complementation operator \(C\) also commutes with the
\(U(2f)\) action. Since the harmonic decomposition is multiplicity-free,
Schur's lemma implies that both operators act by scalars on each
\(\Harm_\mu\). We denote these scalars as
\[
    \tau_\mu(\alpha)
    \qquad\text{and}\qquad
    c_\mu\in\{+1,-1\},
\]
respectively. Therefore
\begin{equation}\label{eq:signed-harmonics-overview}
    K_\alpha(g_z-Cg_z)
    =
    \sum_{\mu\ne0}
    \tau_\mu(\alpha)(1-c_\mu)\Pi_\mu g_z.
\end{equation}
The constant component is absent since \(C\1=\1\).

Equation \eqref{eq:signed-harmonics-overview} separates the two parts of the
argument. The multiplier \(\tau_\mu(\alpha)(1-c_\mu)\) depends only on the
query construction, and \(\Pi_\mu g_z\) measures how much information about
the latent subspace the measurement has placed in harmonic component \(\mu\).

For comparison, at constant confidence, property~\ref{item:planting-harmonics} of
\cref{prop:planting-overview}, together with the embedding
\(L^2(\nu)\subseteq L^1(\nu)\) and
orthogonality of the harmonic spaces, gives
\begin{equation}\label{eq:const-conf-bias-bd}
    \frac12
    \|K_\alpha(g_z-Cg_z)\|_{L^1(\nu)}
    \leq
    \left(
        \sum_{m\geq1}\alpha^{2m}
        \sum_{|\mu|=m}\|\Pi_\mu g_z\|_2^2
    \right)^{1/2}.
\end{equation}
This already controls the relevant distinguishability in the
constant-confidence regime. For high confidence, however, it is not enough to keep
the bias bounded away from one, since we need a quantitative lower bound on the
remaining testing error. We therefore work with the affinity, equivalently
the R\'enyi \(1/2\)-divergence, between the conditional query laws
\[
    p_+(\,\cdot\mid z)=K_\alpha g_z,
    \qquad
    p_-(\,\cdot\mid z)=K_\alpha Cg_z.
\]

\subsection{The source-side bound}

We next control how concentrated a posterior can be after an arbitrary
collective measurement on \(T\) copies of the mixed state. Recall that, conditioned on
a transcript \(Z=z\), the measurement produces a posterior density
\(g_z\) of the latent subspace \(V\) relative to Haar measure. 

Since any planted state is supported on \(\cH\), we may compress the
original POVM to the active space \(\cH^{\otimes T}\).
By Schur--Weyl duality~\citep{fulton2013representation,keyl2001spectrum},
\(\rho_P^{\otimes T}\) is block diagonal in the
Schur decomposition. We may therefore refine the measurement outcome
with the Schur label \(\lambda\vdash T\) and then consider the measurement
within the corresponding irreducible block. Since \(\rho_P\) has rank \(R\), only
partitions with at most \(R\) nonzero rows occur. We denote the resulting
posterior density by \(g_{\lambda,z}\).
The refinement can only increase distinguishability. Hence a lower 
bound for the refined
measurement provides a lower bound for the original one. We present an
explicit justification in \cref{sec:source}.

The source-side quantity we want to control is the second moment 
of the posterior density, \(\|g_{\lambda,z}\|_2^2\),
or equivalently,
\[
    B(g_{\lambda,z}):=\log\|g_{\lambda,z}\|_2^2.
\]

The following proposition provides the desired control.
\begin{proposition}[Source-side posterior bound]\label{prop:overview-source}
Assume \(R\leq f/2\). For every partition \(\lambda\vdash T\) with at
most \(R\) nonzero rows and every measurement outcome \(z\) of positive
probability in the corresponding Schur block,
\[
    B(g_{\lambda,z})
    =
    \log\|g_{\lambda,z}\|_2^2
    \leq
    \frac{2c_R(\lambda)}{f},
\]
where
\[
    c_R(\lambda)
    :=
    \sum_{i=1}^R
    \lambda_i(\lambda_i+R+1-2i)
\]
is the quadratic \(U(R)\) Casimir parameter. Moreover, under the Schur distribution
generated by \(\rho_P^{\otimes T}\),
\begin{equation}\label{eq:casimir-overview}    
    \E_\lambda c_R(\lambda)
    =
    RT+\frac{T(T-1)}{R}.
\end{equation}
\end{proposition}

The important point of \cref{prop:overview-source} is that
the bound holds uniformly over the measurement effect inside
each Schur block. We can therefore average only at the end over the actual
Schur distribution, using the exact identity \eqref{eq:casimir-overview},
rather than taking a worst-case bound over \(\lambda\).

We provide the proof in \cref{sec:source}. Its main ingredient is the symmetry of
Schur blocks. After conditioning on \(\lambda\), the posterior associated
with an arbitrary measurement effect can be expressed in terms of the
corresponding Schur representation. Computing its second moment over the
random subspace \(V\) then reduces, by unitary invariance, to a comparison of
dimensions of Schur modules. Crucially, the resulting estimate is independent
of the measurement effect corresponding to \(z\), which leads to the uniform bound
\cref{prop:overview-source}.

The standard identity \eqref{eq:casimir-overview} is obtained by averaging
the quadratic Casimir over the Schur distribution of \(\rho_P^{\otimes T}\).

\subsection{The query-side bound}

We next state the query-side estimate.
Recall that the Gaussian-lift kernel factors as
\[
    K_\alpha = J^*T_\alpha J,
\]
where \(J\) lifts functions on the Grassmannian to Hermitian Gaussian
matrices, \(T_\alpha\) is Gaussian Ornstein--Uhlenbeck noise, and \(J^*\)
projects back to the Grassmannian.

\begin{proposition}[Query-side contraction]
\label{prop:query-overview}

Let \(g\in L^2(\nu)\) be a probability density with respect to \(\nu\), and set
\[
    B(g)=\log\|g\|_2^2.
\]
For the Gaussian-lift planting kernel of
\cref{prop:planting-overview},
\begin{equation}\label{eq:generating-overview}
    \|K_\alpha g\|_2^2
    =
    1+\sum_{\mu\ne0}
        \tau_\mu(\alpha)^2\|\Pi_\mu g\|_2^2
    \leq
    \exp\left(
        \frac{2\alpha^2}{1+\alpha^2}B(g)
    \right).
\end{equation}
Consequently,
\begin{equation}\label{eq:KL-overview}
    \KL(K_\alpha g\|1)
    \leq
    2\beta_*^{-2}\eta^2 B(g),
\end{equation}
where \(\KL(p\|1)=\int p\log p\dd\nu\).

For the two query laws corresponding to the opposite planted signs, 
we also have
\begin{equation}\label{eq:Renyi-overview}
    \Ren(K_\alpha g\|K_\alpha Cg)
    \leq
    24\beta_*^{-2}\eta^2B(g),
\end{equation}
where
\[
    \Aff(p,q)=\int\sqrt{pq}\dd\nu,
    \qquad
    \Ren(p\|q)=-2\log\Aff(p,q).
\]

\end{proposition}

Since \(J:L^p(\nu)\to L^p(\gamma_{2f})\) is isometric and \(J^*\) is conditional expectation,
ordinary Gaussian hypercontractivity gives
\[
 \|K_\alpha g\|_{L^2(\nu)}
 \leq
 \|T_\alpha Jg\|_{L^2(\gamma_{2f})}
 \leq
 \|Jg\|_{L^{1+\alpha^2}(\gamma_{2f})}
 =
 \|g\|_{L^{1+\alpha^2}(\nu)},
\]
which yields \eqref{eq:generating-overview}. Since
\[
    \KL(p\|1)\leq\log\|p\|_2^2,
\]
we obtain
\[
    \KL(K_\alpha g\|1)
    \leq
    \frac{2\alpha^2}{1+\alpha^2}B(g).
\]
The calibration \(\beta_*\alpha\leq\eta\) then yields
\eqref{eq:KL-overview}. 

The R\'enyi estimate \eqref{eq:Renyi-overview} compares the two different
query laws \(K_\alpha g\) and \(K_\alpha Cg\), and requires an additional
ingredient. We split the Gaussian noise into two stages. Forward
hypercontractivity first reduces the concentration of the two lifted
densities; reverse hypercontractivity for the remaining noise then gives
a lower bound on their affinity. Finally, projecting back to the
Grassmannian by conditional expectation can only increase affinity.
The proof is given in \cref{thm:Renyi-contraction}.

\subsection{Proof of \texorpdfstring{\cref{thm:sign-overview}}{Theorem \ref{thm:sign-overview}}}
\label{sec:proof-assembly}

We now prove \cref{thm:sign-overview} from
\cref{prop:overview-source,prop:query-overview}.

Since the sign is chosen after the measurement is finished, the joint
distribution of the transcript \((\lambda,z)\) is the same for both signs. 
Conditioned on the transcript \((\lambda,z)\), the two query densities 
are \(K_\alpha g_{\lambda,z}\) and \(K_\alpha Cg_{\lambda,z}\).

Since this joint distribution is common to the two hypotheses,
we may express their affinity as
\begin{align}
 \Aff(\mathsf P_+,\mathsf P_-)
 &=
 \E_{\lambda,z}
 \Aff(K_\alpha g_{\lambda,z},K_\alpha Cg_{\lambda,z})
 \notag\\
 &=
 \E_{\lambda,z}
 \exp\left[
   -\frac12
   \Ren(K_\alpha g_{\lambda,z}\|K_\alpha Cg_{\lambda,z})
 \right]
 \notag\\
 &\geq
 \exp\left[
   -\frac12
   \E_{\lambda,z}
   \Ren(K_\alpha g_{\lambda,z}\|K_\alpha Cg_{\lambda,z})
 \right],
 \label{eq:affinity-average-overview}
\end{align}
where we have used Jensen's inequality in the last step.

Using the query-side bound from \cref{prop:query-overview} and the
source-side bound from \cref{prop:overview-source}, we obtain
\begin{align*}
 \E_{\lambda,z}\Ren(K_\alpha g_{\lambda,z}\|K_\alpha Cg_{\lambda,z})
 &\leq
 C\eta^2\E_{\lambda,z} B(g_{\lambda,z})\\
 &\leq
 C\frac{\eta^2}{f}\E_\lambda c_R(\lambda)\\
 &=
 C\frac{\eta^2}{f}
 \left(
   RT+\frac{T(T-1)}R
 \right).
\end{align*}
Therefore,
\begin{equation}\label{eq:budget-overview}
 \Ren(\mathsf P_+\|\mathsf P_-)
 \leq
 C\frac{\eta^2}{f}
 \left(
   RT+\frac{T(T-1)}R
 \right).
\end{equation}

Since we sample the sign uniformly,
\[
 p_e^*
 =
 \frac{1-\TV(\mathsf P_+,\mathsf P_-)}{2}.
\]
Using the classical Fuchs--van de Graaf inequality
\citep{fuchs2002cryptographic}
\[
 \TV(p,q)\leq\sqrt{1-\Aff(p,q)^2},
\]
we obtain
\[
 p_e^*
 \geq
 \frac14\Aff(\mathsf P_+,\mathsf P_-)^2
 =
 \frac14
 \exp\left(-\Ren(\mathsf P_+\|\mathsf P_-)\right).
\]
Together with \eqref{eq:budget-overview}, this proves
\eqref{eq:game-error-overview}.

If a strategy has error at most \(0<\delta\le2/5\), then
\[
 \Ren(\mathsf P_+\|\mathsf P_-)
 \geq
 -\log\bigl(4\delta(1-\delta)\bigr)
 \geq
 \frac1{100}\log(1/\delta).
\]
To obtain the last inequality, we use the following reasoning:
if \(\delta\leq1/16\), then
\(-\log(4\delta(1-\delta))\geq\frac12\log(1/\delta)\);
if \(1/16\leq\delta\leq2/5\), the numerator is at least
\(\log(25/24)\), while \(\log(1/\delta)\leq\log16\), and
\(\log(25/24)/\log16>1/100\).
Hence
\[
 RT+\frac{T^2}{R}
 \gtrsim
 \frac{f\log(1/\delta)}{\eta^2}.
\]
Solving for \(T\),
\[
 T
 =
 \Omega\left(
   \min\left\{
     \frac{f\log(1/\delta)}{\eta^2R},
     \frac{\sqrt{Rf\log(1/\delta)}}{\eta}
   \right\}
 \right),
\]
which gives \eqref{eq:game-general-overview}. 
If
\[
    \eta^2R^3\lesssim f\log(1/\delta),
\]
then
\[
    \frac{f\log(1/\delta)}{\eta^2R}
    \gtrsim
    \frac{\sqrt{Rf\log(1/\delta)}}{\eta},
\]
so the second term in \eqref{eq:game-general-overview} is the smaller one.
Therefore
\begin{equation}
 T
 =
 \Omega\left(
   \frac{\sqrt{Rf\log(1/\delta)}}{\eta}
 \right),
\end{equation}
which proves \eqref{eq:fixedR-overview}.

\subsection{Choice of the effective rank}
\label{sec:overview-ranges}

We now choose the rank of the hard instance. Throughout this subsection we
work in the moderate-error regime \eqref{eq:overview-window}, established in
\cref{subsec:moderate-error-overview}. Recall that
\(f=\lfloor F/2\rfloor=\Theta(F)\) and that
\(\eta=a\eps\) for a fixed universal constant \(a\).

For a fixed rank \(1\leq R\leq f/2\), from the fixed-rank game bound
\eqref{eq:game-general-overview}, we have
\begin{equation}\label{eq:fixed-rank-two-branches-overview}
    T
    =
    \Omega\left(
        \min\left\{
            \frac{f\ell}{\eta^2R},
            \frac{\sqrt{Rf\ell}}{\eta}
        \right\}
    \right).
\end{equation}
The two terms are equal at the balancing scale
\[
    R_{\mathrm{bal}}
    :=
    \left(\frac{f\ell}{\eta^2}\right)^{1/3}.
\]
For \(R>R_{\mathrm{bal}}\), the first term is the smaller one and decreases
with \(R\), whereas for \(R<R_{\mathrm{bal}}\) the square-root term is the
smaller one and increases with \(R\). Thus the largest lower bound obtainable
from \eqref{eq:fixed-rank-two-branches-overview}, subject to \(R\leq r\), is
attained up to constants by choosing
\begin{equation}\label{eq:effective-rank-overview}
    R
    \asymp
    \min\left\{
        r,
        \left(\frac{f\ell}{\eta^2}\right)^{1/3}
    \right\}.
\end{equation}

We first verify that this choice satisfies the auxiliary conditions needed
for the fixed-rank game bound and for the reduction from the classical
shadows protocol to the planted-sign game \cref{def:sign-game}. 
By construction,
\begin{equation}\label{eq:branch-condition-overview}
    \eta^2R^3\lesssim f\ell.
\end{equation}

We also need \(R\leq f/2\), as required by the source-side estimate. Since
\(R\lesssim R_{\mathrm{bal}}\) and
\[
    R_{\mathrm{bal}}^3
    =
    \frac{f\ell}{\eta^2},
\]
the moderate-error assumption \eqref{eq:overview-window}, together with
\(f=\Theta(F)\) and \(\eta=\Theta(\eps)\), leads to
\[
    R_{\mathrm{bal}}^3
    \lesssim
    c_* fF\min\{r,F\}
    \leq
    Cc_* f^3.
\]
Hence
\[
    R
    \lesssim
    R_{\mathrm{bal}}
    \lesssim
    c_*^{1/3}f.
\]
Choosing \(c_*\) sufficiently small therefore ensures that
\[
    R\leq f/2.
\]

It remains to verify that the planted answer is visible with the required
confidence. By the concentration estimate in
\eqref{eq:shadow-to-game}, it is enough to guarantee
\begin{equation}\label{eq:visibility-overview}
    \ell=\log\frac{1}{\delta}
    \lesssim
    Rf\eta^2.
\end{equation}
There are two cases. If \(R\asymp r\), then
\[
    \frac{\ell}{Rf\eta^2}
    \lesssim
    \frac{c_*\eps^2F\min\{r,F\}}
         {rf\eta^2}
    \lesssim
    c_*.
\]
If instead \(R\asymp R_{\mathrm{bal}}\), then, using
\(\ell=\eta^2R_{\mathrm{bal}}^3/f\),
\[
    \frac{\ell}{Rf\eta^2}
    \asymp
    \frac{R_{\mathrm{bal}}^2}{f^2}
    \lesssim
    c_*^{2/3}.
\]
Thus \eqref{eq:visibility-overview} holds in either case after choosing
\(c_*\) sufficiently small. In particular, the planted-answer failure term
in \eqref{eq:shadow-to-game} is at most \(\delta/10\). Hence a shadow
protocol with failure probability at most \(\delta\) induces a strategy for
the planted-sign game with error at most
\[
    \delta' := \frac{11}{10}\delta.
\]
Since \(\delta\leq 1/3\), \(\delta'\leq11/30<2/5\), and
\[
    \log(1/\delta')\asymp \log(1/\delta)=\ell.
\]

Since \(\log(1/\delta')\asymp\ell\), the branch condition
\eqref{eq:branch-condition-overview} also gives
\[
    \eta^2R^3\lesssim f\log(1/\delta')
\]
after adjusting the universal constant. Thus the chosen rank lies in the
square-root regime of \cref{thm:sign-overview}. Applying
\eqref{eq:fixedR-overview} with error parameter \(\delta'\) therefore gives
\begin{equation}\label{eq:square-branch}
    T
    =
    \Omega\left(
        \frac{\sqrt{Rf\ell}}{\eta}
    \right).
\end{equation}

Substituting \eqref{eq:effective-rank-overview}, we obtain
\begin{equation}\label{eq:geometric-after-rank-overview}
    T
    =
    \Omega\left(
        \min\left\{
            \frac{\sqrt{rf\ell}}{\eta},
            \frac{(f\ell)^{2/3}}{\eta^{4/3}}
        \right\}
    \right)
    =
    \Omega\left(
        \min\left\{
            \frac{\sqrt{rF\ell}}{\eps},
            \frac{(F\ell)^{2/3}}{\eps^{4/3}}
        \right\}
    \right).
\end{equation}

Combining \eqref{eq:geometric-after-rank-overview} with the precision bound
\(T=\Omega(\ell/\eps^2)\), we get
\begin{equation}\label{eq:lb-overview}
    T
    =
    \Omega\left(
        \min\left\{
            \frac{\sqrt{rF\ell}}{\eps},
            \frac{(F\ell)^{2/3}}{\eps^{4/3}}
        \right\}
        +
        \frac{\ell}{\eps^2}
    \right).
\end{equation}
Together with the parameter reduction in
\cref{subsec:moderate-error-overview}, this proves the desired
single-observable lower bound in all parameter regimes.

\subsection{Extension to multiple observables}
\label{sec:overview-multiple}

We extend the single-observable game by hiding the planted 
query among \(M-1\) independent decoys. The player must identify 
both the location of the planted query and its sign.

\begin{definition}[Multiple-observable planted-sign game]
\label{def:multiple-game}
Fix \(f,R,T,M\) and \(\alpha\). The dealer and the player perform the following
experiment.

\begin{enumerate}[label=(\roman*)]
    \item The dealer samples
    \[
        V\sim\nu,
        \qquad
        P\sim\Gr(R,V),
        \qquad
        \rho_P=\frac{\Pi_P}{R},
    \]
    exactly as in \cref{def:sign-game}.

    \item The dealer provides the player with \(T\) copies of \(\rho_P\). 
    The player performs an arbitrary joint measurement and records a 
    classical outcome \(Z\). No quantum system is retained.

    \item After the measurement has been completed, the dealer samples
    independently
    \[
        J\sim\operatorname{Unif}[M],
        \qquad
        b\sim\operatorname{Unif}\{+1,-1\}.
    \]
    As before, we denote \(V_+=V\) and \(V_-=V^\perp\).

    \item At the hidden location \(J\), the dealer samples the planted query
    \[
        W_J\sim K_\alpha(V_b,\cdot).
    \]
    At every other location \(i\neq J\), the dealer independently samples
    \[
        W_i\sim \nu, \qquad \nu = \mathrm{Haar}(\Gr(f,2f)).
    \]
    The dealer reveals the list of observables
    \[
        A_{W_1},\ldots,A_{W_M},
        \qquad
        A_{W_i} = 2\Pi_{W_i}-I_{\cH}.
    \]

    \item Using \(Z\) and the revealed list, the player outputs
    \((\widehat J,\widehat b)\in[M]\times\{\pm1\}\). The player wins if
    \[
        (\widehat J,\widehat b)=(J,b).
    \]
\end{enumerate}
\end{definition}

By construction, the query list itself contains no information about 
the hidden pair \((J,b)\). Indeed, from property~\ref{item:planting-geometry} of
\cref{prop:planting-overview}
it follows that \(K_\alpha\) is Haar-stationary,
\[
    \int_{\Gr(f,2f)}
        K_\alpha(V,\mathrm{d}W)\dd\nu(V)
    =
    \mathrm{d}\nu(W).
\]
After averaging over the latent subspace \(V\), the planted query \(W_J\) 
has distribution \(\nu\) for every fixed \((J,b)\). Since the remaining 
queries are independent Haar samples, the unconditional law of the 
query list, conditioned only on \((J,b)\), is
\[
    (W_1,\ldots,W_M)\mid(J,b)\sim \nu^{\otimes M}.
\]
Thus the player can identify the special coordinate
only through its correlation with the source state encoded in the classical
transcript \(Z\).

We next show that a successful shadow protocol results in
a successful strategy for \cref{def:multiple-game}. For the planted coordinate,
property~\ref{item:planting-margin} of \cref{prop:planting-overview} gives
\[
    bX_P(W_J)\geq \frac{\eta}{4}
\]
except with probability at most
\[
    C\exp(-cRf\eta^2).
\]
For an independent Haar decoy \(W_i\),
Grassmannian concentration, obtained by lifting to Haar unitaries
\citep[Corollary~17]{meckes2013spectral}, gives
\[
    \Pr\!\left[
        |X_P(W_i)|>\frac{\eta}{16}
    \right]
    \leq
    C\exp(-cRf\eta^2).
\]
By a union bound, if
\begin{equation}\label{eq:multiple-visibility-overview}
    \log(2M)\lesssim Rf\eta^2,
\end{equation}
then, after choosing the implicit constant sufficiently small, with probability
at least \(14/15\) the planted coordinate is well-separated from all the decoys:
\[
    bX_P(W_J)\geq\frac{\eta}{4},
    \qquad
    \max_{i\neq J}|X_P(W_i)|\leq\frac{\eta}{16}.
\]
For \(\eta\) a sufficiently large constant multiple of \(\eps\),
simultaneous \(\eps\)-accurate estimates of all \(M\) observables then determine
both \(J\) and \(b\): the planted coordinate is the unique estimate of large
absolute value, and its sign determines \(b\). Since the shadow protocol succeeds
with probability at least \(2/3\), the resulting game strategy succeeds with
probability at least \(3/5\), and hence has error probability at most \(2/5\).

It remains to show that the player cannot identify \((J,b)\) from too few
copies. The analogue of \cref{thm:sign-overview} is the following.
\begin{theorem}\label{thm:multiple-game-overview}

Set \(f\geq 2\), \(1\leq R\leq f/2\), and \(M\geq 2\). Let
\(0<\eta\leq\eta_0\), and choose \(\alpha\) according to
\cref{prop:planting-overview}.

There exists a universal constant \(C>0\) such that every \(T\)-copy player
strategy in \cref{def:multiple-game} has error probability
\begin{equation}\label{eq:multiple-error-overview}
    p_e
    :=
    \Pr\!\left[
        (\widehat J,\widehat b)\neq(J,b)
    \right]
    \geq
    1-
    \frac{
        C\frac{\eta^2}{f}
        \left(
            RT+\frac{T(T-1)}{R}
        \right)
        +\log 2
    }{
        \log(2M)
    }.
\end{equation}

Consequently, achieving error probability at most \(2/5\) requires
\begin{equation}\label{eq:multiple-general-overview}
    T
    =
    \Omega\left(
        \min\left\{
            \frac{f\log(2M)}{\eta^2R},
            \frac{\sqrt{Rf\log(2M)}}{\eta}
        \right\}
    \right).
\end{equation}

In particular, for \(\eta^2R^3\lesssim f\log(2M)\):
\begin{equation}\label{eq:multiple-overview-result}
    T
    =
    \Omega\left(
        \frac{\sqrt{Rf\log(2M)}}{\eta}
    \right).
\end{equation}

\end{theorem}

We provide the proof in \cref{sec:multiple}. The source-side input is the same
posterior estimate used in the single-query argument. The difference is on
the query side. Conditioned on a transcript, one of the \(M\) coordinates has
density either \(K_\alpha g\) or \(K_\alpha Cg\), while all the others have
Haar density \(1\). The Gaussian hypercontractive estimate bounds the
relative entropy of the planted coordinate from Haar. Using the information-radius
identity~\citep[Chapter~2]{cover2006elements}, we obtain
\[
    I(J,b;Z,W_1,\ldots,W_M)
    \leq
    C\frac{\eta^2}{f}
    \left(
        RT+\frac{T(T-1)}{R}
    \right).
\]
Recovering \((J,b)\) requires learning one of \(2M\) possible hypotheses.
By Fano's inequality~\citep[Sec.~2.10]{cover2006elements}, the right-hand side then has to be of order
\(\log(2M)\), which yields \eqref{eq:multiple-general-overview}.

The remainder of the argument is the same rank optimization as in the
single-observable case, with
\[
    \ell=\log(2M)
\]
in place of \(\log(1/\delta)\). Together with the multiple-observable
precision bound, this gives
\[
    T
    =
    \Omega\left(
        \min\left\{
            \frac{\sqrt{rF\log(2M)}}{\eps},
            \frac{(F\log(2M))^{2/3}}{\eps^{4/3}}
        \right\}
        +
        \frac{\log(2M)}{\eps^2}
    \right)
\]
under the moderate-list assumption of \cref{thm:master}.

Finally, the high-confidence and multiple-observable lower bounds hold
simultaneously. Taking their maximum and using
\[
    \ell
    =
    \max\left\{
        \log(2M),\log(1/\delta)
    \right\}
\]
results in \cref{thm:master}.
\section{Discussion}\label{sec:discussion}

We established lower bounds on the sample complexity of fully coherent
classical shadows for mixed states of bounded rank. Our lower bound nearly
matches the best currently proved upper bound. The earlier analysis of the
debiased Keyl estimator gives the relevant second-moment bounds and the
corresponding classical shadows protocol
\cite[Theorem~1.9 and Lemma~7.3]{pelecanos2025debiased}.
More recently, Pelecanos, Spilecki, Tang, and Wright proved a substantially
stronger concentration bound for the related \(\mathrm{Mix}(\mathrm{AGPS})\)
estimator, by controlling the moment-generating function of
\(\Tr(O\widehat\rho)\)
\cite[Proposition~4.7]{pelecanos2026keylwerner}.
Combining these results yields an upper bound that differs from ours only by
a factor of \(\ell^{1/3}\) in the saturated geometric regime.

It is natural to expect that this remaining gap is technical. In particular,
the newer work explicitly discusses extending its concentration analysis to
the stronger \(\mathrm{Mix}^{+}(\mathrm{AGPS})\) estimator and leaves this
question open
\cite[Footnote~6]{pelecanos2026keylwerner}.
The relevant question for our setting is whether such an extension yields
the saturated scaling \((F\ell)^{2/3}/\eps^{4/3}\). We conjecture that this
scaling is attainable, which would make our lower bound optimal up to universal
constants.

A natural direction for further research would be to understand classical shadows
under limited coherence. Recent results in quantum tomography 
and related learning problems show
that restricting the number of copies that may be measured jointly can lead to
nontrivial tradeoffs between coherence and sample complexity
\citep{chen2024entanglementtradeoff,pelecanos2025debiased,keskin2026tight}. It would be
interesting to determine the corresponding tradeoff for the low-rank mixed-state
classical shadows problem. More precisely, one may ask for the optimal sample
complexity when each measurement is allowed to act jointly on at most \(c\) copies
of the unknown state.

The Gaussian-lift noise kernel may also be of independent interest.
It defines a reversible equivariant Markov kernel on the Grassmannian
whose spectral attenuation is controlled by Cartan degree through
the minimum Gaussian-chaos degree of each Grassmannian representation.
Additional harmonic properties of the kernel will be explored in
future work.%
\section{AI use disclosure}
The author used ChatGPT 5.6 Sol Plus and ChatGPT 6 Astra Plus, as
well as Claude Opus 5.5 to a lesser extent,
during the preparation of this manuscript. The results were obtained under
the author's close guidance, through protracted back-and-forth
interaction. The idea to mimic the proof of the BHM lower bound,
as well as the formulation of the corresponding game, belongs to
the author. The author also proposed probabilistic planting (via
the heat kernel at the initial step) to obtain the desired
attenuation of the harmonic multipliers. The hypercontractivity 
transfer technique was suggested as an option by ChatGPT
in response to a request to search for known results and techniques
for applying hypercontractivity on a Grassmannian. 
The author suggested the Gaussian lift via the more standard 
QR decomposition, and
ChatGPT 6 Astra later discovered that the Hermitian Gaussian lift is
better suited for the problem in this paper. ChatGPT 6 Astra 
helped with the representation theory arguments and was heavily used
in the proof of the concentration property \cref{prop:Gaussian-coupling-concentration}.
The author is especially thankful for it pointing out the delicate details,
such as the application of the MTP\(_2\) argument \citep{karlin1980classes}
and the need for the Chafa{\"i}-Lehec extension \citep{chafai2020poincare} 
of the Brascamp-Lieb inequality.
Claude Opus 5.5 was used for additional proofreading.

The author verified all results and derivations. ChatGPT 5.6 and 6
were used in drafting and proofreading the manuscript, however, the
exposition was iteratively and heavily edited by the author.
The author takes full responsibility for the content of the manuscript.%
\section{The Gaussian-lift planting kernel}\label{sec:gaussian}

Our construction of the planting kernel uses the coupling mechanism described in
\citet[Sec.~1.5.1]{filmus2024hypercontractivity}: we lift functions from the
Grassmannian to a Gaussian probability space, apply Ornstein--Uhlenbeck noise,
and project back by conditional expectation. Related constructions for the
symmetric group were studied in
\citet{keevash2023sharp,filmus2024hypercontractivity}. The closest analogue of
our construction is the orthogonal-group construction of
\citet[Sec.~8]{ellis2024productmixing}, which couples Gaussian and Haar matrices
through Gram--Schmidt before transferring Gaussian noise.

The feature specific to our setting is the choice of lift. We use the spectral
map
\[
    \pi:\Herm_f^\circ(2f)\to\Gr(f,2f)
\]
defined in \cref{subsec:planting-kernel}, which sends a Hermitian matrix to its
top-half spectral subspace. As we prove below, this map is equivariant under
unitary conjugation, so the transferred operator is automatically
\(U(2f)\)-equivariant. Moreover,
\[
    \pi(-H)=\pi(H)^\perp,
\]
so Gaussian reflection realizes the two signs of the planting. The coupling
mechanism itself is general, whereas the negation--complement correspondence
is a special feature of the balanced Grassmannian and is particularly 
well-matched to the sign symmetry of our hard instance.

The general coupling argument supplies hypercontractivity
(\cref{lem:Gaussian-lift-hypercontractivity}), but the spectral
estimates needed here depend on the representation-theoretic structure of our
particular lift. We show that a Grassmannian harmonic of Cartan degree \(|\mu|\)
can occur only in Gaussian chaos of degree at least \(|\mu|\), which yields the
uniform multiplier bound
\[
    0\leq \tau_\mu(\alpha)\leq \alpha^{|\mu|}.
\]
Thus higher Cartan degrees are increasingly suppressed. We separately show
that the degree-one multiplier is bounded below by a universal multiple of
\(\alpha\), which allows us to calibrate the kernel to a planted signal of
order \(\eps\).

\subsection{Construction and Haar marginals}

Throughout this section, \(N:=2f\).
Let \(\gamma_N\) denote the standard Gaussian measure on \(\Herm(N)\)
viewed as a real Euclidean space:
\[
\gamma_N(\mathrm{d}H)
=
(2\pi)^{-N^2/2}
\exp\!\left(-\frac12\Tr(H^2)\right)\dd H,
\]
where \(\mathrm{d}H\) is the Lebesgue measure induced by the Hilbert--Schmidt inner
product. Equivalently, if \(H\sim\gamma_N\), then
\[
\E\!\left[\Tr(AH)\Tr(BH)\right]
=
\Tr(AB)
\]
for all \(A,B\in\Herm(N)\).

For \(H\in\Herm(N)\) with simple spectrum, write its spectral decomposition as
\[
H=\sum_{i=1}^N x_i(H)\,|u_i(H)\rangle\langle u_i(H)|,
\qquad
x_1(H)>\cdots>x_N(H).
\]
Define
\[
\pi(H)
:=
\sum_{i=1}^{f}|u_i(H)\rangle\langle u_i(H)|.
\]
We use the same symbol \(\pi(H)\) for the corresponding point of
\(\Gr(f,N)\). Thus \(\pi(H)\) is the spectral projector onto the
eigenspaces associated with the largest \(f\) eigenvalues of \(H\).

The spectrum of \(H\sim\gamma_N\) is simple almost surely~\citep{mehta2004random}.
The Gaussian measure \(\gamma_N\) is invariant under unitary conjugation.
Moreover, for every \(U\in U(N)\),
\begin{equation}\label{eq:pi-equivariance}
    \pi(UHU^\dagger)=U\pi(H)U^\dagger.
\end{equation}
It follows that the law of \(\pi(H)\) is invariant under the natural
\(U(N)\)-action on \(\Gr(f,N)\). Since the Grassmannian admits a unique
\(U(N)\)-invariant probability measure, \(\pi(H)\) is Haar distributed.
We denote this measure by \(\nu\).

Now observe that if
\[
x_1(H)>\cdots>x_N(H),
\]
then
\[
-x_N(H)>\cdots>-x_1(H)
\]
are the ordered eigenvalues of \(-H\). Therefore
\[
\pi(-H)
=
\sum_{i=f+1}^{N}|u_i(H)\rangle\langle u_i(H)|.
\]
Since \(N=2f\),
\[
\pi(-H)=I-\pi(H),
\]
or equivalently, identifying projectors with their ranges,
\[
\pi(-H)=\pi(H)^\perp.
\]

\begin{remark}
Notice that this reflection property distinguishes the spectral lift \(\pi\) from the standard
Gaussian QR construction of Haar frames~\citep{mezzadri2007generate}.
Taking the span of the first \(f\) columns of a Gaussian matrix also gives
a Haar subspace, but that span is unchanged by Gaussian negation. The
spectral lift \(\pi\) instead turns negation into the complementation needed for
the signed planting. 
\end{remark}

\begin{remark}\label{rem:traceless-reduction}
The construction depends only on the traceless component of the Gaussian
matrix. Indeed, if
\[
H_0:=H-\frac{\Tr H}{N}I,
\]
then
\[
\pi(H)=\pi(H_0),
\]
and the same remains true after applying Ornstein--Uhlenbeck noise.
We defer the elementary verification of this fact to
\cref{app:traceless-gaussian-reduction}.
Nevertheless, it is more convenient to formulate
the Gaussian lift and the Ornstein--Uhlenbeck operator 
on the full Euclidean space \(\Herm(N)\).
\end{remark}

Define the pullback
\[
(Jh)(H):=h(\pi(H)),
\]
and let \(T_\alpha\) denote the Ornstein--Uhlenbeck operator on
\(\Herm(N)\) in Mehler's representation \citep[Sec.~2.7]{bakry2014analysis}:
\[
(T_\alpha F)(H)
:=
\E_Z\!\left[
F\!\left(\alpha H+\sqrt{1-\alpha^2}\,Z\right)
\right],
\qquad
0\leq \alpha<1,
\]
where \(Z\sim\gamma_N\) is independent of \(H\).

Since \(\pi(H)\sim\nu\) for \(H\sim\gamma_N\), the pullback
\[
J:L^p(\nu)\longrightarrow L^p(\gamma_N)
\]
is an isometry for every \(1\leq p\leq\infty\). As a map from
\(L^2(\nu)\) to \(L^2(\gamma_N)\), its adjoint
is given by
\[
(J^*F)(V)
=
\E\!\left[F(H)\mid \pi(H)=V\right].
\]
In particular,
\[
J^*J=I.
\]

Identifying Grassmannian points with their orthogonal projectors, we use
the same notation for the unitary action on both function spaces:
\[
\bigl(\mathcal{U}_U F\bigr)(X):=F(U^\dagger XU),
\qquad
U\in U(N),
\]
where \(X\) is either a rank-\(f\) orthogonal projector or a Hermitian matrix,
according to the domain of \(F\).
The pullback \(J\) intertwines the two representations. Indeed, by
\eqref{eq:pi-equivariance},
\[
\bigl(J\mathcal{U}_U h\bigr)(H)
=
h\!\left(U^\dagger\pi(H)U\right)
=
h\!\left(\pi(U^\dagger HU)\right)
=
\bigl(\mathcal{U}_U Jh\bigr)(H).
\]
Thus the natural \(U(N)\)-representations on \(L^2(\nu)\) and
\(L^2(\gamma_N)\) satisfy
\[
J\mathcal{U}_U=\mathcal{U}_U J.
\]

We define the Gaussian-lift planting kernel by
\begin{equation}\label{eq:Gaussian-kernel}
K_\alpha:=J^*T_\alpha J.
\end{equation}
Thus \(K_\alpha\) is the Markov operator on \(\Gr(f,N)\) obtained by
applying Gaussian Ornstein--Uhlenbeck noise in the lifting space
\(\Herm(N)\) and projecting back to the Grassmannian.
We use the resulting one-step kernel. Our argument requires
neither a semigroup property nor the explicit knowledge of the spherical
transform of \(K_\alpha\).

Equivalently, the kernel \(K_\alpha\) has the following sampling
description. Given \(V\in\Gr(f,N)\), sample a Hermitian Gaussian
matrix conditioned on having \(V\) as its top-\(f\) eigenspace,
\[
    H\sim\gamma_N(\,\cdot\mid\pi(H)=V),
\]
draw an independent \(Z\sim\gamma_N\), and set
\begin{equation}\label{eq:Gaussian-coupling}
W
=
\pi\!\left(
\alpha H+\sqrt{1-\alpha^2}\,Z
\right).
\end{equation}
The distribution of \(W\) is \(K_\alpha(V,\cdot)\).

Since a prescribed subspace \(V\) has Haar measure zero, the
conditional Gaussian law above is initially defined only for
Haar-almost every \(V\). We can specify it for every \(V\) using
the spectral decomposition of a Hermitian Gaussian matrix.

Let \(E_0=\operatorname{span}\{e_1,\ldots,e_f\}\), and choose
\(U_V\in U(N)\) such that \(U_VE_0=V\). Independently sample
the ordered eigenvalues \(\mathbf x\) of a matrix with law
\(\gamma_N\) and a Haar unitary \(S\in U(f)\times U(f)\),
the stabilizer of \(E_0\). Then
\begin{equation}\label{eq:conditional-Gaussian-version}
H=U_VSD_xS^\dagger U_V^\dagger,
\qquad D_x=\diag(\mathbf x),
\end{equation}
has the desired conditional distribution.

Indeed, the ordered eigenvalues of a Hermitian Gaussian matrix
are independent of its Haar-distributed eigenbasis. Conditioning
on the top-\(f\) eigenspace fixes only the decomposition
\(V\oplus V^\perp\). The eigenbasis remains Haar-distributed
within each subspace. This is precisely the distribution in
\eqref{eq:conditional-Gaussian-version}. Moreover, changing
\(U_V\) amounts to multiplying it on the right by an element
of \(U(f)\times U(f)\), which does not change the law of \(H\)
by Haar invariance. Thus the construction is independent of
the chosen representative and is unitarily equivariant.
A measurable choice of \(U_V\) defines the kernel for every \(V\).

\begin{lemma}[Transferred hypercontractivity]
\label{lem:Gaussian-lift-hypercontractivity}
Let \(1<p\leq q<\infty\) and \(0\leq\alpha<1\) satisfy
\(\alpha^2\leq(p-1)/(q-1)\). Then, for every \(h\in L^p(\nu)\),
\[
\|K_\alpha h\|_{L^q(\nu)}\leq\|h\|_{L^p(\nu)}.
\]
\end{lemma}
\begin{proof}
Conditional expectation is an \(L^q\)-contraction by Jensen's inequality,
\(J\) is an \(L^p\)-isometry, and Gaussian hypercontractivity
\citep{Gro75} gives
\[
\|K_\alpha h\|_{L^q(\nu)}
=\|J^*T_\alpha Jh\|_{L^q(\nu)}
\leq\|T_\alpha Jh\|_{L^q(\gamma_N)}
\leq\|Jh\|_{L^p(\gamma_N)}
=\|h\|_{L^p(\nu)}.
\]
\end{proof}
The posterior specialization used later is proved in
\cref{lem:Gaussian-posterior-contraction}.

We now prove the kernel properties stated in \cref{prop:planting-overview},
except for the margin bound, which is deferred to \cref{sec:planting-tail}.

\begin{proposition}[Properties of the Gaussian-lift planting kernel]
\label{prop:Gaussian-planting}
Let \(0\leq\alpha<1\). The kernel \(K_\alpha\) has the
following properties.

\begin{enumerate}[label=(\roman*)]

\item
\label{item:Gaussian-invariance}
\emph{Invariance and reversibility.}
The kernel \(K_\alpha\) is \(U(N)\)-equivariant and reversible
with respect to Haar measure \(\nu\):
\[
\nu(\mathrm{d}V)K_\alpha(V,\mathrm{d}W)
=
\nu(\mathrm{d}W)K_\alpha(W,\mathrm{d}V).
\]
In particular, \(\nu\) is stationary for \(K_\alpha\), and
the corresponding operator on \(L^2(\nu)\) is self-adjoint.
Moreover, \(K_\alpha\) commutes with orthogonal
complementation,
\[
K_\alpha C=CK_\alpha,
\qquad
(Ch)(V):=h(V^\perp).
\]

\item
\label{item:Gaussian-mean-signal}
\emph{Mean planted signal.}
Let \(P\subseteq V\), set
\[
\rho_P:=\frac{\Pi_P}{R},
\qquad
V_+:=V,
\qquad
V_-:=V^\perp,
\]
and, conditional on \(b\in\{\pm1\}\), sample
\[
W\sim K_\alpha(V_b,\cdot).
\]
For the observable \(A_W:=2\Pi_W-I_N\), the planted answer
\[
X_P(W):=\Tr(A_W\rho_P)
=
\frac{2}{R}\Tr(\Pi_P\Pi_W)-1
\]
satisfies
\begin{equation}
\label{eq:planted-mean}
\E\!\left[
X_P(W)
\,\middle|\,
V,P,b
\right]
=
b\,\tau_{(1)}(\alpha),
\end{equation}
where \(\tau_{(1)}(\alpha)\) is the multiplier of
\(K_\alpha\) on the degree-one Grassmannian harmonics.

\item
\label{item:Gaussian-harmonic-multipliers}
\emph{Harmonic attenuation and calibration.}
On every Cartan harmonic component \(\Harm_\mu\) of
degree \(|\mu|\), the operator \(K_\alpha\) acts by
a scalar \(\tau_\mu(\alpha)\), satisfying
\begin{equation}
\label{eq:multiplier-envelope}
0\leq\tau_\mu(\alpha)\leq\alpha^{|\mu|}.
\end{equation}
The degree-one multiplier satisfies
\begin{equation}
\label{eq:eta-calibration}
\beta_*\alpha
\leq
\tau_{(1)}(\alpha)
\leq
\alpha,
\qquad
\beta_*:=\frac{9}{128}.
\end{equation}
Moreover, \(\tau_{(1)}(\alpha)\) is continuous and
strictly increasing for \(\alpha \in [0, 1)\) with 
\[ 
    \tau_{(1)}(0) = 0, 
    \qquad 
    \lim_{\alpha\uparrow1}\tau_{(1)}(\alpha)=1.
\]
\end{enumerate}
\end{proposition}

\begin{proof}

\noindent\emph{Proof of \ref{item:Gaussian-invariance}.}
Recall that
\[
K_\alpha=J^*T_\alpha J.
\]
The operators \(J\), \(J^*\), and \(T_\alpha\) are positivity
preserving and unital. Hence \(K_\alpha\) is a Markov
operator. Moreover, since \(T_\alpha\) is self-adjoint on
\(L^2(\gamma_N)\),
\[
K_\alpha^*
=
(J^*T_\alpha J)^*
=
J^*T_\alpha J
=
K_\alpha.
\]
Thus \(K_\alpha\) is self-adjoint on \(L^2(\nu)\).
Reversibility (detailed balance) is equivalent to 
self-adjointness by a standard argument. 

Reversibility immediately implies stationarity 
of the Haar measure. Indeed, integrating the detailed-balance 
identity over \(V\),
\[
\int_{\Gr(f,N)}
\nu(\mathrm{d}V)K_\alpha(V,\mathrm{d}W)
=
\nu(\mathrm{d}W)
\int_{\Gr(f,N)}K_\alpha(W,\mathrm{d}V)
=
\nu(\mathrm{d}W),
\]
where the last equality uses that \(K_\alpha\) is Markov.

We next verify unitary equivariance. Let \(\Gamma_V\) denote the version
of the conditional law fixed in \eqref{eq:conditional-Gaussian-version}.
Suppose
\[
H\sim\Gamma_V,
\qquad
Z\sim\gamma_N
\]
independently. For any fixed \(U\in U(N)\), unitary invariance of
\(\gamma_N\) and
\[
\pi(UHU^\dagger)
=
U\pi(H)U^\dagger
\]
imply
\[
UHU^\dagger\sim\Gamma_{UVU^\dagger},
\qquad
UZU^\dagger\sim\gamma_N.
\]
The corresponding output of the planting started from \(UVU^\dagger\)
is therefore
\[
\begin{aligned}
\pi\!\left(
\alpha UHU^\dagger
+
\sqrt{1-\alpha^2}\,UZU^\dagger
\right)
&=
\pi\!\left(
U\bigl(\alpha H+\sqrt{1-\alpha^2}\,Z\bigr)U^\dagger
\right)\\
&=
U\pi\!\left(
\alpha H+\sqrt{1-\alpha^2}\,Z
\right)U^\dagger.
\end{aligned}
\]
Hence
\[
K_\alpha(\Ad_U(V),\cdot)
=
K_\alpha(UVU^\dagger,\cdot)
=
(\Ad_U)_*K_\alpha(V,\cdot),
\]
which is the definition of \(U(N)\)-equivariance.

Finally, consider orthogonal complementation. Since
\[
\pi(-H)=\pi(H)^\perp,
\]
if \(H\sim\Gamma_V\), then
\[
-H\sim\Gamma_{V^\perp}.
\]
By symmetry of the Gaussian law, \(-Z\sim\gamma_N\). Hence a planting started 
from \(V^\perp\) can be coupled with one
started from \(V\) by using \((-H,-Z)\) instead of \((H,Z)\).
Concretely, if the planting from \(V\) produces
\[
W
=
\pi\!\left(
\alpha H+\sqrt{1-\alpha^2}\,Z
\right),
\]
then, using the same realization of \(H\) and \(Z\), the planting from
\(V^\perp\) produces
\[
\begin{aligned}
W'
&=
\pi\!\left(
-\alpha H-\sqrt{1-\alpha^2}\,Z
\right)\\
&=
\pi\!\left(
-\bigl(\alpha H+\sqrt{1-\alpha^2}\,Z\bigr)
\right)\\
&=
W^\perp.
\end{aligned}
\]
Thus the two output distributions are related by orthogonal
complementation.
In particular, for every Borel set \(B\subseteq\Gr(f,N)\),
\[
K_\alpha(V^\perp,B)=K_\alpha(V,\{W:W^\perp\in B\}).
\]
Consequently, for any bounded measurable \(h\),
\[
(K_\alpha Ch)(V)
=
(CK_\alpha h)(V),
\]
and hence
\[
K_\alpha C=CK_\alpha.
\]

\medskip
\noindent\emph{Proof of \ref{item:Gaussian-harmonic-multipliers}.}
Recall the multiplicity-free harmonic decomposition
of the Grassmannian
\citep{HelgasonGGA,goodman2009symmetry}:
\[
L^2(\nu)
=
\C\1
\oplus
\widehat{\bigoplus}_{\mu\neq0}\Harm_\mu.
\]
By unitary equivariance and Schur's lemma,
\(K_\alpha\) acts on each \(\Harm_\mu\) by
a scalar \(\tau_\mu(\alpha)\).

We first establish the harmonic attenuation bound.
The Gaussian space
\[
(\Herm(N),\gamma_N)
\simeq
(\R^{N^2},\gamma_{N^2})
\]
has the orthogonal Wiener-chaos decomposition
\citep[Sec.~2.7.1]{bakry2014analysis}
\[
L^2(\gamma_N)
=
\widehat{\bigoplus}_{a=0}^{\infty}\mathcal C_a,
\]
where \(\mathcal C_a\) is the \(a\)-th homogeneous
Gaussian chaos. The Ornstein--Uhlenbeck operator
acts on it by
\[
T_\alpha F=\alpha^a F,
\qquad F\in\mathcal C_a.
\]

Fix \(h\in\Harm_\mu\).
We claim that its Gaussian lift \(Jh\) has no
chaos components below degree \(|\mu|\).

The conjugation representation on Hermitian matrices
complexifies to
\[
\Herm(N)_{\C}
\simeq
\C^N\otimes(\C^N)^*.
\]
Hence the degree-\(a\) homogeneous polynomials,
and equivalently the Gaussian chaos \(\mathcal C_a\),
can contain only representations whose highest
weights have total positive weight at most \(a\):
each tensor factor contributes at most one.

On the other hand, the Cartan--Helgason classification
\citep{HelgasonGGA,goodman2009symmetry} gives the
highest weight of \(\Harm_\mu\) as
\[
(\mu_1,\ldots,\mu_f,-\mu_f,\ldots,-\mu_1),
\]
whose total positive weight is \(|\mu|\).
Thus \(\Harm_\mu\) cannot occur in
\(\mathcal C_a\) for \(a<|\mu|\).

Indeed, the lift \(J\) is \(U(N)\)-equivariant,
as follows from \eqref{eq:pi-equivariance}, and
projection onto \(\mathcal C_a\) also commutes with
the unitary action. Consequently,
\[
\operatorname{Proj}_{\mathcal C_a}J:
\Harm_\mu\longrightarrow\mathcal C_a
\]
is an intertwining map, and must vanish when
\(a<|\mu|\).

We may therefore write
\[
Jh=\sum_{a\geq |\mu|}F_a,
\qquad
F_a:=\operatorname{Proj}_{\mathcal C_a}Jh
\in\mathcal C_a,
\]
with orthogonal summands. Using the isometry of
\(J\) and the diagonal action of \(T_\alpha\),
we obtain
\[
\begin{aligned}
\tau_\mu(\alpha)\|h\|_{L^2(\nu)}^2
&=
\langle h,K_\alpha h\rangle_{L^2(\nu)}\\
&=
\langle Jh,T_\alpha Jh\rangle_{L^2(\gamma_N)}\\
&=
\sum_{a\geq |\mu|}
\alpha^a\|F_a\|_{L^2(\gamma_N)}^2.
\end{aligned}
\]
Hence
\begin{equation}\label{eq:multiplier-chaos-expansion}
\tau_\mu(\alpha)
=
\sum_{a\geq |\mu|}
\alpha^a
\frac{\|F_a\|_{L^2(\gamma_N)}^2}
     {\|h\|_{L^2(\nu)}^2}.
\end{equation}
The coefficients are nonnegative and sum to one.
Since \(0\leq\alpha<1\), every term satisfies
\(\alpha^a\leq\alpha^{|\mu|}\), proving
\[
0\leq\tau_\mu(\alpha)\leq\alpha^{|\mu|},
\]
as asserted in \eqref{eq:multiplier-envelope}.

\medskip
\noindent\emph{Degree-one calibration.}
For traceless Hermitian \(A\), define
\[
h_A(V):=\Tr\!\left(A(2\Pi_V-I_N)\right).
\]
These functions span \(\Harm_{(1)}\). Writing
\[
A_V:=2\Pi_V-I_N,
\]
the scalar action of \(K_\alpha\) gives
\[
K_\alpha h_A=\tau_{(1)}(\alpha)h_A,
\]
or equivalently,
\[
\E[A_W\mid V]
=
\tau_{(1)}(\alpha)A_V,
\qquad W\sim K_\alpha(V,\cdot).
\]

In particular, for an unconditioned correlated
Gaussian pair
\[
H'=\alpha H+\sqrt{1-\alpha^2}\,Z,
\]
we obtain
\begin{equation}\label{eq:signal-Gaussian-overlap}
\begin{aligned}
\tau_{(1)}(\alpha)
&=
\frac1N\E\Tr\!\left(
A_{\pi(H)}A_{\pi(H')}
\right)\\
&=
\frac4N\E\Tr\!\left(
\Pi_{\pi(H)}\Pi_{\pi(H')}
\right)-1.
\end{aligned}
\end{equation}
Here we conditioned on \(\pi(H)\) and used
\(A_V^2=I_N\).

We now establish a dimension-independent lower
bound on \(\tau_{(1)}(\alpha)\). The key is to
show that a constant fraction of the lift \(Jh_A\)
lies in the first Gaussian chaos. Fix a nonzero
traceless Hermitian \(A\), and define
\begin{equation}\label{eq:betaN-definition}
\beta_N
:=
\frac{
\|\operatorname{Proj}_{\mathcal C_1}(Jh_A)\|_{L^2(\gamma_N)}^2
}{
\|Jh_A\|_{L^2(\gamma_N)}^2
}.
\end{equation}

We first compute the denominator. Since \(J\)
is an isometry, the Haar projector second-moment
identity, proved in
\cref{app:Gaussian-moment-identities}, gives
\begin{equation}\label{eq:Haar-degree-one-inner-product}
\langle h_A,h_B\rangle_{L^2(\nu)}
=
\frac{N}{N^2-1}\Tr(AB)
\end{equation}
for traceless Hermitian \(A,B\). In particular,
\[
\|Jh_A\|_{L^2(\gamma_N)}^2
=
\frac{N}{N^2-1}\Tr(A^2).
\]

To compute the numerator, let
\[
G:=H-\frac{\Tr H}{N}I_N
\]
be the traceless component of \(H\).
As explained in \cref{rem:traceless-reduction},
\(G\) is a standard Gaussian on \(\Herm_0(N)\),
independent of the scalar component, and
\(\pi(G)=\pi(H)\). Thus
\[
Jh_A(H)=h_A(\pi(G))
\]
depends only on \(G\).

Its first-chaos projection therefore has the form
\[
\operatorname{Proj}_{\mathcal C_1}(Jh_A)(H)
=
\Tr(B_A G)
\]
for some traceless Hermitian \(B_A\).
We determine \(B_A\) by testing against arbitrary
linear functions \(\Tr(BG)\).

By conditional unitary invariance, the conditional
mean of \(G\) given \(\pi(G)=V\) must have the form
\[
\E[G\mid\pi(G)=V]
=
a\Pi_V+b(I_N-\Pi_V).
\]
Since \(G\) is traceless and \(N=2f\), we have
\(b=-a\). Hence
\begin{equation}\label{eq:cond-mean}
\E[G\mid\pi(G)=V]
=
c_N(2\Pi_V-I_N).
\end{equation}

Write
\[
x_1(G)\geq\cdots\geq x_N(G)
\]
for the ordered eigenvalues of \(G\).
Taking the trace of \eqref{eq:cond-mean}
against \(\Pi_V\), and using independence of
the ordered eigenvalues and Haar eigenbasis,
gives
\[
\frac N2 c_N
=
\E\sum_{i=1}^{N/2}x_i(G).
\]
Thus
\begin{equation}\label{eq:cN-eigenvalue-formula}
c_N
=
\frac2N\E\sum_{i=1}^{N/2}x_i(G).
\end{equation}

Now, for any traceless Hermitian \(B\),
the tower property and \eqref{eq:cond-mean} give
\[
\begin{aligned}
\left\langle Jh_A,\Tr(BG)\right\rangle_{L^2(\gamma_N)}
&=
\E\!\left[
h_A(\pi(G))\Tr(BG)
\right]\\
&=
c_N\E\!\left[
h_A(\pi(G))h_B(\pi(G))
\right]\\
&=
c_N\frac{N}{N^2-1}\Tr(AB),
\end{aligned}
\]
where the last equality uses
\eqref{eq:Haar-degree-one-inner-product}.

On the other hand, the standard traceless Gaussian
covariance identity, derived in
\cref{app:Gaussian-moment-identities}, is
\begin{equation}\label{eq:traceless-Gaussian-covariance}
\E\!\left[
\Tr(AG)\Tr(BG)
\right]
=
\Tr(AB).
\end{equation}

The defining property of orthogonal projection
therefore implies
\[
\Tr(B_A B)
=
c_N\frac{N}{N^2-1}\Tr(AB)
\]
for every traceless Hermitian \(B\), and hence
\[
B_A=c_N\frac{N}{N^2-1}A.
\]
Consequently,
\[
\operatorname{Proj}_{\mathcal C_1}(Jh_A)(G)
=
c_N\frac{N}{N^2-1}\Tr(AG).
\]

Using \eqref{eq:traceless-Gaussian-covariance}
once more, we find
\[
\|\operatorname{Proj}_{\mathcal C_1}(Jh_A)\|_{L^2(\gamma_N)}^2
=
c_N^2\frac{N^2}{(N^2-1)^2}\Tr(A^2).
\]
Dividing by the denominator computed above,
we obtain
\begin{equation}\label{eq:betaN-cN}
\beta_N
=
\frac{c_N^2N}{N^2-1}.
\end{equation}

It remains to lower-bound \(\beta_N\) uniformly
in \(N\). Since \(G\) is traceless,
\[
\sum_{i=1}^{N/2}x_i(G)
\geq\frac14\Tr|G|.
\]
Indeed, writing
\(s=\sum_{x_i(G)>0}x_i(G)=\frac12\Tr|G|\),
if at least \(N/2\) eigenvalues are positive,
the largest \(N/2\) have sum at least \(s/2\).
Otherwise, the additional nonpositive eigenvalues
contribute at worst \(-s/2\), giving the same bound.

Moreover, Schatten interpolation and
H\"older's inequality imply
\[
\E\Tr|G|
\geq
\frac{
(\E\Tr G^2)^{3/2}
}{
(\E\Tr G^4)^{1/2}
}.
\]
For the standard traceless Hermitian Gaussian,
\begin{equation}\label{eq:traceless-G-second-moment}
\E\Tr G^2=N^2-1,
\end{equation}
while, by
\eqref{eq:traceless-Gaussian-fourth-moment},
\begin{equation}\label{eq:traceless-G-fourth-moment}
\E\Tr G^4
=
2N^3-5N+\frac3N
\leq2N^3.
\end{equation}

Combining these estimates with
\eqref{eq:cN-eigenvalue-formula} and
\eqref{eq:betaN-cN}, we obtain
\[
\beta_N
\geq
\frac{(N^2-1)^2}{8N^4}
\geq
\frac9{128}
=:\beta_*.
\]

Returning to the chaos expansion
\eqref{eq:multiplier-chaos-expansion} with
\(\mu=(1)\), the contribution of the first chaos
is \(\beta_N\alpha\). Since all other terms
are nonnegative,
\[
\beta_*\alpha
\leq
\tau_{(1)}(\alpha)
\leq
\alpha,
\]
which proves \eqref{eq:eta-calibration}.

Finally, since
\[
Jh_A(-H)=-Jh_A(H),
\]
only odd Gaussian chaoses occur in its expansion.
Thus
\[
\tau_{(1)}(\alpha)
=
\sum_{j\geq0}w_{N,2j+1}\alpha^{2j+1},
\]
where
\[
w_{N,2j+1}\geq0,
\qquad
\sum_{j\geq0}w_{N,2j+1}=1,
\qquad
w_{N,1}=\beta_N\geq\beta_*.
\]
Here
\[
w_{N,a}
=
\frac{
\|\operatorname{Proj}_{\mathcal C_a}(Jh_A)\|_{L^2(\gamma_N)}^2
}{
\|Jh_A\|_{L^2(\gamma_N)}^2
}.
\]
The coefficients are independent of the choice
of nonzero traceless \(A\), by irreducibility.

This expansion shows that
\(\tau_{(1)}(\alpha)\) is continuous and strictly
increasing on \([0,1)\), with
\[
\tau_{(1)}(0)=0,
\qquad
\lim_{\alpha\uparrow1}\tau_{(1)}(\alpha)=1.
\]
This completes the proof of
\ref{item:Gaussian-harmonic-multipliers}.

\medskip
\noindent\emph{Proof of \ref{item:Gaussian-mean-signal}.}
As established above, the degree-one harmonics
are spanned by
\[
h_A(V)=\Tr(A(2\Pi_V-I_N)),
\qquad \Tr A=0,
\]
and \(K_\alpha\) acts on them by
\(\tau_{(1)}(\alpha)\). Consequently,
\begin{equation}\label{eq:mean-projector}
\E[2\Pi_W-I_N\mid V]
=
\tau_{(1)}(\alpha)(2\Pi_V-I_N),
\qquad W\sim K_\alpha(V,\cdot).
\end{equation}

Now fix \(P\subseteq V\), with
\(\rho_P=\Pi_P/R\), and recall
\[
V_+=V,
\qquad
V_-=V^\perp.
\]
Since \(P\subseteq V\) and
\(P\perp V^\perp\),
\[
\Tr\!\left(
\rho_P(2\Pi_{V_b}-I_N)
\right)
=
b.
\]
Applying \eqref{eq:mean-projector} to \(V_b\),
we obtain
\[
\begin{aligned}
\E[X_P(W)\mid V,P,b]
&=
\tau_{(1)}(\alpha)
\Tr\!\left(
\rho_P(2\Pi_{V_b}-I_N)
\right)\\
&=
b\,\tau_{(1)}(\alpha),
\end{aligned}
\]
proving \eqref{eq:planted-mean}.

Equivalently, the expected overlap with the
query subspace is
\begin{equation}\label{eq:exact-signal-normalization}
\begin{aligned}
\E[\Tr(\rho_P\Pi_W)\mid V,P,b]
&=
\frac1R\E[\Tr(\Pi_P\Pi_W)\mid V,P,b]\\
&=
\frac12\bigl(1+b\tau_{(1)}(\alpha)\bigr).
\end{aligned}
\end{equation}
\end{proof}

\paragraph{Calibration of the planting signal.}
For any prescribed \(0<\eta<1\), the continuity
and strict monotonicity of \(\tau_{(1)}\) give a
unique \(0<\alpha<1\) such that
\[
\tau_{(1)}(\alpha)=\eta.
\]
The bounds \eqref{eq:eta-calibration} imply
\begin{equation}\label{eq:alpha-eta-comparison}
\eta\leq\alpha\leq\beta_*^{-1}\eta.
\end{equation}
We henceforth use this calibrated value of
\(\alpha\). In particular, the mean planted
signal in \eqref{eq:planted-mean} becomes
\[
\E[X_P(W)\mid V,P,b]=b\eta.
\]

\paragraph{Sign-independent marginals.}
As an immediate consequence of the invariance properties, the planting
does not reveal the hidden sign through either marginal. The incident
pair \((P,V)\) is unitarily invariant, so the marginal law of \(P\) is
Haar on \(\Gr(R,N)\). Since orthogonal complementation preserves
\(\nu\), the starting subspace \(V_b\) is Haar distributed for either
value of \(b\). By stationarity of \(K_\alpha\),
\[
\Pr[W\in E\mid b]
=
\int_{\Gr(f,N)}K_\alpha(V,E)\dd\nu(V)
=
\nu(E)
\]
for every Borel set \(E\subseteq\Gr(f,N)\). Hence the marginal laws of
both \(P\) and \(W\) are independent of \(b\).
\section{Concentration of the planted answer}\label{sec:planting-tail}

In this section we prove the remaining property~\ref{item:planting-margin}
of \cref{prop:planting-overview}. Recall that after the calibration
\(\tau_{(1)}(\alpha)=\eta\), for every incident pair \(P\subseteq V\),
\[
\E\!\left[\Tr(A_W\rho_P)\,\middle|\,V,P,b\right]
=
b\eta.
\]
It remains to show that the planted answer is separated from zero with
high probability.

\begin{theorem}[Planted-tail estimate]\label{thm:PT}
There are universal constants \(\alpha_0,c,C>0\) such that for
\(0<\alpha\leq\alpha_0\), \(N=2f\), and \(1\leq R\leq f/2\), the
Hermitian Gaussian planting satisfies
\begin{equation}\label{eq:planted-margin-target}
\Pr\!\left[
    b\Tr(A_W\rho_P)<\frac{\eta}{4}
    \,\middle|\,
    V,P,b
\right]
\leq
C\exp(-cRf\eta^2)
\end{equation}
for every incident pair \(P\subseteq V\). Here \(\eta=\eta(\alpha)\),
and the constants are independent of \(N,R\), and the incident pair.
\end{theorem}

However, we cannot apply Gaussian concentration directly.
The obstacle is that the spectral projection \(H\mapsto\pi(H)\) is not uniformly Lipschitz. 
Its sensitivity to perturbations is controlled by the spectral gap \(x_f(H)-x_{f+1}(H)\),
which can be arbitrarily small \citep{davis1970rotation}. We avoid this 
obstacle by working instead with the relative eigenbasis, defined below, 
of the correlated Hermitian Gaussian pair underlying \(K_\alpha\).

Recall from \cref{sec:gaussian} that
\[
H'
=
\alpha H+\sqrt{1-\alpha^2}\,Z,
\qquad
H,Z\in\Herm(N),
\]
where \(H\) and \(Z\) are independent standard Hermitian Gaussian
matrices. Almost surely, \(H\) and \(H'\) have simple spectra. We express
the pair in spectral coordinates as
\[
H
=
U_0D_xU_0^\dagger,
\qquad
H'
=
U_0UD_yU^\dagger U_0^\dagger,
\]
where \(\mathbf x=(x_1,\ldots,x_N)\) and
\(\mathbf y=(y_1,\ldots,y_N)\) are in decreasing order, and
\(D_x=\diag(\mathbf x)\), \(D_y=\diag(\mathbf y)\).
The absolute eigenbasis \(U_1\) of \(H'\) satisfies
\(U_1=U_0U\), so \(H'=U_1D_yU_1^\dagger\) and \(U=U_0^\dagger U_1\).
Here \(U_0\in U(N)\) denotes the absolute eigenbasis of \(H\),
while \(U\in U(N)\) denotes the relative eigenbasis of \(H'\) with
respect to \(H\). Since the individual eigenvectors are defined only up
to complex phases, we choose these phases independently and uniformly
from \(S^1\), thereby obtaining a \(U(N)\)-valued representative of
\(U\).

The eigenvalues \(\mathbf x\) and \(\mathbf y\) are taken in decreasing order, 
so they both belong to the open Weyl chamber
\[
\mathcal W_N
:=
\{\mathbf x\in\R^N:x_1>x_2>\cdots>x_N\}.
\]

The following propositions are the auxiliary concentration statements that
we will need.
\begin{proposition}[Logarithmic Sobolev inequality for the Hermitian Gaussian coupling]
\label{prop:Gaussian-coupling-LSI}
Let
\[
PU(N)=U(N)/U(1)\cong SU(N)/Z_N
\]
be equipped with the quotient Riemannian metric induced by
the bi-invariant metric on \(SU(N)\) defined by
\[
g_S(SX,SY)
=
\operatorname{Re}\operatorname{Tr}(X^\dagger Y),
\qquad X,Y\in\mathfrak{su}(N).
\]
Let \(\mu_\alpha\) denote the law of the relative eigenbasis \(U\),
and let \(\bar\mu_\alpha\) be its pushforward to \(PU(N)\).

There exist universal constants \(\alpha_0,C>0\) such that,
whenever \(0\leq\alpha\leq\alpha_0\), the measure
\(\bar\mu_\alpha\) satisfies
\[
\operatorname{Ent}_{\bar\mu_\alpha}(f^2)
\leq
\frac{C}{N}
\int_{PU(N)}\|\nabla f\|^2\,d\bar\mu_\alpha
\]
for every smooth real-valued function \(f:PU(N)\to\R\).
\end{proposition}

\begin{proposition}[Concentration for the Hermitian Gaussian coupling]
\label{prop:Gaussian-coupling-concentration}
Under the assumptions of
Proposition~\ref{prop:Gaussian-coupling-LSI},
let \(F:U(N)\to\R\) be a real-valued function satisfying
\[
F(e^{i\theta}U)=F(U)
\qquad
\text{for all }U\in U(N),\ \theta\in\R.
\]
Suppose that \(F\), viewed as a function on \(PU(N)\),
is \(L\)-Lipschitz with respect to the quotient Riemannian
distance defined above.

Then there exists a universal constant \(c>0\) such that
\begin{equation}\label{eq:Gaussian-coupling-concentration}
\Pr\!\left[
    |F(U)-\E F(U)|\geq t
\right]
\leq
2\exp\!\left(
    -cN\frac{t^2}{L^2}
\right),
\qquad t\geq0.
\end{equation}
\end{proposition}

\begin{proof}
This follows from Proposition~\ref{prop:Gaussian-coupling-LSI}
by the Herbst argument \citep{herbst1975letter,bakry2014analysis}.
\end{proof}

We now outline the proof of
\cref{prop:Gaussian-coupling-LSI}
and the tools that we use.

We first introduce the measures used in the proof. Let
\(p_\alpha(H,H')\) denote the joint density of the correlated Gaussian
pair. Since this law is invariant under simultaneous unitary
conjugation, its expression in the spectral coordinates
\((\mathbf x,\mathbf y,U,U_0)\) is independent of the absolute eigenbasis \(U_0\).
Integrating out \(U_0\) therefore gives a joint density
\[
q_\alpha(\mathbf x,\mathbf y,U)
\]
for the two ordered eigenvalue lists and the relative eigenbasis. Thus
\(q_\alpha\) is the spectral-coordinate marginal of \(p_\alpha\) after
the common absolute orientation has been forgotten.

Integrating out the two eigenvalue lists gives the marginal
density of the relative eigenbasis with respect to Haar measure:
\begin{equation}\label{eq:angular-partition-function}
Z_\alpha(U)
:=
\int_{x_1>\cdots>x_N}
\int_{y_1>\cdots>y_N}
q_\alpha(\mathbf x,\mathbf y,U)\dd\mathbf y\dd\mathbf x.
\end{equation}
Thus the law \(\mu_\alpha\) of \(U\) satisfies
\[
\mathrm{d}\mu_\alpha(U)=Z_\alpha(U)\dd U.
\]
Since \(Z_\alpha\) is invariant under multiplication by scalar
phases, it also defines the density of the pushforward
\(\bar\mu_\alpha\) on \(PU(N)\) with respect to normalized Haar measure.
We denote
\[
\Phi(U):=\log Z_\alpha(U),
\]
viewing \(\Phi\) also as a function on \(PU(N)\).

The proof of \cref{prop:Gaussian-coupling-LSI} proceeds by
controlling the curvature of this angular measure. We first use the
Weyl integration formula \citep{mehta2004random} to derive an explicit
expression for \(q_\alpha(\mathbf x,\mathbf y,U)\). For fixed \(U\),
we then study the conditional law of \((\mathbf x,\mathbf y)\).
Positive association for multivariate totally positive measures
of order \(2\) (MTP\(_2\)) \citep{karlin1980classes,fallat2017total}
allows us to compare the conditional eigenvalue distributions with
explicitly tractable distributions, yielding uniform control of the
extreme eigenvalues. We explain this comparison below. The
Brascamp--Lieb Poincar\'e inequality \citep{brascamp1976extensions}
controls the conditional variance term arising when differentiating
\(\Phi(U)\). Since the eigenvalues are restricted to ordered Weyl
chambers and the potential is singular at eigenvalue collisions,
we use the extension of this inequality obtained by regularization
\citep{chafai2020poincare}.

These estimates yield the problem-specific curvature bound
\[
\Hess\Phi\preceq C\alpha N\,g,
\]
where \(g\) is the quotient metric on \(PU(N)\) defined in
\cref{prop:Gaussian-coupling-LSI}. Equivalently, we may work on
\(SU(N)\) with the unnormalized Hilbert--Schmidt metric, for which 
\citep{meckes2013spectral,meckes2019random}
\[
\Ric_{SU(N)}=\frac N2\,g.
\]
Thus, for sufficiently small \(\alpha\), the Bakry--\'Emery
curvature satisfies
\[
\Ric-\Hess\Phi\succeq cN\,g.
\]
The Bakry--\'Emery curvature criterion
\citep{bakry1985diffusions,bakry2014analysis} then yields a
logarithmic Sobolev inequality with constant \(O(N^{-1})\),
proving \cref{prop:Gaussian-coupling-LSI}.

\subsection{Concentration for the Gaussian coupling}

\subsubsection{The angular density}\label{sec:angular-density}

We now derive the joint density
\(q_\alpha(\mathbf x,\mathbf y,U)\) in spectral
coordinates and the resulting integral representation
of the angular density \(Z_\alpha(U)\). Set
\[
\sigma^2:=1-\alpha^2,
\qquad
k:=\frac{\alpha}{\sigma^2}.
\]
Since \(H\) and \(H'\) are standard Hermitian Gaussians with correlation
\(\alpha\), their joint density with respect to Lebesgue measure on
\(\Herm(N)\times\Herm(N)\) is
\[
p_\alpha(H,H')
=
C_{\alpha,N}
\exp\!\left[
-\frac{
    \Tr H^2+\Tr(H')^2-2\alpha\Tr(HH')
}{2\sigma^2}
\right],
\]
where, in the Hilbert--Schmidt Lebesgue normalization of
\cref{sec:gaussian},
\[
C_{\alpha,N}=(2\pi)^{-N^2}(\sigma^2)^{-N^2/2}.
\]
Indeed, the real dimension of \(\Herm(N)\) is \(N^2\), and the map
\((H,Z)\mapsto(H,H')\) has Jacobian \(\sigma^{N^2}\).

Applying the Weyl integration formula to the two Hermitian matrices
\citep{mehta2004random}, write
\[
H=U_0D_x U_0^\dagger,
\qquad
H'=U_1D_yU_1^\dagger,
\qquad
D_x=\diag(\mathbf x),
\quad
D_y=\diag(\mathbf y).
\]
Since the eigenvalues are ordered, and since the eigenvector phases have
been randomized uniformly as above, the corresponding volume elements
may be written as
\[
\mathrm{d}H
=
c_N^{\mathrm W}\Delta(\mathbf x)^2\dd\mathbf x\dd U_0,
\qquad
\mathrm{d}H'
=
c_N^{\mathrm W}\Delta(\mathbf y)^2\dd\mathbf y\dd U_1,
\qquad
\Delta(\mathbf x)
:=
\prod_{i<j}(x_i-x_j),
\]
where \(\mathrm{d}U_0\) and \(\mathrm{d}U_1\) denote normalized Haar
measure on \(U(N)\), with the arbitrary eigenvector phases averaged
uniformly. For ordered eigenvalues, the normalization constant is
\[
c_N^{\mathrm W}=\frac{(2\pi)^{N^2/2}}{\mathcal I_N},
\qquad
\mathcal I_N
:=\int_{x_1>\cdots>x_N}
\Delta(\mathbf x)^2e^{-\|\mathbf x\|^2/2}\dd\mathbf x
=(2\pi)^{N/2}\prod_{j=0}^{N-1}j!.
\]
The integral is obtained by expressing the Vandermonde \(\Delta(\mathbf x)\) 
as a determinant of monic Hermite polynomials, the squared norms of which 
are \(\sqrt{2\pi}\,j!\). Restriction to the ordered chamber removes the
factor \(N!\) from integration over \(\R^N\)
\citep[Secs.~2 and~5.1]{fyodorov2004introduction}. 

The quadratic terms depend only on the eigenvalues:
\[
\Tr H^2=\|\mathbf x\|^2,
\qquad
\Tr(H')^2=\|\mathbf y\|^2.
\]
The coupling term is
\[
\Tr(HH')
=
\Tr\!\left(
D_x U_0^\dagger U_1 D_y U_1^\dagger U_0
\right).
\]
Introducing the relative eigenbasis
\[
U:=U_0^\dagger U_1,
\]
this becomes
\[
\Tr(HH')
=
\Tr\!\left(
D_x U D_y U^\dagger
\right).
\]

Thus, in the spectral variables \((\mathbf x,\mathbf y,U,U_0)\), the joint measure
takes the form
\[
\begin{aligned}
p_\alpha(H,H')\dd H\dd H'
={}&
\widetilde C_{\alpha,N}\,
\Delta(\mathbf x)^2\Delta(\mathbf y)^2\\
&\times
\exp\!\left[
-\frac{\|\mathbf x\|^2+\|\mathbf y\|^2}{2\sigma^2}
+
k\Tr\!\left(
D_x U D_y U^\dagger
\right)
\right]
\dd\mathbf x\dd\mathbf y\dd U\dd U_0,
\end{aligned}
\]
where we used the bi-invariance of Haar measure to pass from
\((U_0,U_1)\) to \((U_0,U)\). The density is independent of the absolute
orientation \(U_0\), so integrating \(U_0\) out gives
\begin{equation}\label{eq:angular-joint}
q_\alpha(\mathbf x,\mathbf y,U)
=
\widetilde C_{\alpha,N}\,
\Delta(\mathbf x)^2\Delta(\mathbf y)^2
\exp\!\left[
-\frac{\|\mathbf x\|^2+\|\mathbf y\|^2}{2\sigma^2}
+
k\Tr\!\left(
D_x\,
UD_yU^\dagger
\right)
\right].
\end{equation}
The density is invariant under \(U\mapsto D_0UD_1\)
for diagonal unitaries \(D_0,D_1\), reflecting the
arbitrary eigenvector phases. In particular,
\(Z_\alpha(e^{i\theta}U)=Z_\alpha(U)\), so the angular
density descends to \(PU(N)\).

The spectral normalizing constant in \eqref{eq:angular-joint} is
\[
\widetilde C_{\alpha,N}
=C_{\alpha,N}(c_N^{\mathrm W})^2
=\frac{(\sigma^2)^{-N^2/2}}{\mathcal I_N^2}.
\]
Consequently \(\int Z_\alpha(U)\dd U=1\). As a check, at \(\alpha=0\),
the two ordered eigenvalue densities are independent and
\(Z_0(U)=1\).

Integrating the two eigenvalue lists gives the density of the
relative eigenbasis with respect to Haar measure:
\begin{equation}\label{eq:angular-density-explicit}
Z_\alpha(U)
=
\int_{x_1>\cdots>x_N}
\int_{y_1>\cdots>y_N}
q_\alpha(\mathbf x,\mathbf y,U)\dd\mathbf y\dd\mathbf x.
\end{equation}
Hence, for fixed \(U\), the conditional density of the two eigenvalue
lists is
\begin{equation}\label{eq:conditional-spectral-density}
q_\alpha(\mathbf x,\mathbf y\mid U)
=
\frac{q_\alpha(\mathbf x,\mathbf y,U)}{Z_\alpha(U)}.
\end{equation}

The next two steps establish the spectral estimates
needed to bound the Riemannian Hessian of
\[
\Phi(U):=\log Z_\alpha(U).
\]
By the phase invariance established above, \(\Phi\)
may be regarded as a function on \(PU(N)\), equipped
with the quotient Riemannian metric defined in
\cref{prop:Gaussian-coupling-LSI}.

\subsubsection{Uniform control of the conditional spectra}

The derivatives of the angular density \(Z_\alpha(U)\) depend on
the extreme eigenvalues of the two Hermitian matrices. We therefore
need to control their conditional second moments uniformly over
the relative eigenbasis.

For fixed \(U\), let \(\Pr_U\) and \(\E_U\) denote probability and
expectation with respect to the conditional density
\(q_\alpha(\mathbf x,\mathbf y\mid U)\) from
\eqref{eq:conditional-spectral-density}.

\begin{lemma}\label{lem:conditional-spectral-moment}
For \(0\leq\alpha\leq1/2\),
\begin{equation}\label{eq:conditional-spectral-moment}
\sup_{U\in U(N)}
\E_U\!\left[
    \|\mathbf x\|_\infty^2+\|\mathbf y\|_\infty^2
\right]
\leq C_0N
\end{equation}
for a universal constant \(C_0\).
\end{lemma}

The proof proceeds by conditioning on one eigenvalue list 
\(\mathbf y = (y_1,\ldots,y_N)\).
The resulting law of the other list contains a linear external
field, which generally depends on the index \(i\) of the eigenvalue.
For a constant external field, however, the law reduces to
a translated GUE spectrum. We will compare the actual field
with two constant fields using a monotonicity property of
multivariate totally positive measures of order \(2\)
(MTP\(_2\)) \citep{karlin1980classes,fallat2017total}. 
We first record this property.

For \(x,y\in\R^n\), define their coordinatewise minimum and
maximum by
\[
(x\wedge y)_i:=\min\{x_i,y_i\},
\qquad
(x\vee y)_i:=\max\{x_i,y_i\}.
\]
Recall that a set \(D\subseteq\R^n\) is called a sublattice if
\[
x,y\in D
\qquad\Longrightarrow\qquad
x\wedge y,\ x\vee y\in D.
\]

\begin{lemma}[Monotonicity under an external field]
\label{lem:MTP2-external-field}
Let \(D\subseteq\R^n\) be a sublattice, and let \(r\) be a
probability density on \(D\), extended by zero outside \(D\).
Assume that \(r\) is MTP\(_2\), meaning that
\[
r(x\wedge y)\,r(x\vee y)
\geq r(x)\,r(y)
\qquad
\text{for all }x,y\in D.
\]
The zero extension is then also MTP\(_2\) on \(\R^n\),
since \(D\) is a sublattice.

For \(h\in\R^n\), define the exponentially tilted probability
density
\[
r_h(x)
:=
\frac{1}{Z_h}
\exp\!\left(\frac{h\cdot x}{\sigma^2}\right)r(x),
\]
whenever the normalizing constant \(Z_h\) is finite, and let
\(\E_h\) denote expectation with respect to \(r_h\).

Then, for any bounded coordinatewise increasing function
\(\varphi:D\to\R\),
\[
h\leq h'
\qquad\Longrightarrow\qquad
\E_h\varphi
\leq
\E_{h'}\varphi,
\]
where the inequality between \(h\) and \(h'\) is coordinatewise.
\end{lemma}

\begin{proof}
Multiplication by a linear exponential tilt preserves the MTP\(_2\)
property. Hence increasing functions are positively associated under
\(r_h\):
\[
\Cov_h(f,g) \geq 0
\]
for bounded coordinatewise increasing functions \(f\) and \(g\), by the
continuous association theorem for MTP\(_2\) measures
\citep{karlin1980classes}; see also \citet[Eq.~(3.5)]{fallat2017total}.

If \(h\leq h'\), the function
\[
T(x):=\exp\!\left(\frac{(h'-h)\cdot x}{\sigma^2}\right)
\]
is coordinatewise increasing and satisfies
\(\E_h T=Z_{h'}/Z_h<\infty\).
Apply positive association to \(\varphi\) and \(\min\{T,M\}\),
then let \(M\to\infty\). Since \(\varphi\) is bounded,
\[
\E_{h'}\varphi
=\frac{\E_h[\varphi T]}{\E_h T}
\geq\E_h\varphi.
\]
\end{proof}

\begin{proof}[Proof of \cref{lem:conditional-spectral-moment}]
Fix \(U\) and set
\[
Q_{ij}:=|U_{ij}|^2.
\]
The matrix \(Q\) is doubly stochastic. The eigenvalue variable
\(\mathbf x\) ranges over the ordered Weyl chamber
\(\mathcal W_N\), which is a sublattice under coordinatewise
minimum and maximum.
All conditional polynomial moments are finite: since
\(\|Q\|_{2\to2}\leq1\) and \(\alpha<1\), the quadratic confinement
in \eqref{eq:angular-joint} bounds the density by a polynomial multiplied
by an integrable Gaussian.

Conditional on \(\mathbf y\), the density of \(\mathbf x\) under \(\Pr_U\),
obtained from \eqref{eq:angular-joint}, is
\begin{equation}\label{eq:eigen-external-field}
p_h(\mathbf x)
\propto
\Delta(\mathbf x)^2
\exp\!\left[
-\frac{\|\mathbf x\|^2-2h\cdot\mathbf x}{2\sigma^2}
\right],
\qquad
h=\alpha Q\mathbf y.
\end{equation}
Since \(Q\) is doubly stochastic,
\[
|h_i|
=
\alpha |(Q\mathbf y)_i|
\leq
\alpha\|\mathbf y\|_\infty.
\]
Set
\[
a:=\alpha\|\mathbf y\|_\infty.
\]
Then
\begin{equation}\label{eq:external-field-sandwich}
-a\1
\leq
h
\leq
a\1
\end{equation}
coordinatewise.

This is the point at which
\cref{lem:MTP2-external-field} becomes useful. Were the external field in
\eqref{eq:eigen-external-field} a constant, the resulting
eigenvalue law would be completely explicit: for \(h=c\1\),
completing the square results in
\[
\mathbf x
\stackrel{d}{=}
\sigma\mathbf x(G)+c\1,
\]
where \(G\) is a standard Hermitian Gaussian matrix. Thus
\eqref{eq:external-field-sandwich} allows us to compare the
nonconstant field \(h=\alpha Q\mathbf y\) with two translated GUE laws,
provided that the eigenvalue law is monotone in \(h\).

To verify this monotonicity, remove the linear tilt from
\eqref{eq:eigen-external-field}. The logarithm of the remaining density
on \(\mathcal W_N\) is, up to an additive constant,
\[
2\sum_{i<j}\log(x_i-x_j)
-
\frac{\|\mathbf x\|^2}{2\sigma^2}.
\]
For \(i\neq j\),
\[
\frac{\partial^2}{\partial x_i\partial x_j}
\left(
2\sum_{a<b}\log(x_a-x_b)
-
\frac{\|\mathbf x\|^2}{2\sigma^2}
\right)
=
\frac{2}{(x_i-x_j)^2}
\geq0.
\]
To justify MTP\(_2\) on the chamber, including its support restriction,
we can check the lattice inequality directly. For each \(i<j\), the
factor \((x_i-x_j)^2\) is MTP\(_2\) on the chamber.
Indeed, for two pairs \((a,b)\) and \((c,d)\) with \(a>b\) and \(c>d\),
comparable pairs give equality. In the remaining case, after interchanging
the pairs if necessary, \(a\leq c\) and \(b\geq d\), and
\[
(a-d)(c-b)-(a-b)(c-d)=(c-a)(b-d)\geq0.
\]
All gaps here are positive, so the inequality remains true after
squaring. Multiplying over \(i<j\), and observing that the product
of one-coordinate Gaussian factors gives equality in the lattice
inequality, proves MTP\(_2\). Extension by zero outside the sublattice
\(\mathcal W_N\) preserves the property. Thus,
\cref{lem:MTP2-external-field} applies.

We can now apply the comparison principle to the extreme
eigenvalues. The quantity we ultimately wish to bound,
\(\|\mathbf x\|_\infty^2\), is not coordinatewise increasing,
so the monotonicity lemma does not apply to it directly.
Instead, we control its tails through the largest and
smallest eigenvalues. Since the eigenvalues are ordered,
\[
\|\mathbf x\|_\infty
=
\max\{x_1,-x_N\}.
\]
The event \(\{x_1>s\}\) is coordinatewise increasing,
whereas \(\{x_N<s\}\) is coordinatewise decreasing.
The comparison lemma therefore provides, using
\eqref{eq:external-field-sandwich},
\begin{equation}\label{eq:h-to-const-pos-concentration}
\Pr_h[x_1>s]
\leq
\Pr_{a\1}[x_1>s],
\end{equation}
and
\begin{equation}\label{eq:h-to-const-neg-concentration}
\Pr_h[x_N<s]
\leq
\Pr_{-a\1}[x_N<s].
\end{equation}

For the constant fields \(h=\pm a\1\), the corresponding
eigenvalue laws are those of \(\sigma G\pm aI_N\). Therefore the
comparison bounds \eqref{eq:h-to-const-pos-concentration} and 
\eqref{eq:h-to-const-neg-concentration} imply, for any \(t\geq0\),
\[
\begin{aligned}
\Pr_U\!\left[
    x_1>a+t
    \,\middle|\,
    \mathbf y
\right]
& \leq
\Pr\!\left[
    \sigma x_1(G)>t
\right],
\\
\Pr_U\!\left[
    -x_N>a+t
    \,\middle|\,
    \mathbf y
\right]
& \leq
\Pr\!\left[
    -\sigma x_N(G)>t
\right].
\end{aligned}
\]
Combining the two bounds yields
\begin{equation}\label{eq:conditional-extreme-tail}
\Pr_U\!\left[
    \|\mathbf x\|_\infty>a+t
    \,\middle|\,
    \mathbf y
\right]
\leq
2\Pr\!\left[
    \sigma\|G\|_{\op}>t
\right].
\end{equation}

We first derive the desired spectral moment bound in terms of
the operator norm of a standard Hermitian Gaussian matrix.
Denoting \(a=\alpha\|\mathbf y\|_\infty\), integration of
\eqref{eq:conditional-extreme-tail} yields
\[
\E_U\!\left[
    \bigl(\|\mathbf x\|_\infty-a\bigr)_+^2
    \,\middle|\,
    \mathbf y
\right]
\leq
2\sigma^2\E\|G\|_{\op}^2,
\]
where \(t_+:=\max\{t,0\}\).
Since
\[
\|\mathbf x\|_\infty
\leq
a+\bigl(\|\mathbf x\|_\infty-a\bigr)_+,
\]
Minkowski's inequality yields
\[
\left(
\E_U\!\left[
    \|\mathbf x\|_\infty^2
    \,\middle|\,
    \mathbf y
\right]
\right)^{1/2}
\leq
\alpha\|\mathbf y\|_\infty
+
\sqrt{2\E\|G\|_{\op}^2},
\]
where we used \(\sigma\leq1\).

It remains to remove the conditioning on \(\mathbf y\).
Taking the \(L^2(\Pr_U)\) norm and applying Minkowski's
inequality once more, we obtain
\[
s_x
\leq
\alpha s_y+\sqrt{2\E\|G\|_{\op}^2},
\]
where
\[
s_x
:=
\left(
\E_U\|\mathbf x\|_\infty^2
\right)^{1/2},
\qquad
s_y
:=
\left(
\E_U\|\mathbf y\|_\infty^2
\right)^{1/2}.
\]
Interchanging the two eigenvalue lists, the same argument provides
\[
s_y
\leq
\alpha s_x+\sqrt{2\E\|G\|_{\op}^2},
\]
since \(Q^{\mathsf T}\) is also doubly stochastic.

Adding the inequalities for \(s_x\) and \(s_y\) and using \(\alpha\leq1/2\),
we bound
\[
s_x+s_y
\leq
\frac{2\sqrt{2\E\|G\|_{\op}^2}}{1-\alpha}
\leq
4\sqrt{2\E\|G\|_{\op}^2}.
\]
Consequently,
\[
\E_U\!\left[
    \|\mathbf x\|_\infty^2+\|\mathbf y\|_\infty^2
\right]
=
s_x^2+s_y^2
\leq
32\E\|G\|_{\op}^2.
\]

Finally, for the stated Hermitian Gaussian normalization,
an elementary net argument provides
\[
\E\|G\|_{\op}^2\leq C_GN
\]
with a universal constant \(C_G\).
Indeed, take a \(1/4\)-net \(\mathcal N\) of the complex unit
sphere with \(|\mathcal N|\leq9^{2N}\). Then
\(\|G\|_{\op}\leq2\max_{v\in\mathcal N}|v^\dagger Gv|\),
and each \(v^\dagger Gv\) is a real standard Gaussian.
The union bound and integration of the Gaussian tail yield
\[
\E\|G\|_{\op}^2
\leq
8\log(2|\mathcal N|)+8
\leq64N.
\]
Thus \eqref{eq:conditional-spectral-moment} holds with
\(C_0=32C_G\), for example \(C_0=2048\).

\end{proof}

\subsubsection{Curvature of the angular density}

\begin{proof}[Proof of \cref{prop:Gaussian-coupling-LSI}]
We use the standard Bakry--\'Emery curvature criterion
\citep{bakry1985diffusions,bakry2014analysis}. The Bakry--\'Emery curvature 
criterion states that if a probability measure
\[
\mathrm{d}\nu=e^\Phi\dd\mathrm{vol}_g
\]
on a Riemannian manifold satisfies
\[
\Ric-\Hess\Phi
\succeq
\rho\,g
\]
for some \(\rho>0\), then it satisfies the
logarithmic Sobolev inequality at scale \(\rho^{-1}\). In the
normalization we use,
\[
\mathrm{Ent}_\nu(f^2)
\leq
\frac{2}{\rho}
\int \|\nabla f\|^2\dd\nu.
\]

Thus, in order to prove
\cref{prop:Gaussian-coupling-LSI}, it is enough to establish the
Bakry--\'Emery curvature lower bound of order \(N\) for the angular
measure \(Z_\alpha(U)\).

In our setting, the law \(\mu_\alpha\) of the relative
eigenbasis has density
\[
\mathrm{d}\mu_\alpha(U)
=
e^{\Phi(U)}\dd U,
\qquad
\Phi(U)=\log Z_\alpha(U),
\]
where \(\mathrm{d}U\) denotes normalized Haar measure on \(U(N)\).

Since both \(Z_\alpha\) and the functions considered in
\cref{prop:Gaussian-coupling-LSI} are invariant under scalar phases,
we may work on \(SU(N)\). Indeed, normalized Haar measure on \(U(N)\)
is the law of \(e^{i\Theta}S\), where \(S\) is Haar on \(SU(N)\)
and \(\Theta\) is independent and uniform on \([0,2\pi/N)\)
\citep[Lemma~16]{meckes2013spectral}.
Consequently, expectations, entropies, and Dirichlet energies
of phase-invariant functions under \(\mu_\alpha\) agree with
those computed on \(SU(N)\) with density \(Z_\alpha(S)\)
relative to Haar measure.
The case \(N=1\) is trivial, so we assume \(N\geq2\) below.

We equip \(SU(N)\) with the unnormalized Hilbert--Schmidt metric
\[
g_U(XU,YU)
:=
\Re\Tr(X^\dagger Y),
\qquad
X,Y\in\mathfrak{su}(N).
\]
For this normalization, \citet[Prop.~5.13]{meckes2019random} gives
\[
\Ric_{SU(N)}
=
\frac{N}{2}\,g.
\]
Therefore, it remains to
control the curvature contributed by the density \(e^\Phi\). 
It suffices to prove
\begin{equation}\label{eq:angular-Hessian-target}
\Hess\Phi
\preceq
C_1\alpha N\,g.
\end{equation}
Indeed, this would result in
\[
\Ric-\Hess\Phi
\succeq
N\left(
    \frac12-C_1\alpha
\right)g,
\]
which is bounded below by \(Ng/4\) whenever
\(0\leq\alpha\leq\alpha_0\), with
\(\alpha_0\leq\min\{1/2,(4C_1)^{-1}\}\). The Bakry--\'Emery criterion
would then yield a logarithmic Sobolev constant \(O(N^{-1})\).

To prove \eqref{eq:angular-Hessian-target}, it is enough to control the
second derivative of \(\Phi\) along unit-speed geodesics. Indeed, for
\(U\in SU(N)\) and \(X\in\mathfrak{su}(N)\) with
\[
\|X\|_F=1,
\]
the curve
\[
U(t):=e^{tX}U
\]
is a unit-speed geodesic for the bi-invariant Hilbert--Schmidt metric.
The standard geodesic characterization of the Riemannian Hessian gives
\[
\left.
\frac{\mathrm{d}^2}{\mathrm{d}t^2}
\Phi(U(t))
\right|_{t=0}
=
\Hess\Phi_U(XU,XU).
\]
Consequently, \eqref{eq:angular-Hessian-target} follows once we show,
uniformly in \(U\) and in every \(X\in\mathfrak{su}(N)\) with
\(\|X\|_F=1\), that
\begin{equation}\label{eq:angular-geodesic-target}
\left.
\frac{\mathrm{d}^2}{\mathrm{d}t^2}
\Phi(e^{tX}U)
\right|_{t=0}
\leq
C_1\alpha N.
\end{equation}

We therefore fix \(U\) and such a unit tangent direction \(X\), and
study the function \(t\mapsto\Phi(e^{tX}U)\).

For fixed eigenvalue lists \(\mathbf x\) and \(\mathbf y\), denote
\[
D_x:=\diag(\mathbf x),
\qquad
B:=UD_yU^\dagger.
\]
Along the geodesic \(U(t)=e^{tX}U\), the only part of the exponent in
\eqref{eq:angular-joint} that depends on \(t\) is the coupling term
\[
S(t)
:=
k\Tr\!\left(
D_x e^{tX} B e^{-tX}
\right).
\]
Its first two derivatives at \(t=0\) are
\[
S'(0)
=
k\Tr(D_x[X,B]),
\qquad
S''(0)
=
k\Tr(D_x[X,[X,B]]).
\]

Since
\[
Z_\alpha(U(t))
=
\int
\exp(S(t))\dd\zeta(\mathbf x,\mathbf y),
\]
where \(\mathrm{d}\zeta\) collects all factors in
\eqref{eq:angular-joint} that are independent of \(t\), differentiating
the logarithm of this partition function gives
\begin{equation}\label{eq:angular-Hessian-identity}
\left.
\frac{\mathrm{d}^2}{\mathrm{d}t^2}
\Phi(e^{tX}U)
\right|_{t=0}
=
\E_U[S''(0)]
+
\Var_U(S'(0)).
\end{equation}
Here \(\E_U\) and \(\Var_U\) refer to the conditional eigenvalue law
\(q_\alpha(\mathbf x,\mathbf y\mid U)\). Differentiation under the integral is
justified by dominated convergence: for fixed \(\alpha<1\), the
quadratic confinement remains uniformly positive, while the
differentiated integrands are polynomial factors times an integrable
Gaussian envelope.

We estimate these two terms separately. The expectation
term can be bounded directly using
\cref{lem:conditional-spectral-moment}. For the variance term, 
we use the strong convexity of the
conditional potential \(\mathcal V_U\) via the
Brascamp--Lieb Poincar\'e inequality.

We begin with the expectation term.
Using
\[
S''(0)
=
k\Tr(D_x[X,[X,B]]),
\]
and the cyclicity of trace,
\[
S''(0)
=
k\Tr([D_x,X][X,B]).
\]
Hence, by Cauchy--Schwarz,
\[
|S''(0)|
\leq
k\|[D_x,X]\|_F\|[X,B]\|_F.
\]
Since \(\|X\|_F=1\),
\[
\|[D_x,X]\|_F
\leq
2\|D_x\|_{\op},
\qquad
\|[X,B]\|_F
\leq
2\|B\|_{\op},
\]
and therefore
\[
|S''(0)|
\leq
4k\|D_x\|_{\op}\|B\|_{\op}.
\]
Using
\(
\|D_x\|_{\op}
=
\|\mathbf x\|_\infty
\) and
\(
\|B\|_{\op}
=
\|\mathbf y\|_\infty
\),
we obtain
\[
\bigl|\E_U[S''(0)]\bigr|
\leq
4k\,
\E_U\!\left[
    \|\mathbf x\|_\infty\|\mathbf y\|_\infty
\right].
\]
By Cauchy--Schwarz and
\cref{lem:conditional-spectral-moment},
\begin{equation}\label{eq:angular-second-derivative-mean}
\bigl|\E_U[S''(0)]\bigr|
\leq
2C_0kN.
\end{equation}

It remains to control the variance term
\(\Var_U(S'(0))\). We use the strong convexity of the
conditional eigenvalue law together with an extension of the Brascamp--Lieb
Poincar\'e inequality \citep{brascamp1976extensions,chafai2020poincare}.

For fixed \(U\), the conditional density of \((\mathbf x,\mathbf y)\) has the
Gibbs form
\[
q_\alpha(\mathbf x,\mathbf y\mid U)
\propto
\exp\!\left(
-\mathcal V_U(\mathbf x,\mathbf y)
\right)
\]
on \(\mathcal W_N\times\mathcal W_N\), where, up to an additive
constant,
\[
\mathcal V_U(\mathbf x,\mathbf y)
=
\frac{1}{2\sigma^2}
\left(
\|\mathbf x\|^2+\|\mathbf y\|^2
-2\alpha\mathbf x^{\mathsf T}Q\mathbf y
\right)
-2\log\Delta(\mathbf x)
-2\log\Delta(\mathbf y),
\]
and
\[
Q_{ij}:=|U_{ij}|^2.
\]

We first establish a uniform lower bound on the Hessian of
\(\mathcal V_U\). The quadratic part has Hessian
\[
\frac{1}{\sigma^2}
\begin{pmatrix}
I&-\alpha Q\\
-\alpha Q^{\mathsf T}&I
\end{pmatrix}.
\]
Since \(Q\) is doubly stochastic,
\[
\|Q\|_{2\to2}\leq1.
\]
Therefore, for \(x,y\in\R^N\),
\[
\begin{aligned}
\begin{pmatrix}x\\y\end{pmatrix}^{\mathsf T}
\begin{pmatrix}
I&-\alpha Q\\
-\alpha Q^{\mathsf T}&I
\end{pmatrix}
\begin{pmatrix}x\\y\end{pmatrix}
&=
\|x\|^2+\|y\|^2-2\alpha x^{\mathsf T}Qy\\
&\geq
(1-\alpha)\bigl(\|x\|^2+\|y\|^2\bigr).
\end{aligned}
\]
Using \(\sigma^2=1-\alpha^2\), this gives
\[
\frac{1}{\sigma^2}
\begin{pmatrix}
I&-\alpha Q\\
-\alpha Q^{\mathsf T}&I
\end{pmatrix}
\succeq
\frac{1}{1+\alpha}I.
\]

The remaining terms in \(\mathcal V_U\) only improve convexity:
\[
-2\log\Delta(\mathbf x)
\qquad\text{and}\qquad
-2\log\Delta(\mathbf y)
\]
are convex on the ordered Weyl chamber. Hence
\begin{equation}\label{eq:conditional-potential-convexity}
\nabla^2\mathcal V_U
\succeq
\frac{1}{1+\alpha}I.
\end{equation}

We use the following version of the Brascamp--Lieb Poincar\'e
inequality, which allows potentials that are infinite outside
a convex domain \citep{brascamp1976extensions,chafai2020poincare}.
If a probability measure \(\nu\) on \(\R^d\) has density
proportional to \(e^{-V}\), where
\(V:\R^d\to(-\infty,+\infty]\) is lower semicontinuous and
\[
V(z)-\frac{m}{2}\|z\|^2
\]
is convex for some \(m>0\), then
\[
\Var_\nu(G)
\leq
\frac{1}{m}\int\|\nabla G\|^2\dd\nu
\]
for every smooth function \(G\) for which the function and
its gradient are square-integrable.

We apply this result to the conditional potential
\(\mathcal V_U\), extended by \(+\infty\) outside
\(\mathcal W_N\times\mathcal W_N\).
The extension is lower semicontinuous because the
logarithmic terms diverge at eigenvalue collisions.
The Weyl chamber is convex, and
\eqref{eq:conditional-potential-convexity} implies that
the extended potential is \(m\)-strongly convex with
\(m=(1+\alpha)^{-1}\). The conditional measure is
normalizable by Gaussian confinement.

The regularization argument in
\citet[Sec.~3.1, proof of Theorem~1.1]{chafai2020poincare}
therefore applies and provides
\begin{equation}\label{eq:conditional-eigen-Poincare}
\Var_U(G)
\leq
(1+\alpha)
\E_U\!\left[
    \|\nabla_{\mathbf x,\mathbf y}G\|^2
\right].
\end{equation}
In particular, it applies to \(G=S'(0)\), which is a
quadratic polynomial in \((\mathbf x,\mathbf y)\) and
hence, together with its gradient, is square-integrable
under the conditional measure.

We apply \eqref{eq:conditional-eigen-Poincare} to
\[
G=S'(0)
=
k\Tr(D_x[X,B]).
\]
Since \(D_x=\diag(\mathbf x)\),
\[
\nabla_{\mathbf x} S'(0)
=
k\,\diag([X,B]).
\]
Similarly, differentiating through
\(B=UD_yU^\dagger\) gives
\[
\nabla_{\mathbf y} S'(0)
=
k\,\diag\!\left(
U^\dagger[D_x,X]U
\right).
\]
Therefore
\[
\begin{aligned}
\|\nabla_{\mathbf x} S'(0)\|^2
& \leq
k^2\|[X,B]\|_F^2
\leq
4k^2\|B\|_{\op}^2, 
\\ 
\|\nabla_{\mathbf y} S'(0)\|^2
& \leq
k^2\|[D_x,X]\|_F^2
\leq
4k^2\|D_x\|_{\op}^2.
\end{aligned}
\]
Using again
\(
\|D_x\|_{\op}
=
\|\mathbf x\|_\infty
\) and
\(
\|B\|_{\op}
=
\|\mathbf y\|_\infty,
\)
together with \cref{lem:conditional-spectral-moment}, we obtain
\begin{equation}\label{eq:angular-first-derivative-variance}
\Var_U(S'(0))
\leq
4(1+\alpha)C_0k^2N.
\end{equation}

Combining
\eqref{eq:angular-second-derivative-mean} and
\eqref{eq:angular-first-derivative-variance} with
\eqref{eq:angular-Hessian-identity}, we get
\[
\left.
\frac{\mathrm{d}^2}{\mathrm{d}t^2}
\Phi(e^{tX}U)
\right|_{t=0}
\leq
C_0\bigl(2k+4(1+\alpha)k^2\bigr)N.
\]
Since \(k=\alpha/(1-\alpha^2)\), we have
\[
2k+4(1+\alpha)k^2
=
\frac{2\alpha}{(1-\alpha)^2}
\leq 8\alpha,
\qquad 0\leq\alpha\leq1/2.
\]
Hence
\begin{equation}\label{eq:angular-geodesic-bound}
\left.
\frac{\mathrm{d}^2}{\mathrm{d}t^2}
\Phi(e^{tX}U)
\right|_{t=0}
\leq
C_1\alpha N
\end{equation}
uniformly in \(U\) and every unit tangent direction \(X\), with
\(C_1=8C_0\).

Since \(U\) and the unit tangent direction \(X\) were arbitrary,
\eqref{eq:angular-geodesic-bound} establishes
\eqref{eq:angular-Hessian-target}:
\[
\Hess\Phi\preceq C_1\alpha N\,g.
\]
Combining this with
\[
\Ric_{SU(N)}=\frac N2\,g,
\]
we obtain
\[
\Ric-\Hess\Phi
\succeq
N\left(\frac12-C_1\alpha\right)g.
\]
For
\[
0\leq\alpha\leq\alpha_0
:=
\min\left\{\frac12,\frac{1}{4C_1}\right\},
\]
the Bakry--\'Emery curvature is therefore bounded below by
\(Ng/4\). The Bakry--\'Emery criterion yields
\[
\operatorname{Ent}_{\bar\mu_\alpha}(f^2)
\leq
\frac{8}{N}
\int_{PU(N)}
\|\nabla f\|^2\,d\bar\mu_\alpha,
\]
which proves \cref{prop:Gaussian-coupling-LSI}.

\end{proof}

\subsection{Proof of the planted tail estimate}

\begin{proof}[Proof of \cref{thm:PT}]
We first consider the positive planting, \(b=+1\), and fix an
incident pair \(P\subseteq V\). Denote
\[
X_P(W):=\Tr(A_W\rho_P)
=\frac1R\Tr(\Pi_P A_W),
\qquad
W\sim K_\alpha(V,\cdot),
\]
where \(\rho_P=\Pi_P/R\) and \(A_W=2\Pi_W-I_N\).
Our goal is to bound the probability of the complement of
\begin{equation}\label{eq:planted-good-event}
\mathcal G:=\left\{X_P(W)\geq\frac{\eta}{4}\right\}.
\end{equation}

\paragraph{Reduction to a Haar-random source plane.}
Although the theorem is conditional on \(P\), we may average over a
Haar-random \(R\)-plane inside \(V\). Indeed, the stabilizer
\[
G_V:=\{S\in U(N):SV=V\}
\]
acts transitively on \(\Gr(R,V)\), and \(K_\alpha(V,\cdot)\) is invariant
under this stabilizer. If \(SP=P'\), then
\(X_{P'}(SW)=X_P(W)\). Consequently, the conditional failure probability
is the same for any \(P\subseteq V\). Hence,
\begin{equation}\label{eq:restore-P-randomness}
\Pr\!\left[X_P(W)<\frac\eta4\,\middle|\,V,P\right]
=
\Pr_{P',W}\!\left[X_{P'}(W)<\frac\eta4\,\middle|\,V\right],
\end{equation}
where \(P' \sim \Gr(R,V)\) is Haar and is independent of
\(W\sim K_\alpha(V,\cdot)\) conditional on \(V\).
We use this randomized-\(P\) representation until the end of the proof.

The conditional mean over the source subspace \(P\) is
\begin{equation}\label{eq:source-overlap-mean}
\begin{aligned}
M(V,W)
&:=\E_P[X_P(W)\mid V,W]\\
&=\frac1f\Tr(\Pi_VA_W)
=\frac2f\Tr(\Pi_V \Pi_W)-1,
\end{aligned}
\end{equation}
where we have used
\(\E_P[\Pi_P\mid V,W]=(R/f)\Pi_V\).
If \(M(V,W)\geq\eta/2\) and
\(X_P(W)-M(V,W)\geq-\eta/4\), then \(\mathcal G\) occurs.
Therefore, by the union bound,
\begin{equation}\label{eq:two-scale-bad-event}
\begin{aligned}
\Pr(\mathcal G^c\mid V)
\leq{}&
\Pr\!\left[M(V,W)<\frac\eta2\,\middle|\,V\right]\\
&+\Pr\!\left[X_P(W)-M(V,W)<-\frac\eta4
\,\middle|\,V\right].
\end{aligned}
\end{equation}
We control these terms separately.

\paragraph{Fluctuations of the source subspace.}
Fix \(V,W\), and regard
\[
B:=\Pi_VA_W\Pi_V\big|_V
\]
as an operator on \(V\). Then \(\|B\|_{\op}\leq1\) and
\(X_P(W)=R^{-1}\Tr(\Pi_PB)\).
Fix \(P_0\in\Gr(R,V)\) and express \(P=SP_0\), with
\(S\) Haar on \(SU(f)\), acting on \(V\).
Equip \(SU(f)\) with the unnormalized Hilbert--Schmidt metric.
For \(Y\in\mathfrak{su}(f)\) with \(\|Y\|_F=1\) and
\(P(t)=e^{tY}P\), we have
\[
\begin{aligned}
\left|\left.\frac{\mathrm d}{\mathrm dt}X_{P(t)}(W)
\right|_{t=0}\right|
&=\frac1R\left|\Tr\bigl(Y[\Pi_P,B]\bigr)\right|\\
&\leq\frac1R\|[\Pi_P,B]\|_F
\leq\frac{2}{\sqrt R}.
\end{aligned}
\]
Thus the lifted function \(S\mapsto X_{SP_0}(W)\) is
\(2/\sqrt R\)-Lipschitz. The Haar concentration on \(SU(f)\)
\citep{meckes2013spectral,meckes2019random} states that an
\(L\)-Lipschitz function satisfies
\begin{equation}\label{eq:Lipschitz-Haar-concentration}
\Pr\bigl[|F(S)-\E F(S)|\geq t\bigr]
\leq 2\exp\!\left(-c\frac{ft^2}{L^2}\right).
\end{equation}
Since \(\E_P[X_P(W)\mid V,W]=M(V,W)\),
\begin{equation}\label{eq:incident-concentration}
\Pr\!\left[
|X_P(W)-M(V,W)|\geq t\,\middle|\,V,W
\right]
\leq 2e^{-cRft^2}.
\end{equation}
The estimate is uniform in \(W\). Taking \(t=\eta/4\) and
averaging over \(W\) yields
\begin{equation}\label{eq:source-fluctuation-tail}
\Pr\!\left[
X_P(W)-M(V,W)<-\frac\eta4\,\middle|\,V
\right]
\leq 2e^{-cRf\eta^2}.
\end{equation}

\paragraph{Fluctuations of the planted overlap.}
The degree-one calibration provides
\begin{equation}\label{eq:overlap-mean}
\E[M(V,W)\mid V]=\eta.
\end{equation}
To apply \cref{prop:Gaussian-coupling-concentration}, we express
\(M(V,W)\) in terms of the relative eigenbasis \(U\) of the coupled
Hermitian Gaussian matrices.

The spectral-coordinate calculation in \cref{sec:angular-density} 
shows that \(U_0\) is Haar
and independent of \((\mathbf x,\mathbf y,U)\) with the joint density 
\(q_\alpha\). To specify the conditional law at every starting
eigenbasis, fix \(u_0\in U(N)\), sample the ordered GUE spectrum
\(\mathbf x\), and set \(H=u_0D_xu_0^\dagger\). Then
\[
u_0^\dagger H'u_0
=\alpha D_x+\sigma u_0^\dagger Zu_0.
\]
By unitary invariance of the independent GUE noise,
\(u_0^\dagger Zu_0\) has the same law as \(Z\), independently of
\(\mathbf x\). With the randomized eigenvector phases used above,
the resulting law of \((\mathbf x,\mathbf y,U)\) is therefore
\(q_\alpha\) for every \(u_0\), not merely almost everywhere.

Now fix \(V\) and choose \(U_V\) such that \(U_VE_0=V\).
The conditional query law \(K_\alpha(V,\cdot)\) is obtained by taking
\(U_0=U_VS\), where \(S\) is Haar on \(U(f)\times U(f)\) and
independent of \(U\sim\mu_\alpha\), and setting
\(W=U_VSUE_0\). Set \(D_0:=\diag(I_f,-I_f)\).
Since \(S\) commutes with \(D_0\), it cancels
from the overlap, resulting in
\begin{equation}\label{eq:overlap-angular-form}
M(V,W)
=\frac1N\Tr(D_0UD_0U^\dagger).
\end{equation}
This representation holds for every fixed \(V\).

Define \(\psi(U):=N^{-1}\Tr(D_0UD_0U^\dagger)\).
It is invariant under scalar phases and so descends to \(PU(N)\).
For a unit-speed geodesic \(U(t)=e^{tY}U\) with
\(Y\in\mathfrak{su}(N)\) and \(\|Y\|_F=1\),
\[
\begin{aligned}
\left|\left.\frac{\mathrm d}{\mathrm dt}
\psi(U(t))\right|_{t=0}\right|
&=\frac1N\left|\Tr\bigl([D_0,UD_0U^\dagger]Y\bigr)\right|\\
&\leq\frac1N\|[D_0,UD_0U^\dagger]\|_F
\leq\frac2{\sqrt N},
\end{aligned}
\]
since both \(D_0\) and \(UD_0U^\dagger\) are Hermitian unitaries.
Applying \cref{prop:Gaussian-coupling-concentration} to \(\psi\)
and using \eqref{eq:overlap-mean}, we obtain
\begin{equation}\label{eq:overlap-concentration}
\Pr\!\left[
|M(V,W)-\eta|\geq t\,\middle|\,V
\right]
\leq 2e^{-cN^2t^2}.
\end{equation}
In particular,
\begin{equation}\label{eq:overlap-lower-tail}
\Pr\!\left[M(V,W)<\frac\eta2\,\middle|\,V\right]
\leq 2e^{-cN^2\eta^2}.
\end{equation}

\paragraph{Conclusion.}
Combining \eqref{eq:two-scale-bad-event},
\eqref{eq:source-fluctuation-tail}, and
\eqref{eq:overlap-lower-tail}, we bound
\[
\Pr(\mathcal G^c\mid V)
\leq 2e^{-cN^2\eta^2}+2e^{-cRf\eta^2}
\leq C e^{-c'Rf\eta^2},
\]
since \(N=2f\) and \(R\leq f/2\) imply \(N^2\geq8Rf\).
Equation~\eqref{eq:restore-P-randomness} restores the bound for
every fixed incident pair \(P\subseteq V\).

Finally, under the negative planting,
\(W\sim K_\alpha(V^\perp,\cdot)\).
The kernel commutes with complementation and
\(A_{W^\perp}=-A_W\), so the positive-sign result applies to
\(-\Tr(A_W\rho_P)\).
Thus, for \(b\in\{\pm1\}\),
\[
\Pr\!\left[
 b\Tr(A_W\rho_P)<\frac\eta4
\,\middle|\,V,P,b
\right]
\leq C e^{-cRf\eta^2}.
\]
\end{proof}
\section{The source-side posterior bound}\label{sec:source}

In this section we analyze how an arbitrary collective measurement 
of \(\rho^{\otimes T}_P\) concentrates 
the posterior distribution of the latent subspace \(V\).

Recall the posterior formulation from \cref{sec:posterior-bias}: 
for a measurement outcome \(z\) of positive probability, let
\[
q_z
=
\int
\E_{P\mid V}
\Tr\!\left(M_z\rho_P^{\otimes T}\right)
\dd\nu(V),
\]
where \(\nu\) is Haar measure on \(\Gr(f,2f)\). 
The posterior density of \(V\) relative to \(\nu\) is
\[
g_z(V)
=
\frac{
\E_{P\mid V}
\Tr\!\left(M_z\rho_P^{\otimes T}\right)
}{
q_z
},
\qquad
\int g_z(V)\dd\nu(V)=1,
\qquad
g_z \in L^2(\nu).
\]

Since \(\rho_P^{\otimes T}\) is block diagonal in the Schur 
decomposition, we may refine the transcript by recording 
the Schur label \(\lambda\). 
We therefore denote the refined transcript by \((\lambda,z)\) and 
the corresponding posterior density by \(g_{\lambda,z}\).

The goal of this section is to bound the second moment of the 
posterior density with respect to the Haar prior \(\nu\), 
\emph{i.e.}, its squared \(L^2(\nu)\)-norm
\[
    \|g_{\lambda,z}\|_2^2
    =
    \int_{\Gr(f,2f)}
    g_{\lambda,z}(V)^2\dd\nu(V).
\]
It will be convenient to work with its logarithm,
\[
    B(g_{\lambda,z})
    :=
    \log\|g_{\lambda,z}\|_2^2.
\]

The argument has three steps. First, after conditioning on the Schur 
label \(\lambda\), we derive an explicit expression for \(g_{\lambda,z}\). 
The key observation is that, conditioned on \(V\), the 
hidden \(R\)-dimensional subspace \(P\) is Haar-random inside \(V\), 
so its Schur-block state can be averaged exactly. 
Second, we bound \(\|g_{\lambda,z}\|_2\), uniformly over 
the measurement outcome. Finally, we compute the exact average of 
the resulting quadratic Casimir parameter over the Schur label \(\lambda\).

\subsection{Schur--Weyl reduction and inclusion absorption}

For a partition \(\lambda\vdash T\), let \([\lambda]\) denote
the Specht module, let \(V_{\lambda,n}\) denote the irreducible
polynomial \(U(n)\)-module indexed by \(\lambda\), and denote
\[
    D_{\lambda,n}:=\dim V_{\lambda,n}.
\]
Using Schur--Weyl duality~\citep{fulton2013representation,keyl2001spectrum}, we decompose
\[
    (\C^{2f})^{\otimes T}
    =
    \bigoplus_{\substack{\lambda\vdash T\\ \ell(\lambda)\leq 2f}}
    V_{\lambda,2f}\otimes[\lambda].
\]

For a subspace \(S\subseteq\C^{2f}\) of dimension \(\dim S\geq \ell(\lambda)\), let
\(\Pi_{\lambda,S}\) denote the orthogonal projector from \(V_{\lambda,2f}\)
onto the natural copy of \(V_{\lambda,S}\).

Recall that the source state is
\[
    \rho_P=\frac{P}{R},
    \qquad P\in\Gr(R,V).
\]

Since \(\rho_P\) is maximally mixed on its \(R\)-dimensional support,
\(\rho_P^{\otimes T}\) has the Schur decomposition
\begin{equation}\label{eq:schur-weyl-decomp}
\rho_P^{\otimes T}
=
\bigoplus_{\substack{\lambda\vdash T\\ \ell(\lambda)\leq R}}
p_\lambda\,
\frac{\Pi_{\lambda,P}}{D_{\lambda,R}}
\otimes
\frac{I_{[\lambda]}}{\dim[\lambda]},
\end{equation}
where
\begin{equation}\label{eq:Schur-probability}
p_\lambda
=
\frac{\dim[\lambda]\,D_{\lambda,R}}{R^T},
\qquad
\ell(\lambda)\leq R.
\end{equation}
The distribution of the Schur label depends
only on the flat spectrum of \(\rho_P\) and not on the orientation of \(P\).

Let \(\Pi_\lambda^{\mathrm{Schur}}\) denote the projector onto
\(V_{\lambda,2f}\otimes[\lambda]\), and denote
\[
    M_{z,\lambda}
    :=
    \Pi_\lambda^{\mathrm{Schur}} M_z \Pi_\lambda^{\mathrm{Schur}} .
\]
Since \(\sum_\lambda\Pi_\lambda^{\mathrm{Schur}}=I\),
\(\{M_{z,\lambda}\}_{\lambda,z}\) is a POVM.
Every \(\rho^{\otimes T}\) commutes with the Schur projectors, hence
\[
\sum_\lambda\Tr(M_{z,\lambda}\rho^{\otimes T})
=\Tr(M_z\rho^{\otimes T}).
\]
Thus forgetting \(\lambda\) recovers the original transcript.
By the data processing inequality, revealing \(\lambda\) can only decrease affinity
and increase distinguishability. Thus, 
a lower bound for the refined measurement 
transfers to the unrefined one.

From \eqref{eq:schur-weyl-decomp}, it follows that
\[
\begin{aligned}
    \Pr(z\mid P,\lambda)
    &=
    \Tr\!\left[
        M_{z,\lambda}
        \left(
            \frac{\Pi_{\lambda,P}}{D_{\lambda,R}}\otimes
            \frac{I_{[\lambda]}}{\dim[\lambda]}
        \right)
    \right]  \\
    &=
    \frac{\Tr\!\left(E_{z,\lambda}\Pi_{\lambda,P}\right)}
         {D_{\lambda,R}},
\end{aligned}
\]
where
\[
    E_{z,\lambda}
    :=
    \frac{1}{\dim[\lambda]}
    \Tr_{[\lambda]} M_{z,\lambda}.
\]

Moreover,
\[
    \sum_z E_{z,\lambda}=I_{V_{\lambda,2f}},
\]
so \(\{E_{z,\lambda}\}_z\) is a POVM on \(V_{\lambda,2f}\).
Thus, for fixed \(\lambda\) and \(z\), we may work with the corresponding effect
\[
    E:=E_{z,\lambda}\ge0
\]
on \(V_{\lambda,2f}\).

\begin{lemma}[Inclusion absorption]\label{lem:absorption}
If \(P\) is Haar in \(\Gr(R,V)\), then
\begin{equation}\label{eq:absorption}
\E_{P\mid V}
\frac{\Pi_{\lambda,P}}{D_{\lambda,R}}
=
\frac{\Pi_{\lambda,V}}{D_{\lambda,f}}.
\end{equation}
\end{lemma}

\begin{proof}
The operator
\[
    \E_{P\mid V}
    \frac{\Pi_{\lambda,P}}{D_{\lambda,R}}
\]
is supported on \(V_{\lambda,V}\). The Haar distribution of \(P\subseteq V\)
is \(U(V)\)-invariant, so this operator commutes with the irreducible
\(U(V)\)-action on \(V_{\lambda,V}\). By Schur's lemma, it is therefore
a scalar multiple of \(\Pi_{\lambda,V}\). Its trace is one, and since
\(\Tr\Pi_{\lambda,V}=D_{\lambda,f}\), the scalar is \(1/D_{\lambda,f}\).
\end{proof}
Lemma~\ref{lem:absorption} therefore allows us to average the likelihood
\(\Pr(z\mid P,\lambda)\) over the conditional distribution of \(P\) given \(V\),
leaving a likelihood that depends only on \(V\).

We can now express the posterior \(g_{\lambda,z}\) explicitly. 
Fix \(\lambda\) and \(z\), and let \(E=E_{z,\lambda}\) be 
the corresponding effect on \(V_{\lambda,2f}\). 
Conditioned on \(V\) and \(\lambda\), by Lemma~\ref{lem:absorption},
\[
    \Pr(z\mid V,\lambda)
    =
    \frac{\Tr(E\Pi_{\lambda,V})}{D_{\lambda,f}}.
\]

Averaging over Haar \(V\in\Gr(f,2f)\) and using
\[
    \int_{\Gr(f,2f)}
    \Pi_{\lambda,V}\dd V
    =
    \frac{D_{\lambda,f}}{D_{\lambda,2f}}I,
\]
we obtain
\[
    \Pr(z\mid\lambda)
    =
    \frac{\Tr E}{D_{\lambda,2f}}.
\]
Since the prior distribution of \(V\) is Haar, Bayes' rule gives
\begin{equation}\label{eq:Schur-posterior}
g_{\lambda,z}(V)
=
\frac{D_{\lambda,2f}}{D_{\lambda,f}}
\frac{\Tr(E\Pi_{\lambda,V})}{\Tr E}.
\end{equation}
Thus the posterior produced by an arbitrary collective measurement 
is reduced, within each Schur block, to a scalar density on \(\Gr(f,2f)\).

Taking the spectral decomposition of \(E\),
\[
    E=\sum_a e_a\,|v_a\rangle\!\langle v_a|,
    \qquad e_a\geq 0,
\]
we may express
\[
    g_{\lambda,z}
    =
    \sum_a \frac{e_a}{\Tr E}\,g_{\lambda,v_a},
\]
where
\begin{equation}\label{eq:rank-one-posterior}
g_{\lambda,v}(V)
=
\frac{D_{\lambda,2f}}{D_{\lambda,f}}
\langle v,\Pi_{\lambda,V}v\rangle,
\qquad
|v|=1.
\end{equation}
Hence \(g_{\lambda,z}\) is a convex combination of rank-one posteriors,
and it suffices to prove a bound that is uniform in \(v\).

\subsection{A seed-uniform moment theorem}

Define the quadratic \(U(R)\) Casimir
\begin{equation}\label{eq:Casimir-def}
c_R(\lambda)
:=
\sum_{i=1}^R
\lambda_i(\lambda_i+R+1-2i).
\end{equation}

\begin{theorem}[Uniform source moment]\label{thm:source-moment}
Assume \(R\leq f/2\). For every Schur label \(\lambda\vdash T\) with
\(\ell(\lambda)\leq R\) and every measurement outcome \(z\) of positive
probability, the posterior \(g_{\lambda,z}\) satisfies
\begin{equation}\label{eq:source-moment}
B(g_{\lambda,z})
=
\log\|g_{\lambda,z}\|_2^2
\leq
\sum_{i=1}^R
\frac{\lambda_i^2}{f-i+1}
\leq
\frac{c_R(\lambda)}{f-R+1}
\leq
\frac{2c_R(\lambda)}{f}.
\end{equation}
In particular, the bound is uniform over the POVM effect \(E_{z,\lambda}\).
\end{theorem}

We will use the Littlewood--Richardson decomposition of tensor products of 
polynomial irreducible representations~\citep{fulton1997young,fulton2013representation}. For every pair of partitions \(\lambda,\mu\), there
are finite-dimensional multiplicity spaces
\(M_{\lambda,\mu}^{\nu}\), independent of the ambient vector space \(W\),
such that for every finite-dimensional complex vector space \(W\),
\begin{equation}\label{eq:LR-natural-decomposition}
V_{\lambda,W}\otimes V_{\mu,W}
\cong
\bigoplus_{\nu}
V_{\nu,W}\otimes M_{\lambda,\mu}^{\nu}.
\end{equation}
Here \(V_{\nu,W}=0\) when \(\ell(\nu) > \dim W\), and the decomposition is
natural with respect to linear embeddings of \(W\). The
Littlewood--Richardson coefficient is defined as
\[
    c_{\lambda,\mu}^{\nu}
    :=
    \dim M_{\lambda,\mu}^{\nu}.
\]

In particular,
\begin{equation}\label{eq:LR-tensor-square}
V_{\lambda,W}^{\otimes2}
\cong
\bigoplus_{\nu:\,c_{\lambda,\lambda}^{\nu}>0}
V_{\nu,W}\otimes M_{\lambda,\lambda}^{\nu}.
\end{equation}

\begin{lemma}\label{lem:LR-dimension-ratio}
Suppose

\[
    c_{\lambda,\lambda}^{\nu}
    =
    \dim M_{\lambda,\lambda}^{\nu}
    >0
\]

and \(D_{\nu,f}>0\). Then
\begin{equation}\label{eq:dimension-ratio-bound}
\log\left[
\left(\frac{D_{\lambda,2f}}{D_{\lambda,f}}\right)^2
\frac{D_{\nu,f}}{D_{\nu,2f}}
\right]
\leq
\sum_{i=1}^R
\frac{\lambda_i^2}{f-i+1}.
\end{equation}
\end{lemma}

\begin{proof}
The proof combines the hook--content formula with the Young 
diagram dominance \(\nu\preceq 2\lambda\).

Denote the boxes in the Young diagram of \(\lambda\) as 
\[
    \square_{i,j},
    \qquad
    1\leq i,\quad 1\leq j\leq \lambda_i,
\]
where \(i\) and \(j\) are the row and the column of the box,
and \(\lambda_i\) is the length of the \(i\)-th row. 
The content and hook length of the box are defined as
\[
    c(\square_{i,j})=j-i
\]
and
\[
    h(\square_{i,j})
    =
    \lambda_i-j+\lambda'_j-i+1,
\]
respectively. Here \(\lambda'_j\) is the length of the \(j\)-th column.

The hook--content formula~\citep{fulton1997young,Sagan2001} gives
\[
    D_{\lambda,n}
    =
    \prod_{\square_{i,j}\in\lambda}
    \frac{n+c(\square_{i,j})}{h(\square_{i,j})}.
\]
Hence
\[
    \frac{D_{\lambda,2f}}{D_{\lambda,f}}
    =
    \prod_{\square_{i,j}\in\lambda}
    \frac{2f+c(\square_{i,j})}{f+c(\square_{i,j})},
\]
since the hook-length factors cancel. Similarly,
\[
    \frac{D_{\nu,f}}{D_{\nu,2f}}
    =
    \prod_{\square_{i,j}\in\nu}
    \frac{f+c(\square_{i,j})}{2f+c(\square_{i,j})}.
\]

Define
\[
    \phi(x)
    :=
    \log\frac{2f+x}{f+x},
    \qquad x>-f.
\]
Taking logarithms, we may express
\[
    \log\left[
        \left(\frac{D_{\lambda,2f}}{D_{\lambda,f}}\right)^2
        \frac{D_{\nu,f}}{D_{\nu,2f}}
    \right]
    =
    2\sum_{\square_{i,j}\in\lambda}
    \phi\!\left(c(\square_{i,j})\right)
    -
    \sum_{\square_{i,j}\in\nu}
    \phi\!\left(c(\square_{i,j})\right).
\]
The Littlewood--Richardson condition
\(c_{\lambda,\lambda}^{\nu}>0\) implies
\[
    \nu\preceq 2\lambda
\]
in dominance order~\citep{fulton1997young,fulton2013representation}, where
\[
    2\lambda
    :=
    (2\lambda_1,2\lambda_2,\ldots).
\]
Equivalently, \(2\lambda\) can be obtained from \(\nu\) by a sequence
of moves that transfer a removable box from a lower row to an addable
position in a higher row, while maintaining a valid partition diagram. 
Each such move strictly increases the content of the moved box. 
Since \(\phi\) is decreasing,
\[
    \sum_{\square_{i,j}\in\nu}
    \phi\!\left(c(\square_{i,j})\right)
    \geq
    \sum_{\square_{i,j}\in2\lambda}
    \phi\!\left(c(\square_{i,j})\right).
\]
Moreover,
\[
    \sum_{\square_{i,j}\in2\lambda}
    \phi\!\left(c(\square_{i,j})\right)
    =
    \sum_{i=1}^R
    \sum_{j=1}^{2\lambda_i}
    \phi(j-i).
\]

Splitting each row of \(2\lambda\) into two blocks of length
\(\lambda_i\), we obtain
\[
\begin{aligned}
    &\log\left[
        \left(\frac{D_{\lambda,2f}}{D_{\lambda,f}}\right)^2
        \frac{D_{\nu,f}}{D_{\nu,2f}}
    \right] \\
    &\qquad\leq
    \sum_{i=1}^R
    \sum_{j=1}^{\lambda_i}
    \left[
        \phi(j-i)-\phi(j+\lambda_i-i)
    \right].
\end{aligned}
\]

For \(a\ge0\) and \(x>-f\),
\[
\begin{aligned}
    \phi(x)-\phi(x+a)
    &=
    \int_0^a
    \left(
        \frac{1}{f+x+s}
        -
        \frac{1}{2f+x+s}
    \right)\mathrm{d}s \\
    &\leq
    \frac{a}{f+x}.
\end{aligned}
\]

Taking \(a=\lambda_i\) and \(x=j-i\), and using
\[
    f+j-i\geq f-i+1,
\]
we obtain
\[
    \sum_{j=1}^{\lambda_i}
    \left[
        \phi(j-i)-\phi(j+\lambda_i-i)
    \right]
    \leq
    \frac{\lambda_i^2}{f-i+1}.
\]
Summing over \(i\) proves \eqref{eq:dimension-ratio-bound}.
\end{proof}

\begin{proof}[Proof of Theorem~\ref{thm:source-moment}]
Consider first a rank-one posterior \(g_{\lambda,v}\) from
\eqref{eq:rank-one-posterior}. Introduce the second-moment operator
\[
    \mathcal K_\lambda
    :=
    \int_{\Gr(f,2f)}
    \Pi_{\lambda,V}^{\otimes2}\dd V
\]
on \(V_{\lambda,2f}^{\otimes2}\). By definition,
\begin{equation}\label{eq:posterior-two-copy}
\|g_{\lambda,v}\|_2^2
=
\left(\frac{D_{\lambda,2f}}{D_{\lambda,f}}\right)^2
\langle v^{\otimes2},
\mathcal K_\lambda v^{\otimes2}\rangle.
\end{equation}

Application of \eqref{eq:LR-tensor-square} to
\(W=\C^{2f}\) yields
\begin{equation}\label{eq:LR-ambient}
V_{\lambda,2f}^{\otimes2}
=
\bigoplus_{\nu:\,c_{\lambda,\lambda}^{\nu}>0}
V_{\nu,2f}\otimes M_{\lambda,\lambda}^{\nu}.
\end{equation}
Fix a reference \(f\)-dimensional subspace
\(V_0\subseteq\C^{2f}\). Applying the same decomposition \eqref{eq:LR-tensor-square}
to \(W=V_0\) yields
\begin{equation}\label{eq:LR-subspace}
V_{\lambda,V_0}^{\otimes2}
=
\bigoplus_{\nu:\,c_{\lambda,\lambda}^{\nu}>0}
V_{\nu,V_0}\otimes M_{\lambda,\lambda}^{\nu},
\end{equation}
where \(V_{\nu,V_0}=0\) whenever \(\ell(\nu)>f\).

Because the decomposition
\eqref{eq:LR-natural-decomposition} is natural under the inclusion
\(V_0\hookrightarrow\C^{2f}\), the summand
\(V_{\nu,V_0}\otimes M_{\lambda,\lambda}^{\nu}\) in
\eqref{eq:LR-subspace} is contained in the corresponding summand
\(V_{\nu,2f}\otimes M_{\lambda,\lambda}^{\nu}\) in
\eqref{eq:LR-ambient}. Hence
\(\Pi_{\lambda,V_0}^{\otimes2}\) is block diagonal with respect to
\eqref{eq:LR-ambient}, and its restriction to the
\(\nu\)-isotypic component is
\begin{equation}\label{eq:Pi-restriction}
\left.
\Pi_{\lambda,V_0}^{\otimes2}
\right|_{\nu}
=
\Pi_{\nu,V_0}\otimes I_{M_{\lambda,\lambda}^{\nu}},
\end{equation}
where \(\Pi_{\nu,V_0}\) denotes the orthogonal projector from
\(V_{\nu,2f}\) onto \(V_{\nu,V_0}\).

Now express \(V=UV_0\), with \(U\in U(2f)\), and let
\(U_\lambda\) and \(U_\nu\) denote the actions of \(U\) on
\(V_{\lambda,2f}\) and \(V_{\nu,2f}\), respectively. Then
\[
    \Pi_{\lambda,V}^{\otimes2}
    =
    (U_\lambda\otimes U_\lambda)
    \Pi_{\lambda,V_0}^{\otimes2}
    (U_\lambda\otimes U_\lambda)^\dagger.
\]
Substituting this into \eqref{eq:Pi-restriction}, we 
obtain
\[
    \left.
    \Pi_{\lambda,V}^{\otimes2}
    \right|_{\nu}
    =
    \bigl(
        U_\nu\Pi_{\nu,V_0}U_\nu^\dagger
    \bigr)
    \otimes I_{M_{\lambda,\lambda}^{\nu}}.
\]

Averaging over the Haar-random \(U\in U(2f)\), we obtain
\begin{align}
\left.\mathcal K_\lambda\right|_{\nu}
&=
\left(
\int_{U(2f)}
U_\nu\Pi_{\nu,V_0}U_\nu^\dagger\,\mathrm{d}U
\right)
\otimes I_{M_{\lambda,\lambda}^{\nu}}
\nonumber\\
&=
a_\nu\,
I_{V_{\nu,2f}}
\otimes I_{M_{\lambda,\lambda}^{\nu}},
\label{eq:Klambda-block}
\end{align}
where the second equality follows from Schur's lemma, since
\(V_{\nu,2f}\) is irreducible. Taking the trace over \(V_{\nu,2f}\),
we determine the scalar
\[
    a_\nu
    =
    \frac{D_{\nu,f}}{D_{\nu,2f}}.
\]

Therefore
\begin{equation}\label{eq:Klambda-exact}
\mathcal K_\lambda
=
\bigoplus_{\nu:\,c_{\lambda,\lambda}^{\nu}>0}
\frac{D_{\nu,f}}{D_{\nu,2f}}\,
I_{V_{\nu,2f}}
\otimes
I_{M_{\lambda,\lambda}^{\nu}},
\end{equation}
and consequently
\begin{equation}\label{eq:Klambda-scalar}
\mathcal K_\lambda
\preceq
\max_{\nu:\,c_{\lambda,\lambda}^{\nu}>0}
\frac{D_{\nu,f}}{D_{\nu,2f}}\,I.
\end{equation}

Using \eqref{eq:Klambda-scalar} in
\eqref{eq:posterior-two-copy}, we obtain
\begin{equation}\label{eq:posterior-second-moment-dim}
\|g_{\lambda,v}\|_2^2
\leq
\left(\frac{D_{\lambda,2f}}{D_{\lambda,f}}\right)^2
\max_{\nu:\,c_{\lambda,\lambda}^{\nu}>0}
\frac{D_{\nu,f}}{D_{\nu,2f}}.
\end{equation}
We discard the terms with \(D_{\nu,f}=0\), since they do not contribute.
Applying Lemma~\ref{lem:LR-dimension-ratio} to every remaining term,
we obtain
\begin{equation}\label{eq:source-posterior-pre-Casimir}
\log\|g_{\lambda,v}\|_2^2
\leq
\sum_{i=1}^R
\frac{\lambda_i^2}{f-i+1}.
\end{equation}

Now let \(E=E_{z,\lambda}\) be an arbitrary effect. As shown above,
\(g_{\lambda,z}\) is a convex combination of the rank-one posteriors
\(g_{\lambda,v}\). By convexity of the squared \(L^2(\nu)\) norm,
\[
    \|g_{\lambda,z}\|_2^2
    \leq
    \max_{\|v\|=1}
    \|g_{\lambda,v}\|_2^2.
\]
Thus \eqref{eq:source-posterior-pre-Casimir} also holds for
\(g_{\lambda,z}\).

It remains to compare the right-hand side with the Casimir parameter
\(c_R(\lambda)\). Since
\[
    f-i+1\geq f-R+1,
\]
we have
\[
    \sum_{i=1}^R
    \frac{\lambda_i^2}{f-i+1}
    \leq
    \frac{\sum_{i=1}^R\lambda_i^2}{f-R+1}.
\]

Moreover,
\begin{align*}
c_R(\lambda)-\sum_{i=1}^R\lambda_i^2
&=
\sum_{i=1}^R
(R+1-2i)\lambda_i \\
&=
\sum_{i=1}^{\lfloor R/2\rfloor}
(R+1-2i)
\bigl(\lambda_i-\lambda_{R+1-i}\bigr) \\
&\ge0.
\end{align*}
Indeed, for \(i\leq\lfloor R/2\rfloor\),
\[
    R+1-2i>0,
    \qquad
    \lambda_i\geq\lambda_{R+1-i}.
\]
Hence
\[
    \sum_{i=1}^R\lambda_i^2
    \leq
    c_R(\lambda),
\]
and therefore
\[
    B(g_{\lambda,z})
    \leq
    \frac{c_R(\lambda)}{f-R+1}
    \leq
    \frac{2c_R(\lambda)}{f},
\]
where we have used \(R\leq f/2\).
\end{proof}

\subsection{The exact Schur--Casimir average}

Theorem~\ref{thm:source-moment} bounds the posterior in a fixed
Schur block in terms of \(c_R(\lambda)\). We now average this bound
over the Schur label \(\lambda\).

\begin{proposition}[Schur--Casimir average]\label{prop:schur-casimir-average}
Let \(\lambda\vdash T\) be distributed according to the Schur--Weyl measure

\[
    p_\lambda
    =
    \frac{\dim[\lambda]\,D_{\lambda,R}}{R^T},
    \qquad
    \ell(\lambda)\leq R.
\]

Then the average of the quadratic \(U(R)\) Casimir

\[
    c_R(\lambda)
    =
    \sum_{i=1}^R
    \lambda_i(\lambda_i+R+1-2i)
\]

over the Schur label \(\lambda\) is

\[
    \E_\lambda c_R(\lambda)
    =
    RT+\frac{T(T-1)}{R}.
\]

\end{proposition}

This is a standard Schur--Weyl character calculation; see, e.g.,
\cite{fulton2013representation,Sagan2001}. We include a short
derivation in \cref{app:schur-casimir-average-proof} to fix our
normalization.
\section{The query-side bounds}\label{sec:query}

In the previous section we analyzed how the posterior distribution 
of the latent subspace concentrates after the measurement. 
We now show that the Gaussian planting kernel strongly
contracts this posterior information.

Let \(\nu\) denote Haar probability measure on
\(\Gr(f,2f)\), and let \(g\in L^2(\nu)\) be a probability density:
\[
    g\geq 0,
    \qquad
    \int_{\Gr(f,2f)} g(V)\dd\nu(V)=1.
\]
Recall the notation
\[
    B(g):=\log\|g\|_2^2.
\]
Recall also the complementation operator \((Cg)(V):=g(V^\perp)\).

The Gaussian planting kernel constructed in
Section~\ref{sec:gaussian} factors as
\[
    K_\alpha=J^*T_\alpha J.
\]
Here \(J\) lifts a function on the Grassmannian to the Gaussian space of
Hermitian \(2f\times 2f\) matrices, \(T_\alpha\) is the
Gaussian Ornstein--Uhlenbeck operator, and \(J^*\) is the
conditional expectation given the top-\(f\)-dimensional eigenspace.

We use this factorization in two ways. First, in
\cref{prop:KL-contraction}, ordinary Gaussian hypercontractivity
\cite{Gro75} controls the entire lifted posterior at once. Since
\(J:L^p(\nu)\to L^p(\gamma_N)\) is isometric and
\(J^*:L^2(\gamma_N)\to L^2(\nu)\) is contractive, this
provides a second-moment bound for \(K_\alpha g\), and hence bounds the
divergences between planted queries and their Haar marginals.

For the high-confidence bound in \cref{thm:Renyi-contraction}, we
compare \(K_\alpha g\) and \(K_\alpha Cg\) directly. We split the
Gaussian noise into two stages. Forward hypercontractivity first reduces
the concentration of the two lifted densities. We then apply the
two-function reverse-hypercontractive inequality
\cite[Corollary~6.4]{mossel2013reverse} to the remaining noise to lower-bound their
affinity. Finally, conditional expectation through \(J^*\) can only
increase affinity.

Neither argument requires truncating the Grassmannian harmonic 
decomposition or separately summing over harmonic degrees.

For probability densities \(p\) and \(q\), for \(\alpha\in(0,1)\cup(1,\infty)\),
we denote the R\'enyi divergence of order \(\alpha\) with respect to a common
probability measure \(\mu\) by the standard definition
\citep{vanerven2014renyi}
\[
    D_\alpha(p\|q)
    :=
    \frac{1}{\alpha-1}
    \log
    \int
        p(x)^\alpha q(x)^{1-\alpha}
        \dd\mu(x).
\]
The Kullback--Leibler divergence is obtained in the limit
\[
    D_{\mathrm{KL}}(p\|q)
    :=
    \lim_{\alpha\to1} D_\alpha(p\|q),
\]
while for \(\alpha=\frac12\),
\[
    D_{1/2}(p\|q)
    =
    -2\log
    \int
        \sqrt{p(x)q(x)}
        \dd\mu(x).
\]
Denote the affinity of two probability densities \(p\)
and \(q\) with respect to a common probability measure \(\mu\) by
\[
    \Aff_\mu(p,q)
    :=
    \int \sqrt{p(x)q(x)}\dd\mu(x),
\]
so that
\[
    D_{1/2}(p\|q)=-2\log\Aff_\mu(p,q).
\]

For convenience, we restate \cref{prop:query-overview} here.
\begin{proposition}[Query-side contraction]
\label{prop:query-restatement}

Let \(g\in L^2(\nu)\) be a probability density with respect to \(\nu\), and set
\[
    B(g)=\log\|g\|_2^2.
\]
For the Gaussian-lift planting kernel of
\cref{prop:planting-overview},
\begin{equation}\label{eq:query-restated-second-moment}
    \|K_\alpha g\|_2^2
    =
    1+\sum_{\mu\ne0}
        \tau_\mu(\alpha)^2\|\Pi_\mu g\|_2^2
    \leq
    \exp\left(
        \frac{2\alpha^2}{1+\alpha^2}B(g)
    \right).
\end{equation}
Consequently,
\begin{equation}\label{eq:query-restated-KL}
    D_{\mathrm{KL}}(K_\alpha g\|1)
    \leq
    2\beta_*^{-2}\eta^2 B(g).
\end{equation}
For the two query laws corresponding to the opposite planted signs,
we also have
\begin{equation}\label{eq:query-restated-Renyi}
    D_{1/2}(K_\alpha g\|K_\alpha Cg)
    \leq
    24\beta_*^{-2}\eta^2B(g).
\end{equation}
\end{proposition}

\subsection{Hypercontractive control of the posterior second moment}

We begin with the estimate used in both arguments.

\begin{lemma}[Gaussian posterior contraction]
\label{lem:Gaussian-posterior-contraction}
Let \(g\in L^2(\nu)\) be a probability density and let
\(\xi\in[0,1)\) be any Gaussian Ornstein--Uhlenbeck correlation. Then
\begin{equation}\label{eq:Gaussian-HC-posterior}
    \|T_\xi Jg\|_{L^2(\gamma_N)}
    \leq
    \|g\|_{L^{1+\xi^2}(\nu)}.
\end{equation}
Consequently,
\begin{equation}\label{eq:lifted-moment}
    \log\|T_\xi Jg\|_{L^2(\gamma_N)}^2
    \leq
    \frac{2\xi^2}{1+\xi^2}B(g).
\end{equation}
\end{lemma}

\begin{proof}
The proof uses the coupling-transfer argument of
\citet[Fact~8.13 and proof of Lemma~8.14]{ellis2024productmixing},
applied here to the spectral-projector lift.

We now use Gaussian hypercontractivity \cite{Gro75}:
\[
    \|T_\xi F\|_{L^2(\gamma_N)}
    \leq
    \|F\|_{L^{1+\xi^2}(\gamma_N)}.
\]
The map \(J\) is an \(L^p\)-isometry, 
hence
\[
    \|Jg\|_{L^p(\gamma_N)}
    =
    \|g\|_{L^p(\nu)}
\]
for any \(p \geq 1\). Combining these inequalities, we prove
\eqref{eq:Gaussian-HC-posterior}.

It remains to bound \(\|g\|_{L^{1+\xi^2}(\nu)}\).
Let \(q\in[0,1]\). By H\"older interpolation between
\(L^1(\nu)\) and \(L^2(\nu)\),
\[
    \|g\|_{L^{1+q}(\nu)}
    \leq
    \|g\|_{L^1(\nu)}^{1-\theta}
    \|g\|_2^\theta,
    \qquad
    \theta=\frac{2q}{1+q}.
\]
Since
\[
    \|g\|_{L^1(\nu)}
    =
    \int_{\Gr(f,2f)}g(V)\dd\nu(V)
    =
    1,
\]
and
\[
    \|g\|_2^\theta
    =
    \exp\left(\frac{\theta}{2}B(g)\right),
\]
we obtain
\[
    \log\|g\|_{L^{1+q}(\nu)}^2
    \leq
    \frac{2q}{1+q}B(g).
\]
Taking \(q=\xi^2\) and using \eqref{eq:Gaussian-HC-posterior}, we
obtain \eqref{eq:lifted-moment}.
\end{proof}

We may present the same estimate using the harmonic decomposition.
Recall that by \cref{prop:planting-overview}, 
the planting kernel \(K_\alpha\) acts on \(\Harm_\mu\) by the multiplier 
\(\tau_\mu(\alpha)\) and that the constant component of \(g\) is \(1\).
Then,
\begin{equation}\label{eq:harmonic-second-moment}
    \|K_\alpha g\|_2^2
    =
    1+
    \sum_{\mu\neq0}
        \tau_\mu(\alpha)^2
        \|\Pi_\mu g\|_2^2.
\end{equation}
\subsection{Divergence contraction toward Haar measure}
Recall from \cref{sec:overview} that, conditioned on a transcript corresponding
to the posterior density \(g\) of the latent subspace, the two query
densities are
\[
    p_+ = K_\alpha g,
    \qquad
    p_- = K_\alpha Cg.
\]
Moreover, by Haar stationarity of \(K_\alpha\), the unconditional query 
marginal under either sign is Haar measure. Equivalently, its density with
respect to Haar measure \(\nu\) is the constant function \(1\). In the
multiple-observable game this is also the law of every decoy query.
We therefore need to control the divergences of \(K_\alpha g\) and
\(K_\alpha Cg\) from the Haar density \(1\).

\begin{proposition}[Divergence contraction toward Haar measure]
\label{prop:KL-contraction}
For every probability density \(g\in L^2(\nu)\),
\begin{align}
    D_{1/2}(K_\alpha g\|1)
    &\leq
    D_{\mathrm{KL}}(K_\alpha g\|1)
    \leq
    \frac{2\alpha^2}{1+\alpha^2}B(g)
    \leq
    2\beta_*^{-2}\eta^2B(g),
    \label{eq:KL-contraction}
    \\
    D_{1/2}(K_\alpha Cg\|1)
    &\leq
    D_{\mathrm{KL}}(K_\alpha Cg\|1)
    \leq
    \frac{2\alpha^2}{1+\alpha^2}B(g)
    \leq
    2\beta_*^{-2}\eta^2B(g).
    \label{eq:KL-contraction-complement}
\end{align}
\end{proposition}

\begin{proof}
Let \(p\) be a probability density with respect to \(\nu\). By monotonicity
of R\'enyi divergence in its order~\citep[Theorem~3]{vanerven2014renyi},
\begin{equation}\label{eq:d2dhalf-le-kl-le-d2}    
    D_{1/2}(p\|1)
    \leq
    D_{\mathrm{KL}}(p\|1)
    \leq
    D_2(p\|1).
\end{equation}
Since the reference density is \(1\),
\[
    D_2(p\|1)
    =
    \log
    \int_{\Gr(f,2f)}p(V)^2\dd\nu(V)
    =
    \log\|p\|_2^2.
\]
Taking \(p=K_\alpha g\), using the \(L^2\)-contractivity of \(J^*\)
in \(K_\alpha=J^*T_\alpha J\), and applying
\cref{lem:Gaussian-posterior-contraction},
\[
    D_{\mathrm{KL}}(K_\alpha g\|1)
    \leq
    \frac{2\alpha^2}{1+\alpha^2}B(g).
\]
The degree-one calibration in
\eqref{eq:eta-calibration},
\[
    \beta_*\alpha\leq\eta,
\]
implies
\[
    \alpha^2\leq\beta_*^{-2}\eta^2,
\]
and hence
\[
    D_{\mathrm{KL}}(K_\alpha g\|1)
    \leq
    2\beta_*^{-2}\eta^2B(g).
\]
The bound on \(D_{1/2}(K_\alpha g\|1)\) follows from
\eqref{eq:d2dhalf-le-kl-le-d2}.

For the complemented posterior \(Cg\), by Haar invariance under
\(V\mapsto V^\perp\),
\[
    \int_{\Gr(f,2f)}(Cg)(V)\dd\nu(V)
    =
    \int_{\Gr(f,2f)}g(V)\dd\nu(V)
    =
    1
\]
and
\[
    \|Cg\|_2
    =
    \|g\|_2.
\]
Thus \(B(Cg)=B(g)\), and the same argument proves
\eqref{eq:KL-contraction-complement}.
\end{proof}

\subsection{R\'enyi \texorpdfstring{\(1/2\)}{1/2}-contraction between the two signs}

To obtain a high-confidence lower bound, we need to compare
the query densities \(K_\alpha g\) and \(K_\alpha Cg\) for opposite signs
directly. We now prove that the same posterior budget \(B(g)\) controls
\[
    D_{1/2}(K_\alpha g\|K_\alpha Cg).
\]

The proof uses both forward and reverse Gaussian hypercontractivity.
Denote
\[
    Q_t:=T_{e^{-t}}.
\]
The Gaussian two-function reverse-hypercontractive inequality of
\citet{borell1982positivity}, in the form
\citep[Corollary~6.4]{mossel2013reverse}, implies that, for nonnegative functions
\(a,b\) and \(\tau\geq\log2\),
\begin{equation}\label{eq:reverse-HC}
    \int
        a(H)\,(Q_\tau b)(H)\dd\gamma_N(H)
    \geq
    \|a\|_{L^{1/2}(\gamma_N)}
    \|b\|_{L^{1/2}(\gamma_N)},
\end{equation}
where
\[
    \|a\|_{L^{1/2}(\gamma_N)}
    :=
    \left(
        \int a(H)^{1/2}\dd\gamma_N(H)
    \right)^2.
\]

\begin{theorem}[R\'enyi contraction between opposite signs]
\label{thm:Renyi-contraction}
Suppose \(\alpha\le2^{-1/2}\). Then every probability density
\(g\in L^2(\nu)\) satisfies
\begin{equation}\label{eq:Renyi-contraction}
    D_{1/2}(K_\alpha g\|K_\alpha Cg)
    \leq
    24\alpha^2B(g).
\end{equation}
\end{theorem}

\begin{proof}
For \(\alpha=0\), both
query densities are identically \(1\), and the claim is immediate.
Hence we may assume \(\alpha>0\). Set
\[
    t=-\log\alpha,
    \qquad
    s=\frac12\log2,
    \qquad
    u=t-s.
\]
The assumption \(\alpha\le2^{-1/2}\) guarantees \(u\ge0\).

Define the two intermediate Gaussian densities
\[
    h:=Q_uJg,
    \qquad
    k:=Q_uJ(Cg).
\]
Both are nonnegative and normalized. Indeed, the Ornstein--Uhlenbeck
semigroup preserves nonnegativity and the Gaussian integral, so
\[
    \int h(H)\dd\gamma_N(H)
    =
    \int k(H)\dd\gamma_N(H)
    =
    1.
\]

We first control the second moments \(\|h\|_{L^2(\gamma_N)}\) 
and \(\|k\|_{L^2(\gamma_N)}\) using forward hypercontractivity.
Using the Gaussian-space inequality \eqref{eq:lifted-moment} from
\cref{lem:Gaussian-posterior-contraction} with
\(\xi=e^{-u}=\sqrt2\alpha<1\) for \(u>0\), the identity \(Q_0=I\), 
and that \(J\) is an \(L^2\)-isometry when \(u=0\), we bound
\[
    \log\|h\|_{L^2(\gamma_N)}^2
    \leq
    \frac{2e^{-2u}}{1+e^{-2u}}B(g)
    \leq
    2e^{-2u}B(g).
\]
Using
\[
    e^{-2u}
    =
    e^{-2(t-s)}
    =
    e^{-2t}e^{2s}
    =
    2\alpha^2,
\]
we obtain
\begin{equation}\label{eq:Bh-bound}
    B_h:=
    \log\|h\|_{L^2(\gamma_N)}^2
    \leq
    4\alpha^2B(g).
\end{equation}
Since \(B(Cg)=B(g)\), by the same argument,
\begin{equation}\label{eq:Bk-bound}
    B_k:=
    \log\|k\|_{L^2(\gamma_N)}^2
    \leq
    4\alpha^2B(g).
\end{equation}

We now apply the remaining Ornstein--Uhlenbeck time \(s\). Since the
square root is concave and \(Q_s\) is a Markov operator,
\[
    \sqrt{(Q_sh)(H)}
    \geq
    (Q_s\sqrt h)(H),
    \qquad
    \sqrt{(Q_sk)(H)}
    \geq
    (Q_s\sqrt k)(H)
\]
for \(\gamma_N\)-almost every \(H\). Therefore
\begin{align}
    \Aff_{\gamma_N}(Q_sh,Q_sk)
    &=
    \int
        \sqrt{(Q_sh)(H)(Q_sk)(H)}
        \dd\gamma_N(H)
        \notag\\
    &\geq
    \int
        (Q_s\sqrt h)(H)
        (Q_s\sqrt k)(H)
        \dd\gamma_N(H)
        \notag\\
    &=
    \int
        \sqrt{h(H)}
        (Q_{2s}\sqrt k)(H)
        \dd\gamma_N(H),
    \label{eq:affinity-before-RHC}
\end{align}
where we have used self-adjointness and the semigroup property of
the Ornstein--Uhlenbeck noise in the last equality.

For \(2s=\log2\), \(a = \sqrt{h}\), and \(b = \sqrt{k}\),
the reverse-hypercontractive inequality
\eqref{eq:reverse-HC} implies
\begin{equation}\label{eq:RHC-applied}
    \Aff_{\gamma_N}(Q_sh,Q_sk)
    \geq
    \|\sqrt h\|_{L^{1/2}(\gamma_N)}
    \|\sqrt k\|_{L^{1/2}(\gamma_N)}.
\end{equation}

We next bound these two quasi-norms from below using the second moments.
Let \(a\) be a probability density with respect to \(\gamma_N\).
Decomposing
\[
    a
    =
    (a^{1/4})^{4/7}(a^2)^{3/7}
\]
and applying H\"older's inequality, we obtain
\begin{align*}
    1
    &=
    \int a(H)\dd\gamma_N(H)\\
    &\leq
    \left(
        \int a(H)^{1/4}\dd\gamma_N(H)
    \right)^{4/7}
    \left(
        \int a(H)^2\dd\gamma_N(H)
    \right)^{3/7}.
\end{align*}
Then,
\[
    \left(
        \int a(H)^{1/4}\dd\gamma_N(H)
    \right)^2
    \geq
    \left(
        \int a(H)^2\dd\gamma_N(H)
    \right)^{-3/2}.
\]
Since
\[
    \|\sqrt a\|_{L^{1/2}(\gamma_N)}
    =
    \left(
        \int a(H)^{1/4}\dd\gamma_N(H)
    \right)^2,
\]
we have
\begin{equation}\label{eq:Lhalf-L2-comparison}
    \|\sqrt a\|_{L^{1/2}(\gamma_N)}
    \geq
    \exp\left(
        -\frac32
        \log\|a\|_{L^2(\gamma_N)}^2
    \right).
\end{equation}

Applying \eqref{eq:Lhalf-L2-comparison} to \(h\) and \(k\) in
\eqref{eq:RHC-applied}, we obtain
\[
    \Aff_{\gamma_N}(Q_sh,Q_sk)
    \geq
    \exp\left(
        -\frac32(B_h+B_k)
    \right),
\]
and therefore
\[
    -2\log\Aff_{\gamma_N}(Q_sh,Q_sk)
    \leq
    3(B_h+B_k).
\]
Using \eqref{eq:Bh-bound} and \eqref{eq:Bk-bound}, 
and since 
\[
    Q_sh=Q_tJg,
    \qquad
    Q_sk=Q_tJ(Cg),
\]
we get
\begin{equation}\label{eq:Gaussian-Renyi}
    D_{1/2}(Q_tJg\|Q_tJ(Cg))
    \leq
    24\alpha^2B(g).
\end{equation}

It remains to return from Gaussian space to the Grassmannian. Let
\(a,b\) be nonnegative Gaussian densities. By the conditional
Cauchy--Schwarz inequality,
\[
    (J^*\sqrt{ab})(V)
    \leq
    \sqrt{(J^*a)(V)(J^*b)(V)}
\]
for \(\nu\)-almost every \(V\). Integrating with respect to Haar
measure,
\begin{align*}
    \Aff_\nu(J^*a,J^*b)
    &=
    \int_{\Gr(f,2f)}
        \sqrt{(J^*a)(V)(J^*b)(V)}
        \dd\nu(V)\\
    &\geq
    \int_{\Gr(f,2f)}
        (J^*\sqrt{ab})(V)\dd\nu(V)\\
    &=
    \int
        \sqrt{a(H)b(H)}
        \dd\gamma_N(H)\\
    &=
    \Aff_{\gamma_N}(a,b).
\end{align*}
Thus the conditional expectation can only increase 
affinity and therefore can only decrease \(D_{1/2}\), an instance of
R\'enyi data processing~\citep[Theorem~9]{vanerven2014renyi}.

For \(a=Q_tJg\), \(b=Q_tJ(Cg)\),
and \(K_\alpha=J^*Q_tJ\),
the inequality \eqref{eq:Gaussian-Renyi} implies
\[
    D_{1/2}(K_\alpha g\|K_\alpha Cg)
    \leq
    24\alpha^2B(g).
\]
\end{proof}

\begin{proof}[Proof of Proposition~\ref{prop:query-overview}]
The harmonic expansion in
\eqref{eq:query-restated-second-moment} was obtained in
\eqref{eq:harmonic-second-moment}. By the \(L^2\)-contractivity of \(J^*\)
in \(K_\alpha=J^*T_\alpha J\) and \cref{lem:Gaussian-posterior-contraction},
exponentiating \eqref{eq:lifted-moment}, we obtain
\[
    \|K_\alpha g\|_2^2
    \leq
    \exp\left(
        \frac{2\alpha^2}{1+\alpha^2}B(g)
    \right),
\]
which proves the remaining inequality in
\eqref{eq:query-restated-second-moment}.

The relative-entropy bound
\eqref{eq:query-restated-KL} follows directly from
\cref{prop:KL-contraction}.

Finally, after decreasing the universal constant \(\eta_0\) in
\cref{prop:planting-overview} if necessary, the calibration
\[
    \beta_*\alpha\leq\eta\leq\eta_0
\]
ensures that \(\alpha\le2^{-1/2}\). Hence
\cref{thm:Renyi-contraction} applies. Since
\[
    \alpha\leq\beta_*^{-1}\eta,
\]
we obtain
\[
    D_{1/2}(K_\alpha g\|K_\alpha Cg)
    \leq
    24\beta_*^{-2}\eta^2 B(g),
\]
which proves \eqref{eq:query-restated-Renyi}.
\end{proof}
\section{The multiple-observable extension}\label{sec:multiple}

In this section we prove the lower bound for the case of multiple observables.
The multiple-observable lower bound has two distinct additive terms. The
\emph{geometric term} is obtained by hiding one planted query among \(M-1\)
independent Haar decoys and asking the player to identify both its location and
its sign. The \emph{precision term} is obtained separately: sufficiently many
simultaneous post-measurement queries can be used to reconstruct a low-rank
projector state, reducing the problem to ordinary state tomography. We prove
these two bounds in sequence and then combine them.

\subsection{The hidden-needle information bound}

\begin{theorem}[Needle information budget]\label{thm:needle-budget}
Every \(T\)-copy strategy for the multiple-observable planted-sign game satisfies
\begin{equation}\label{eq:needle-budget}
    I(J,B;\lambda,z,W^M)
    \leq
    C\frac{\eta^2}{f}
    \left(
        RT+\frac{T(T-1)}{R}
    \right).
\end{equation}
\end{theorem}
\begin{proof}
Let \((\lambda,z)\) denote the refined measurement transcript used in the
source-side analysis, and let \(g_{\lambda,z}\) be the corresponding posterior
density of the latent subspace \(V\) relative to \(\nu\). After the measurement,
choose
\[
    (J,B)\sim \operatorname{Unif}([M]\times\{\pm1\}).
\]
Thus \((J,B)\) is independent of \((\lambda,z)\).

Fix a transcript \((\lambda,z)\) and a hypothesis \((j,b)\). Let
\(P_{\lambda,z}^{(j,b)}\) be the conditional law of
\[
    W^M=(W_1,\ldots,W_M)
\]
given \(J=j\) and \(B=b\), and let
\[
    Q:=\nu^{\otimes M}
\]
be the all-decoy law. Since only the \(j\)th coordinate is planted,
\[
    \frac{\mathrm{d}P_{\lambda,z}^{(j,+)}}{\mathrm{d}Q}(W^M)
    =(K_\alpha g_{\lambda,z})(W_j),
    \qquad
    \frac{\mathrm{d}P_{\lambda,z}^{(j,-)}}{\mathrm{d}Q}(W^M)
    =(K_\alpha Cg_{\lambda,z})(W_j).
\]
Consequently,
\[
    \KL\!\left(P_{\lambda,z}^{(j,+)}\middle\|Q\right)
    =\KL(K_\alpha g_{\lambda,z}\|1),
    \qquad
    \KL\!\left(P_{\lambda,z}^{(j,-)}\middle\|Q\right)
    =\KL(K_\alpha Cg_{\lambda,z}\|1).
\]
Since \(C\) preserves Haar measure,
\[
    B(Cg_{\lambda,z})=B(g_{\lambda,z}).
\]
Hence, by \cref{prop:KL-contraction}, uniformly in \((j,b)\),
\begin{equation}\label{eq:needle-one-hypothesis-kl}
    \KL\!\left(P_{\lambda,z}^{(j,b)}\middle\|Q\right)
    \leq C\eta^2 B(g_{\lambda,z}).
\end{equation}

Let
\[
    \overline P_{\lambda,z}
    :=
    \frac1{2M}
    \sum_{j=1}^M\sum_{b\in\{\pm1\}}
    P_{\lambda,z}^{(j,b)}
\]
denote the conditional law of \(W^M\) given the transcript \((\lambda,z)\),
after averaging over the uniformly random hidden pair \((J,B)\). By the
standard representation of mutual information as an average relative
entropy,
\[
    I(J,B;W^M\mid\lambda,z)
    =
    \frac1{2M}
    \sum_{j=1}^M\sum_{b\in\{\pm1\}}
    \KL\!\left(
        P_{\lambda,z}^{(j,b)}
        \middle\|
        \overline P_{\lambda,z}
    \right).
\]
For any reference law \(Q\), the standard KL decomposition
\citep[Chapter~2]{cover2006elements}
\[
    \frac1{2M}
    \sum_{j=1}^M\sum_{b\in\{\pm1\}}
    \KL\!\left(
        P_{\lambda,z}^{(j,b)}
        \middle\|Q
    \right)
    =
    I(J,B;W^M\mid\lambda,z)
    +
    \KL(\overline P_{\lambda,z}\|Q)
\]
therefore implies
\[
    I(J,B;W^M\mid\lambda,z)
    \leq
    \frac1{2M}
    \sum_{j=1}^M\sum_{b\in\{\pm1\}}
    \KL\!\left(
        P_{\lambda,z}^{(j,b)}
        \middle\|Q
    \right).
\]
Dropping the nonnegative \(\KL(\overline P_{\lambda,z}\|Q)\) and applying
\eqref{eq:needle-one-hypothesis-kl} together with
\eqref{eq:source-moment} from \cref{thm:source-moment} yields
\begin{equation}\label{eq:information-radius-bound}
    I(J,B;W^M\mid \lambda,z)
    \leq C\eta^2 B(g_{\lambda,z})
    \leq C\frac{\eta^2 c_R(\lambda)}{f}.
\end{equation}

Since \((J,B)\) is chosen after the measurement, it is independent of the
measurement transcript. Therefore
\[
    I(J,B;\lambda,z,W^M)
    =
    I(J,B;W^M\mid \lambda,z).
\]
Taking the average of \eqref{eq:information-radius-bound} over the transcript
\((\lambda, z)\) and applying \cref{prop:schur-casimir-average} completes 
the proof.
\end{proof}

\subsection{From simultaneous accuracy to needle recovery}

It remains to show that an accurate classical shadows protocol reveals the
hidden coordinate. 

For a fixed source subspace \(P\) and Haar-random \(W\), define
\[
    X_P(W):=\Tr(\rho_P A_W)
    =\frac{2}{R}\Tr(P\Pi_W)-1.
\]
This random variable has mean zero, \(\E[X_P(W)] = 0\), and is \(O(R^{-1/2})\)-Lipschitz on
\(\Gr(f,2f)\).
Fix an \(f\)-plane \(W_0\). Since \(UW_0\) is Haar-distributed
on \(\Gr(f,2f)\) whenever \(U\) is Haar in \(U(2f)\), we may
represent \(X_P(W)\) in distribution as
\[
    \widetilde X_P(U)
    =\frac{2}{R}\Tr(PU\Pi_{W_0}U^\dagger)-1.
\]
The map \(U\mapsto\widetilde X_P(U)\) is
\(4/\sqrt R\)-Lipschitz in Hilbert--Schmidt norm.
Thus \citet[Corollary~17]{meckes2013spectral} yields
\begin{equation}\label{eq:decoy-tail}
    \Pr\!\left[|X_P(W)|>u\right]
    \leq C e^{-cRfu^2}.
\end{equation}

On the other hand, for the planted coordinate \(W_J\), 
from \cref{thm:PT},
\[
    \Pr\!\left[B X_P(W_J)<\eta/4\right]
    \leq C e^{-cRf\eta^2}.
\]
Combining these two estimates and taking a union bound over the \(M-1\) decoys,
we obtain the following visibility event. Whenever
\begin{equation}\label{eq:visibility}
    \log(2M)\leq cRf\eta^2,
\end{equation}
then, after adjusting the universal constant \(c\), with probability at least
\(14/15\),
\begin{equation}\label{eq:needle-separation}
    B X_P(W_J)\geq \eta/4,
    \qquad
    \max_{i\ne J}|X_P(W_i)|\leq \eta/16.
\end{equation}

Set \(\eta=16\eps\). On the event \eqref{eq:needle-separation}, simultaneous
\(\eps\)-accurate estimates \(\widehat X_1,\ldots,\widehat X_M\) satisfy
\[
    |\widehat X_J|\geq 3\eps,
    \qquad
    \max_{i\ne J}|\widehat X_i|\leq 2\eps,
    \qquad
    \operatorname{sgn}(\widehat X_J)=B.
\]
Hence the deterministic rule
\begin{equation}\label{eq:needle-decoder}
    \widehat J\in\arg\max_i|\widehat X_i|,
    \qquad
    \widehat B=\operatorname{sgn}(\widehat X_{\widehat J})
\end{equation}
recovers \((J,B)\). A shadow protocol with failure probability at most \(1/3\)
therefore induces a needle decoder with success probability at least
\(\frac{3}{5}\).

For \(M\) larger than a universal constant, Fano's inequality
\citep[Sec.~2.10]{cover2006elements} implies
\[
    I(J,B;\lambda,z,W^M)
    \geq (1-p_e)\log(2M)-\log 2
    \geq c\log(2M).
\]
Together with \eqref{eq:needle-budget}, this implies
\begin{equation}\label{eq:needle-quadratic-constraint}
    RT+\frac{T(T-1)}{R}
    \geq c\frac{f\log(2M)}{\eta^2}.
\end{equation}
Solving the two possible branches of
\eqref{eq:needle-quadratic-constraint}, we get the fixed-rank geometric bound
\begin{equation}\label{eq:fixed-rank-multiple}
    T=\Omega\!\left(
    \min\left\{
        \frac{f\log(2M)}{\eta^2R},
        \frac{\sqrt{Rf\log(2M)}}{\eta}
    \right\}\right).
\end{equation}
For bounded \(M\), the same conclusion up to universal constants follows from
the single-observable constant-confidence lower bound.

In particular, if
\begin{equation}\label{eq:multiple-square-root-branch}
    \eta^2R^3\lesssim f\log(2M),
\end{equation}
then the square-root term in \eqref{eq:fixed-rank-multiple} is smaller, 
hence
\begin{equation}\label{eq:fixed-rank-multiple-sqrt}
    T=\Omega\!\left(
        \frac{\sqrt{Rf\log(2M)}}{\eta}
    \right).
\end{equation}

\subsection{Precision via a tomography reduction}

The hidden-needle argument yields the geometric lower bound.
The independent precision term follows by constructing a fixed
list of observables whose simultaneous expectation values
determine a low-rank projector state to trace-norm accuracy
\(O(\eps)\), and then invoking the optimal projector-tomography
lower bound of \citet{scharnhorst2025optimal}.
A similar net-based reduction appears in
\citet[Appendix~A]{Grier_2024}.

\begin{proposition}[Multiple-observable precision]
\label{prop:multiple-precision}
For \(d\ge2\), \(1\le r\le d\), \(1\le F\le d\),
\(M\ge1\), \(0<\delta\le1/3\), and sufficiently small
\(\eps>0\), every successful protocol satisfies
\begin{equation}\label{eq:multiple-precision}
    T=\Omega\!\left(
        \frac{\min\{\log(2M),\,d\min\{r,F\}\}}{\eps^2}
    \right).
\end{equation}
\end{proposition}

\begin{proof}
Set
\[
    q:=\min\{r,F\},
    \qquad
    \ell_0:=\min\{\log(2M),c_0dq\},
\]
where \(c_0>0\) is a sufficiently small universal
constant. If \(\ell_0=O(1)\), the claim follows
from the usual two-state testing lower bound.

We first choose an effective dimension \(n\le d\)
and rank \(s\le q\). Let \(r_0,c_1\) be the constants
in \citet[Theorem~5.1]{scharnhorst2025optimal},
and set \(L=\kappa\ell_0\) for a sufficiently small
universal constant \(\kappa>0\). Choose
\[
    (n,s)=
    \begin{cases}
      (\min\{d,\lfloor L\rfloor\},1),
          & L<r_0d,\\[1mm]
      (d,\lfloor L/d\rfloor),
          & L\ge r_0d.
    \end{cases}
\]
For sufficiently small \(\kappa,c_0\), these
choices satisfy
\begin{equation}\label{eq:precision-sn-choice}
    1\le s\le q,\qquad
    2s\le n\le d,\qquad
    sn\asymp\ell_0,
\end{equation}
and, whenever \(s>1\), also
\(r_0\le s\le c_1n\).

Consider the class
\[
    \mathcal A_{s,n}
    :=
    \{O=O^\dagger:
      \rank(O)\le2s,\ \|O\|_{\op}\le1\}
\]
of Hermitian operators on \(\C^n\).
A standard volumetric argument \cite{szarek1998metric} provides
\[
    N(\zeta,\mathcal A_{s,n},\|\cdot\|_{\op})
    \le (1+6/\zeta)^{4ns+2s},
    \qquad 0<\zeta\le1.
\]
Indeed, denote \(O=V\diag(a)V^\dagger\),
where \(V\) has \(2s\) orthonormal columns
and \(a\in[-1,1]^{2s}\). Approximating \(V\)
in operator norm and \(a\) in \(\ell_\infty\)-norm,
each to accuracy \(\zeta/3\), gives an operator
approximation of accuracy \(\zeta\).

Consequently, there is a \(1/4\)-net
\(\mathcal O\subseteq\mathcal A_{s,n}\) satisfying
\[
    |\mathcal O|\le e^{Csn}\le M,
\]
where the last inequality follows by taking
\(\kappa\) sufficiently small.

We work in a fixed copy of \(\C^n\subseteq\C^d\)
and restrict the unknown state to
\[
    \rho_P=P/s,
    \qquad P\in\Gr(s,n).
\]
Extend each \(O\in\mathcal O\) by zero to \(\C^d\),
and define
\[
    \widetilde O
    :=\frac12\left(O-\frac{\Tr O}{d}I_d\right).
\]
Since \(\rank(O)\le2s\) and \(\|O\|_{\op}\le1\),
\[
    \|\widetilde O\|_{\op}\le1,
    \qquad
    \|\widetilde O\|_F^2
    \le\frac14\Tr(O^2)
    \le\frac{s}{2}\le F.
\]
Thus every \(\widetilde O\) belongs to
\(\Obs_d(F)\). We pad the list with repetitions
if necessary to obtain exactly \(M\) observables.

Suppose the shadow protocol succeeds on this
fixed list.
From its estimates \(\widehat y_O\)
of \(\Tr(\widetilde O\rho_P)\), define
\[
    \widehat x_O
    :=2\widehat y_O+\frac{\Tr O}{d}.
\]
Then
\[
    |\widehat x_O-\Tr(O\rho_P)|\le2\eps
    \qquad (O\in\mathcal O).
\]

We now use these estimated expectation values to reconstruct the state. Set
\[
    \widehat P\in
    \arg\min_{Q\in\Gr(s,n)}
    \max_{O\in\mathcal O}
    \left|
        \Tr\left(O\frac{Q}{s}\right)
        -\widehat x_O
    \right|,
    \qquad
    \widehat\rho_P:=\widehat P/s.
\]
Since \(P\) is an admissible candidate, the
minimum is at most \(2\eps\). Therefore
\[
    |\Tr(O(\rho_P-\widehat\rho_P))|
    \le4\eps
    \qquad (O\in\mathcal O).
\]

It remains to convert this uniform control on the net into trace-norm
accuracy. Set \(\Delta:=\rho_P-\widehat\rho_P\).
Since \(\rank(\Delta)\le2s\),
the operator \(\operatorname{sgn}(\Delta)\)
belongs to \(\mathcal A_{s,n}\).
Choose \(O_\Delta\in\mathcal O\) with
\[
    \|O_\Delta-\operatorname{sgn}(\Delta)\|_{\op}
    \le\frac14.
\]
By trace-norm duality,
\[
\begin{aligned}
    \|\Delta\|_1
    &=\Tr(\operatorname{sgn}(\Delta)\Delta)\\
    &\le|\Tr(O_\Delta\Delta)|
      +\frac14\|\Delta\|_1\\
    &\le4\eps+\frac14\|\Delta\|_1.
\end{aligned}
\]
Hence
\begin{equation}\label{eq:tomography-from-net}
    \|\rho_P-\widehat\rho_P\|_1
    \le\frac{16}{3}\eps.
\end{equation}

We have thus obtained a tomography procedure
for rank-\(s\) projector states in dimension \(n\),
with trace-distance error at most \(8\eps/3\)
and success probability at least \(2/3\).
By \citet[Lemma~2.12]{scharnhorst2025optimal},
the success probability can be increased to \(0.99\)
using \(O(T)\) copies, with trace-distance error
at most \(8\eps\).

For \(s\ge r_0\), we apply
\citet[Theorem~5.1]{scharnhorst2025optimal};
for \(s=1\), we use the trace-distance consequence
of \citet[Proposition~3.1]{scharnhorst2025optimal}.
For sufficiently small universal \(\eps\),
both yield
\[
    T=\Omega\left(\frac{sn}{\eps^2}\right)
     =\Omega\left(\frac{\ell_0}{\eps^2}\right).
\]
Since
\[
    \ell_0
    \asymp
    \min\{\log(2M),d\min\{r,F\}\},
\]
this proves the proposition.
\end{proof}

\subsection{Rank optimization and completion of the proof}

We first optimize the lower bound obtained from the multiple-observable game.
Set
\[
    \ell_M:=\log(2M).
\]
After choosing the constant in \eqref{eq:moderate-M} sufficiently small
relative to \(c_0\), the cap in \cref{prop:multiple-precision} is inactive, 
and therefore
\begin{equation}\label{eq:multiple-precision-uncapped}
    T=\Omega\!\left(\frac{\ell_M}{\eps^2}\right).
\end{equation}
As in the parameter-range reduction of \cref{subsec:moderate-error-overview}, 
this already dominates the geometric term unless
\begin{equation}\label{eq:multiple-moderate-error}
    \ell_M
    \leq c_*\eps^2F\min\{r,F\},
\end{equation}
where \(c_*>0\) is a sufficiently small universal constant. We may therefore
assume \eqref{eq:multiple-moderate-error} for the remainder of the geometric
argument.

Recall that \(f=\Theta(F)\) and set \(\eta=16\eps\). Choose the effective rank
\begin{equation}\label{eq:multiple-effective-rank}
    R\asymp
    \min\left\{
        r,
        \left(\frac{f\ell_M}{\eta^2}\right)^{1/3}
    \right\}.
\end{equation}
The same parameter check as in \cref{sec:overview-ranges} shows, 
after decreasing \(c_*\) if necessary, that
\[
    R\leq \frac f2,
    \qquad
    \eta^2R^3\lesssim f\ell_M,
    \qquad
    \ell_M\lesssim Rf\eta^2.
\]
The last inequality is exactly the visibility condition
\eqref{eq:visibility}, while the second places us in the square-root branch of
the fixed-rank bound. Hence \eqref{eq:fixed-rank-multiple-sqrt} yields
\[
    T
    =\Omega\!\left(
        \frac{\sqrt{Rf\ell_M}}{\eta}
    \right).
\]
Substituting \eqref{eq:multiple-effective-rank} and using
\(f=\Theta(F)\) and \(\eta=\Theta(\eps)\) yields
\begin{equation}\label{eq:multiple-geometric-final}
    T=\Omega\!\left(
        \min\left\{
            \frac{\sqrt{rF\ell_M}}{\eps},
            \frac{(F\ell_M)^{2/3}}{\eps^{4/3}}
        \right\}
    \right).
\end{equation}

Combining \eqref{eq:multiple-geometric-final} with the independent precision
bound \eqref{eq:multiple-precision-uncapped}, and using
\(\max\{a,b\}\geq(a+b)/2\), we obtain
\begin{equation}\label{eq:multiple-M-lower-bound}
    T=\Omega\!\left(
        \min\left\{
            \frac{\sqrt{rF\ell_M}}{\eps},
            \frac{(F\ell_M)^{2/3}}{\eps^{4/3}}
        \right\}
        +\frac{\ell_M}{\eps^2}
    \right).
\end{equation}

We can finally prove the multiple-observable version of \cref{thm:master}.
\begin{proof}[Proof of \cref{thm:master}]
Set
\[
    \ell_\delta:=\log(1/\delta),
    \qquad
    \ell:=\max\{\ell_M,\ell_\delta\}.
\]
The preceding argument gives the \(M\)-dependent bound
\eqref{eq:multiple-M-lower-bound}.

For the confidence-dependent obstruction, apply the protocol to a list
consisting of \(M\) identical copies of a single planted observable. The
single-observable high-confidence lower bound, together with the corresponding
two-state precision bound, gives
\[
    T=\Omega\!\left(
        \min\left\{
            \frac{\sqrt{rF\ell_\delta}}{\eps},
            \frac{(F\ell_\delta)^{2/3}}{\eps^{4/3}}
        \right\}
        +\frac{\ell_\delta}{\eps^2}
    \right).
\]

Since the right-hand side is nondecreasing in the logarithmic parameter,
combining this with \cref{eq:multiple-M-lower-bound} and recalling that
\[
    \ell=\max\{\ell_M,\ell_\delta\}
\]
gives
\[
    T=\Omega\!\left(
        \min\left\{
            \frac{\sqrt{rF\ell}}{\eps},
            \frac{(F\ell)^{2/3}}{\eps^{4/3}}
        \right\}
        +\frac{\ell}{\eps^2}
    \right),
\]
which proves \eqref{eq:master-lb}.
\end{proof}

\appendix
\section*{Appendices}
\section{Bias and the optimal sign decision}\label{app:bias}

In this appendix we derive the bias of the 
binary testing problem corresponding to the optimal
decision rule of the sign game \cref{def:sign-game}. This is the
standard equal-prior binary testing identity; see
\citet{fuchs2002cryptographic} for the classical distinguishability measures.
Let \(b \sim \mathrm{Unif}\{-1, +1\}\), and let 
\(\mathsf P_\pm\) denote the 
observation laws conditional on \(b=\pm1\). Recall that,
for a decision rule \(\widehat b\in\{\pm1\}\), 
\[
\Bias(\widehat b):=\E[b\widehat b]
=2\Pr[\widehat b=b]-1.
\]
Suppose that \(\mathsf P_\pm\) have densities \(p_\pm\) with 
respect to a common reference measure. Maximizing the pointwise 
success probability, we obtain the likelihood-ratio rule
\[
\widehat b^{\,*}
=
\operatorname{sgn}(p_+-p_-),
\]
with an arbitrary choice on the set where \(p_+=p_-\). Hence
\[
\sup_{\widehat b}\Pr[\widehat b=b]
=
\frac12+\frac14\int |p_+-p_-|,
\]
and therefore
\begin{equation}\label{eq:bias-TV-app}
\sup_{\widehat b}\Bias(\widehat b)
=
\frac12\int |p_+-p_-|
=
\TV(\mathsf P_+,\mathsf P_-).
\end{equation}

We now specialize this identity to the Grassmannian game. 
Let \(S\) denote the transcript and let \(q(\mathrm{d}s)\) be its 
marginal law, which is the same under both signs. Let \(\nu\) denote the
Haar measure on the balanced Grassmannian. For
\(q\)-almost every \(s\), let \(g_s\) denote the posterior density of the
hidden subspace \(V\) given \(S=s\), so that
\[
\Pr[V\in \mathrm{d}V\mid S=s]
=
g_s(V)\,\nu(\mathrm{d}V).
\]
This posterior is independent of the sign. Define the complement operator by
\[
(Cg)(V):=g(V^\perp).
\]

Conditional on \(S=s\) and \(b=+1\), the planted query \(W\) 
has density \(K_\alpha g_s\) with respect to \(\nu\). 
Indeed, for every bounded test function \(a\),
\[
\begin{aligned}
    \E[a(W)\mid S=s,b=+1]
    &=
    \int g_s(V)(K_\alpha a)(V)\,\nu(\mathrm{d}V)\\
    &=
    \int a(W)(K_\alpha g_s)(W)\,\nu(\mathrm{d}W),
\end{aligned}
\]
where the second equality uses self-adjointness 
of \(K_\alpha\). For \(b=-1\), replacing the planted 
subspace \(V\) by \(V^\perp\) and using invariance 
of \(\nu\) under complementation gives the conditional 
density \(K_\alpha Cg_s\).

Thus, relative to the common measure \(q(\mathrm{d}s)\nu(\mathrm{d}W)\), 
the two observation laws of \((S,W)\) have densities
\[
p_+(s,W)=K_\alpha g_s(W),
\qquad
p_-(s,W)=K_\alpha Cg_s(W).
\]
Applying \eqref{eq:bias-TV-app} and averaging over 
the common transcript \(s\) yields
\begin{equation}\label{eq:conditional-bias}
\sup_{\widehat b}\Bias(\widehat b)
=
\frac12\E_S
\int
\left|
K_\alpha g_S(W)-K_\alpha Cg_S(W)
\right|
\,\nu(\mathrm{d}W).
\end{equation}
In particular, the optimal decision is
\[
\widehat b^{\,*}(S,W)
=
\operatorname{sgn}\!\left(
K_\alpha g_S(W)-K_\alpha Cg_S(W)
\right).
\]
Hence the sign rule used in \cref{sec:overview} is 
the Bayes-optimal decision.
\section{Proof of the exact Schur--Casimir average}
\label{app:schur-casimir-average-proof}

In this appendix we prove \cref{prop:schur-casimir-average}. The argument 
is a standard Schur--Weyl character calculation. See, e.g.,
\cite{fulton2013representation,Sagan2001}.

\begin{proof}
For a partition \(\lambda\vdash T\), define the total content of its Young diagram by
\[
    \kappa_\lambda
    :=
    \sum_{(i,j)\in\lambda}(j-i).
\]
We may express the quadratic \(U(R)\) Casimir parameter in terms of \(\kappa_\lambda\).
First observe:
\begin{align*}
c_R(\lambda)
&=
\sum_{i=1}^R
\lambda_i(\lambda_i+R+1-2i) \\
&=
RT
+
\sum_{i=1}^R
\lambda_i(\lambda_i+1-2i).
\end{align*}
On the other hand, we may express
\[
    2\kappa_\lambda
    =
    2\sum_{i=1}^R
    \sum_{j=1}^{\lambda_i}(j-i)
    =
    \sum_{i=1}^R
    \lambda_i(\lambda_i+1-2i).
\]
Therefore
\begin{equation}\label{eq:Casimir-content}
c_R(\lambda)
=
RT+2\kappa_\lambda.
\end{equation}
The statement of the proposition is true for \(T = 1\), since
in this case \(\kappa_\lambda = 0\). Thus we may assume
\(T \geq 2\).

It remains to compute the expectation of \(\kappa_\lambda\) under the
Schur--Weyl distribution
\[
    p_\lambda
    =
    \frac{\dim[\lambda]\,D_{\lambda,R}}{R^T}.
\]
To do so, consider the sum of all transpositions in the group algebra
\[
    \Omega_T
    :=
    \sum_{1\leq a<b\leq T}(ab)
    \in\C[S_T].
\]

This element is central in \(\C[S_T]\), since it is the sum over the
conjugacy class of transpositions. Hence, on each irreducible
\(S_T\)-module \([\lambda]\), it acts as a scalar. The standard content
formula identifies this scalar with \(\kappa_\lambda\)
\cite{Sagan2001}:
\[
    \Omega_T\big|_{[\lambda]}
    =
    \kappa_\lambda I_{[\lambda]}.
\]
Taking traces on \([\lambda]\), we obtain
\[
    \kappa_\lambda\dim[\lambda]
    =
    \Tr_{[\lambda]}\Omega_T.
\]
All transpositions have the same character value, 
which can be taken to be \(\chi_\lambda((12))\)
and there are \(\binom{T}{2}\) of them, hence
\[
    \Tr_{[\lambda]}\Omega_T
    =
    \binom{T}{2}\chi_\lambda((12)).
\]
Thus, for \(T \geq 2\),
\begin{equation}\label{eq:content-character}
\frac{\chi_\lambda((12))}{\dim[\lambda]}
=
\frac{2\kappa_\lambda}{T(T-1)}.
\end{equation}

We now average the left-hand side of
\eqref{eq:content-character} over 
the Schur--Weyl distribution:
\begin{align}
\E_\lambda
\frac{\chi_\lambda((12))}{\dim[\lambda]}
&=
\frac{1}{R^T}
\sum_{\lambda\vdash T}
D_{\lambda,R}\chi_\lambda((12)).
\label{eq:character-average}
\end{align}

Using Schur--Weyl duality, we decompose 
\[
    (\C^R)^{\otimes T}
    =
    \bigoplus_{\lambda\vdash T}
    V_{\lambda,R}\otimes[\lambda],
\]
where terms with \(\ell(\lambda)>R\) vanish. The symmetric group acts
only on the Specht factor, so the sum on the right-hand side of
\eqref{eq:character-average} is exactly \(\Tr_{(\C^R)^{\otimes T}}(12)\).

This trace can be computed directly in the standard tensor-product
basis. A basis tensor
\[
    e_{i_1}\otimes e_{i_2}\otimes\cdots\otimes e_{i_T}
\]
is fixed by \((12)\) exactly when \(i_1=i_2\). There are \(R\) choices
for this common index and \(R^{T-2}\) choices for the remaining indices.
Hence
\[
    \Tr_{(\C^R)^{\otimes T}}(12)
    =
    R^{T-1}.
\]
Substituting this into \eqref{eq:character-average}, we get
\[
    \E_\lambda
    \frac{\chi_\lambda((12))}{\dim[\lambda]}
    =
    \frac1R.
\]
Combining with \eqref{eq:content-character}, we obtain
\[
    \E_\lambda\,2\kappa_\lambda
    =
    \frac{T(T-1)}{R}.
\]
Finally, taking expectations in \eqref{eq:Casimir-content} results in
\[
    \E_\lambda c_R(\lambda)
    =
    RT+\frac{T(T-1)}{R},
\]
which proves the proposition.

\end{proof}
\section{Reduction to traceless Hermitian Gaussians}
\label{app:traceless-gaussian-reduction}

Let
\[
\Herm_0(N):=\{H\in\Herm(N):\Tr H=0\}.
\]
Any \(H\in\Herm(N)\) may be decomposed orthogonally as
\[
H
=
H_0+\frac{\Tr H}{N}I,
\qquad
H_0:=H-\frac{\Tr H}{N}I,
\]
corresponding to
\[
\Herm(N)
=
\Herm_0(N)
\oplus
\R\,\frac{I}{\sqrt N}.
\]

If \(H\sim\gamma_N\), then the two orthogonal components have 
independent Gaussian distributions. In particular, with
\[
\xi:=\frac{\Tr H}{\sqrt N},
\]
we have
\[
\xi\sim N(0,1),
\qquad
H_0\sim\gamma_N^0,
\]
where \(\gamma_N^0\) is the standard Gaussian measure on
\(\Herm_0(N)\), and \(\xi\) is independent of \(H_0\).

Adding a scalar multiple of the identity shifts all eigenvalues by the
same amount and leaves the eigenspaces unchanged. Hence
\[
\pi(H)=\pi(H_0).
\]

The same reduction is compatible with Ornstein--Uhlenbeck noise. If
\[
H'=\alpha H+\sqrt{1-\alpha^2}\,G,
\qquad
H,G\overset{\mathrm{i.i.d.}}{\sim}\gamma_N,
\]
then the traceless component of \(H'\) is
\[
H_0'
=
\alpha H_0+\sqrt{1-\alpha^2}\,G_0.
\]
Therefore
\[
\pi(H')=\pi(H_0').
\]
Consequently, the Gaussian planting kernel obtained after projection by
\(\pi\) depends only on the traceless Gaussian components.
\section{Gaussian moment identities}
\label{app:Gaussian-moment-identities}

In this appendix, for completeness, we provide proofs of 
two elementary Gaussian moment calculations.

\paragraph{Second moment of the Haar projector.}
Let \(\nu\) be Haar probability measure on \(\Gr(f,N)\) and
\(V\sim\nu\). Denote \(\Pi:=\Pi_V\). This second moment is a special case of Haar
integration~\citep{collins2006integration}; the short derivation below uses only
unitary invariance and Schur--Weyl duality~\citep{fulton2013representation}. Thus,
\[
\E[\Pi\otimes\Pi]
=
a\,I_{N^2}+b\,F,
\]
where \(F(x\otimes y)=y\otimes x\) is the swap operator. 
Taking the trace
and the trace after multiplication by \(F\) yields
\[
\begin{aligned}
f^2
&=
\E[(\Tr\Pi)^2]
=
aN^2+bN,
\\
f
&=
\E[\Tr(\Pi^2)]
=
aN+bN^2.
\end{aligned}
\]
Then
\[
b
=
\frac{f(N-f)}{N(N^2-1)}.
\]
Hence, for traceless Hermitian \(A,B\),
\[
\E\!\left[\Tr(A\Pi)\Tr(B\Pi)\right]
=
\frac{f(N-f)}{N(N^2-1)}\Tr(AB).
\]
Since \(A\) is traceless,
\[
h_A(V)
=
\Tr\!\left(A(2\Pi_V-I_N)\right)
=
2\Tr(A\Pi_V).
\]
Consequently,
\[
\langle h_A,h_B\rangle_{L^2(\nu)}
=
4\frac{f(N-f)}{N(N^2-1)}\Tr(AB).
\]
In the balanced case \(f=N/2\), this simplifies to
\begin{equation}\label{eq:Haar-projector-second-moment}
\langle h_A,h_B\rangle_{L^2(\nu)}
=
\frac{N}{N^2-1}\Tr(AB).
\end{equation}

\paragraph{Moments of the traceless Hermitian Gaussian.}
Let \(H\sim\gamma_N\) be the standard Gaussian on \(\Herm(N)\), and
decompose
\[
H=G+sI_N,
\qquad
s:=\frac{\Tr H}{N}.
\]
Then \(G\) is the standard Gaussian on \(\Herm_0(N)\),
\(s\sim N(0,1/N)\), and \(G\) and \(s\) are independent. In particular,
\[
\E\Tr G^2=N^2-1.
\]

We now compute the fourth moment.
We may express the covariance of \(H\) in 
matrix entries:
\[
\E[H_{ij}H_{kl}]
=
\delta_{il}\delta_{jk}.
\]
By the Gaussian pairing formula (Wick's rule)~\citep{isserlis1918formula},
\[
\begin{aligned}
\E\Tr H^4
&=
\sum_{i,j,k,l}
\E[H_{ij}H_{jk}H_{kl}H_{li}]
\\
&=
N^3+N+N^3
\\
&=
2N^3+N.
\end{aligned}
\]
On the other hand,
\[
\Tr(H^4)
=
\Tr(G^4)
+
4s\Tr(G^3)
+
6s^2\Tr(G^2)
+
Ns^4.
\]
Since \(\Tr G=0\), the term \(4s^3\Tr G\) vanishes.
Taking expectations and using independence and centering,
\[
\E\Tr H^4
=
\E\Tr G^4
+
6\E[s^2]\E\Tr G^2
+
N\E[s^4].
\]
Since
\[
\E[s^2]=\frac1N,
\qquad
\E[s^4]=\frac{3}{N^2},
\]
we obtain
\begin{align}
\E\Tr G^4
&=
2N^3+N
-
\frac{6}{N}(N^2-1)
-
\frac{3}{N}
\nonumber\\
&=
2N^3-5N+\frac3N.
\label{eq:traceless-Gaussian-fourth-moment}
\end{align}
In particular,
\[
\E\Tr G^4\leq 2N^3.
\]

\section{Extension to arbitrarily large lists}
\label{app:large-M}

The moderate-list assumption \eqref{eq:moderate-M} in
\cref{thm:master} can be removed at the price of capping
the logarithmic dependence on \(M\).

Indeed, suppose that
\[
    \log(2M)>c_M d\min\{r,F\},
\]
and let \(M_*\leq M\) be the largest integer satisfying
\[
    \log(2M_*)
    \leq c_M d\min\{r,F\}.
\]
Whenever such an integer exists, we may apply
\cref{thm:master} to a list of \(M_*\) observables and
pad the remaining \(M-M_*\) positions by repetitions
of observables already in the list. Any protocol
that succeeds simultaneously on all \(M\) observables
must also succeed on the original \(M_*\) observables.

Consequently, up to universal constants, the lower
bound \eqref{eq:master-lb} holds for arbitrary \(M\)
with \(\ell\) replaced by
\[
    \ell_{\mathrm{cap}}
    :=
    \max\left\{
        \min\left\{
            \log(2M),\,c_M d\min\{r,F\}
        \right\},
        \log\frac1\delta
    \right\}.
\]
The bounded-parameter case in which no admissible
\(M_*\) exists is covered, after adjusting constants,
by the independent two-state precision bound.

For fixed confidence, the logarithmic dependence
therefore persists up to
\[
    \log M=\Theta\bigl(d\min\{r,F\}\bigr),
\]
at which point the precision term reaches the scale
\[
    \frac{d\min\{r,F\}}{\eps^2}.
\]
Since \(rF\leq d\min\{r,F\}\), the geometric term
is dominated by precision at this scale.
The resulting cap is consistent with the tomography
benchmark \(O(dr/\eps^2)\)
~\citep{10.1145/2897518.2897544}.

We emphasize that the moderate-list assumption
\eqref{eq:moderate-M} is needed to obtain the
uncapped \(\log M/\eps^2\) precision bound.
It is distinct from the moderate-error condition
\eqref{eq:overview-window} used in the geometric
argument. Outside the latter regime, the precision
lower bound already dominates the geometric target
up to universal constants.%

\bibliographystyle{unsrtnat}
\bibliography{references}

\end{document}